\documentclass[10pt,journal,compsoc]{IEEEtran}

\ifCLASSINFOpdf
 \usepackage[pdftex]{graphicx}
  
\else
\fi

\usepackage{amsmath}
\usepackage{amsfonts}
\usepackage{color}
\usepackage{cite}
\usepackage{caption}
\usepackage{url}
\usepackage{multirow}
\usepackage{amssymb}
\usepackage{pifont}
\renewcommand{\arraystretch}{1.0}

\ifCLASSOPTIONcompsoc
 \usepackage[caption=false,font=normalsize,labelfont=sf,textfont=sf]{subfig}
\else
 \usepackage[caption=false,font=footnotesize]{subfig}
\fi

\ifCLASSOPTIONcaptionsoff
 \usepackage[nomarkers]{endfloat}
 \let\MYoriglatexcaption\caption
 \renewcommand{\caption}[2][\relax]{\MYoriglatexcaption[#2]{#2}}
\fi

\begin{document}

\title{System Log Parsing: A Survey}

\author{Tianzhu~Zhang,
	 Han Qiu, 
	 Gabriele Castellano,
	 Myriana Rifai,
	 Chung Shue Chen,
	 and~Fabio Pianese 
\thanks{T. Zhang, C. S. Chen, and F. Pianese are with Nokia Bell Labs, 91620 Nozay, France. (Emails: \{firstname.lastname\}@nokia-bell-labs.com).} 
\thanks{G. Castellano and M. Rifai were affiliated with Nokia Bell Labs during the time of writing. (Email:  gabry.c92x@gmail.com, Myriana.rifai16@gmail.com) }
\thanks{H. Qiu (corresponding author) is with Institute for Network Sciences and Cyberspace, BNRist, Tsinghua University, Beijing 100084, China. (Email: qiuhan@tsinghua.edu.cn)}
}

\IEEEtitleabstractindextext{

\begin{abstract}
Modern information and communication systems have become increasingly challenging to manage.
The ubiquitous system logs contain plentiful information and are thus widely exploited as an alternative source for system management. As log files usually encompass large amounts of raw data, manually analyzing them is laborious and error-prone. Consequently, many research endeavors have been devoted to automatic log analysis. However, these works typically expect structured input and struggle with the heterogeneous nature of raw system logs. Log parsing closes this gap by converting the unstructured system logs to structured records. Many parsers were proposed during the last decades to accommodate various log analysis applications. However, due to the ample solution space and lack of systematic evaluation, it is not easy for practitioners to find ready-made solutions that fit their needs.

This paper aims to provide a comprehensive survey on log parsing. We begin with an exhaustive taxonomy of existing log parsers. Then we empirically analyze the critical performance and operational features for 17 open-source solutions both quantitatively and qualitatively, and whenever applicable discuss the merits of alternative approaches. We also elaborate on future challenges and discuss the relevant research directions. We envision this survey as a helpful resource for system administrators and domain experts to choose the most desirable open-source solution or implement new ones based on application-specific requirements.
\end{abstract}

}

\maketitle

 \ifCLASSOPTIONpeerreview
 \begin{center} \bfseries EDICS Category: 3-BBND \end{center}
 \fi

\IEEEpeerreviewmaketitle

\section{Introduction}

\IEEEPARstart{W}{ith} the proliferation of the Internet of Things (IoT), Cloud/Edge computing, Industry 4.0, and Fifth-generation mobile networks (5G), modern computing and communication systems commonly incorporate a large variety of (distributed) components to provide diversified services with guaranteed performance~\cite{hwang2013distributed}. Consequently, they have become increasingly complex and burdensome to manage. 
System administrators traditionally resort to runtime analysis such as code instrumentation and profiling for execution monitoring and problem diagnosis, but these techniques are non-trivial to configure and can incur non-negligible overhead in production environment~\cite{jiang2008abstracting}. 
Alternatively, many research endeavors seek to explore system logs to accomplish the same tasks in a much less intrusive manner. Indeed, since the advent of the BSD Syslog protocol~\cite{lonvick2001bsd} in the 1980s, information and communications technology (ICT) systems have widely employed log files to keep track of the execution states and system events at runtime.
Log files usually contain rich runtime information that system administrators and domain experts can leverage to perform advanced analytics and are thus deemed a fundamental building block for the development, maintenance, and troubleshooting of the modern systems~\cite{fu2014developers}.

\begin{figure}[!tb]
\begin{center}
\includegraphics[width=0.48\textwidth]{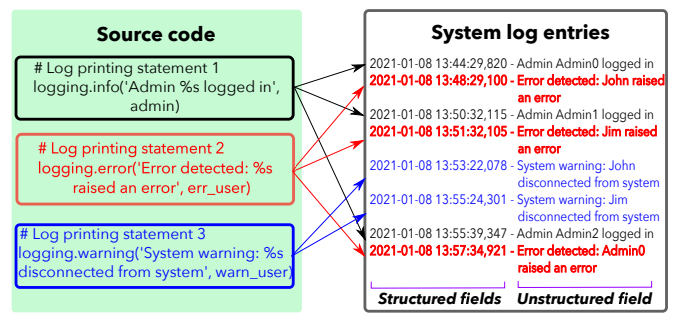}
\caption{A sample snippet of raw log entries along with the corresponding log printing statements in the source code.}
\label{fig:example}
\end{center}
\end{figure}

However, nowadays, as the volume, velocity, and variety of system logs keep exploding, manually inspecting log messages is mostly impractical~\cite{mi2013toward,vaarandi2014using,hamooni2016logmine}.
Existing log management and analytic tools~\cite{splunk,prelert,logstash,sumo,elasticsearch,logrhythm,logsurfer,loggly,logentries,graylog,ossim,braun2016syslog} follow the classic Expert Systems approach~\cite{jackson1986introduction}, which heavily relies on manually composing regular expressions or customized rules to filter the log messages of interest. This approach requires not only a thorough understanding of the system internals but also continuous maintenance over system upgrades~\cite{xu2010system}. For example, composing sufficiently accurate rules for Logsurfer~\cite{logsurfer} can incur a steep overhead~\cite{prewett2003analyzing}. Some network and security service providers even have to operate large data engineering teams to manage the composed rules~\cite{mclean2020adaptive}. 

To tackle this challenge, many end-to-end log analysis frameworks have been proposed. These frameworks employ a multitude of data mining and statistical analysis techniques to extract insights from system events~\cite{tang2010logtree,van2008process,de2005web}.
In particular, with the rise of Artificial Intelligence (AI) and Machine Learning (ML) over the recent decades, there has been considerable activities towards enhancing IT operations analytics using AI/ML techniques (i.e., AIOps), which heavily rely on system logs to collect observational data~\cite{dai2020logram,chu2021prefix,tao2021logstamp}. 
Despite the optimistic outcomes, these frameworks mostly expect the input logs to have a normalized format (e.g., event types, message signatures, vectors, matrices)~\cite{tan2008salsa,hellerstein2002discovering,peng2007event,beschastnikh2011leveraging}. Nonetheless, raw log generally record runtime system events, e.g., operations, warnings, and errors, as single-line or multi-line textual messages whose formats are solely decided by the log printing statements (e.g., {\em log.info(), printf()}). 
An illustrative example of 8 interleaved system log entries originated from 3 separate log printing statements in the program source code is shown in Fig.~\ref{fig:example}. Each entry consists of a timestamp and a free-text message with no event type or message signature. Such an issue is more obvious in large-scale systems~\cite{li2005integrated}. 
Although there have been some efforts towards log format unification~\cite{luotonen1995common,hallam1996extended,team2016apache,bae2020improving,ogle2004canonical,topol2003automating}, most existing systems still generate log entries as unstructured (or semi-structured), free-text messages.

\begin{figure}[!tb]
\begin{center}
\includegraphics[width=0.49\textwidth]{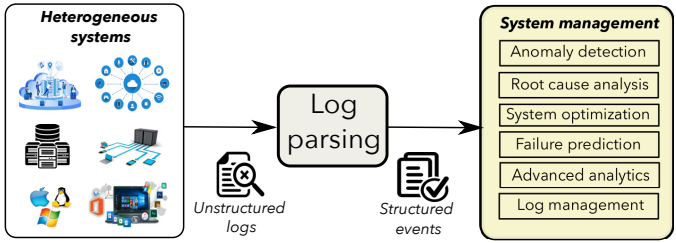}
\caption{The role of log parsing for system management.}
\label{fig:workflow}
\end{center}
\end{figure}

\begin{figure}[!tb]
\begin{center}
\includegraphics[width=0.48\textwidth]{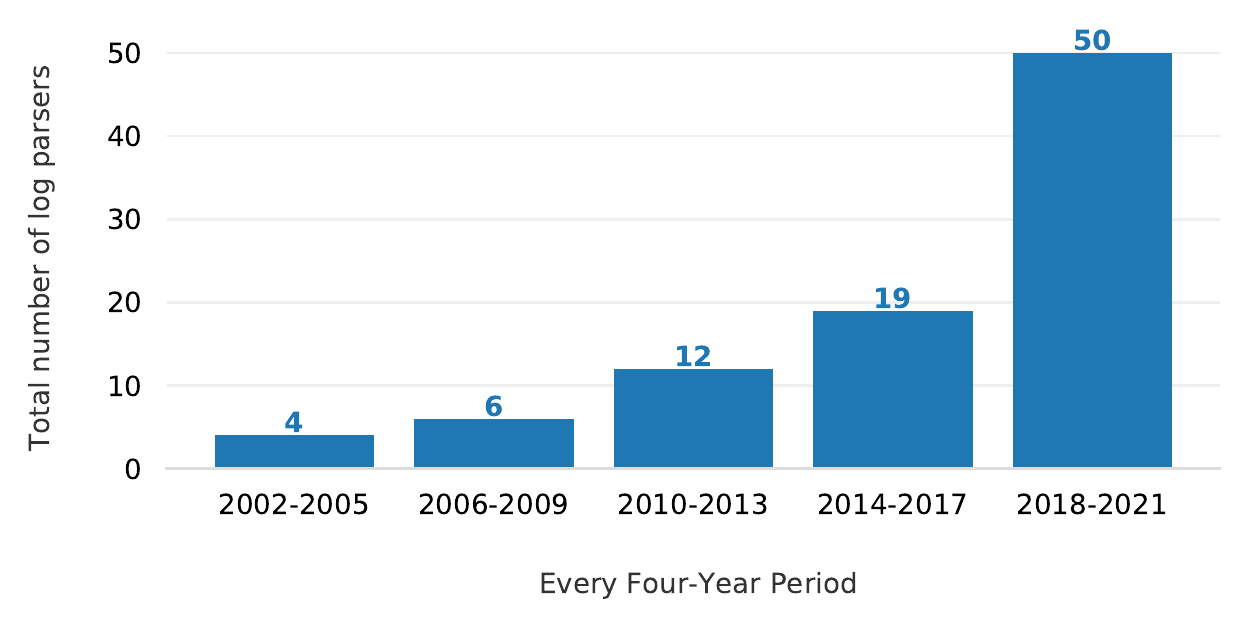}
\caption{Log parsers and the related works proposed during the past two decades.}
\label{fig:trend}
\end{center}
\end{figure}

To close this gap, an extensive collection of research endeavors have been devoted to log parsing, which entails the fundamental step of automatically converting raw log entries into standard system events for high-level log analysis and system management. As depicted in Fig.~\ref{fig:workflow}, raw log entries originate from various real-world systems that are usually comprised of heterogeneous hardware devices (e.g., actuators, sensors, network equipment, end-user terminals) and software components (e.g., operating systems, applications). Log parsing eliminates the need of manually matching and converting every entry of a system log, which can be otherwise extremely laborious and error-prone, and provides a unified data format for varied system management tasks, including anomaly detection~\cite{fu2009execution,oliner2008alert,li2020swisslog,meng2019loganomaly,chen2020logtransfer,du2017deeplog}, root cause analysis~\cite{lin2016log,yuan2010sherlog,kobayashi2017mining}, failure prediction~\cite{zhang2017syslog,salfner2007using,kimura2018proactive}, and end-to-end log analysis~\cite{li2017flap, debnath2018loglens,shang2013assisting}.  
Log parsing can also augment log management systems by converting raw log entries into compact representations and concise message types to save memory and facilitate data queries~\cite{makanju2011lightweight,meng2020summarizing,kobayashi2020amulog,liu2019logzip}.
In recent years, with the rapid expansion of modern ICT systems, log parsing keeps gaining momentum. According to our literature study, as synopsized in Fig.~\ref{fig:trend}, an increasingly large number of log parsers have been proposed during the last two decades, especially over the last four years.
Nonetheless, despite the abundant solutions, their performance characteristics (e.g., parsing accuracy, runtime efficiency) and operational features (e.g., execution mode, accessibility) are still unclear, leading to the duplicated exertions of reinventing the wheel~\cite{he2016evaluation}. 

This paper provides a comprehensive review of existing log parsers and a detailed performance evaluation of open-source solutions. 
Some prior works are related to ours: Landauer et al.~\cite{landauer2020system}, Svacina et al.~\cite{svacina2020vulnerability}, He et al.~\cite{he2020survey}, Bhanage et al.~\cite{bhanage2021infrastructure} and Skopik et al.~\cite{skopik2021online}, and Zheng et al.~\cite{zheng2019survey} investigated the impact of log mining for cybersecurity, reliability engineering, root cause analysis, anomaly detection, and failure prediction respectively. These works only covered specific log parsers relevant to each application domain. Instead, our taxonomy focuses on log parsing and targets all the existing solutions regardless of application domains. 
Besides the taxonomic reviews, some prior works focus on a specific collection of solutions and dive into their internals.
El-Masri et al.~\cite{el2020systematic} qualitatively discussed their performance and operational features of 17 log parsers. Zhu et al.~\cite{zhu2019tools} standardized the quantitative performance analysis process by evaluating 13 log parsers on 16 public datasets. Copstein et al.~\cite{copstein2021log} validated the performance of the exact solutions and investigated their best practices to aid forensic analysis. 
Our work analyzes the performance and operational features of existing log parsers, qualitatively and quantitatively. The novel aspects of this survey with respect to the related work are outlined in Table~\ref{tab:related}. 
\begin{table}[h]
\caption{Comparison with the related works}
\begin{center}
\begin{tabular}{|c|c|cc|}
\hline
\multirow{2}{*}{Prior work} & \multirow{2}{*}{Taxonomy} & \multicolumn{2}{c|}{Performance analysis} \\
& & \it Quantitative & \it Qualitative  \\\hline
Landauer et al.~\cite{landauer2020system} & \checkmark && \\\hline
Svacina et al.~\cite{svacina2020vulnerability}& \checkmark && \\\hline
He et al.~\cite{he2020survey} & \checkmark && \\\hline
Bhanage et al.~\cite{bhanage2021infrastructure} & \checkmark && \\\hline
Skopik et al.~\cite{skopik2021online} & \checkmark && \\ \hline
Zheng et al.~\cite{zheng2019survey} & \checkmark && \\\hline
El-Masri et al.~\cite{el2020systematic} &  &  &  \checkmark \\\hline
Zhu et al.~\cite{zhu2019tools}  &  & \checkmark & \\\hline
Copstein et al.~\cite{copstein2021log}  &  & \checkmark & \\\hline
\hline
\bf Our work & \checkmark & \checkmark & \checkmark \\\hline
\end{tabular}
\end{center}
\label{tab:related}
\end{table}%

The main contributions of this paper are as follows:
\begin{itemize}
\item We review the literature and devise an exhaustive taxonomy of existing solutions based on their log parsing approaches. 
\item We collate the prior research endeavors and our own benchmarking results on 17 open-source log parsers to analyze the performance characteristics and operational features of parsing software empirically.
\item We envision future challenges for log parsing and discuss the possible research directions.
\end{itemize}

This paper is structured as follows: in Sec.~\ref{sec:sum}, we give a general overview of log parsing. In Sec.~\ref{sec:taxonomy}, we present our taxonomy on existing log parsing solutions based on their parsing methodologies. Then we analyze the performance and operational features of the existing log parsers in Sec.~\ref{sec:perf}, and discuss future challenges and research directions in Sec.~\ref{sec:future} before concluding in Sec.~\ref{sec:conclusion}.


\section{Log parsing in a nutshell}\label{sec:sum}
In this section, we give a general definition of system logs and introduce the basic process of log parsing. We list all the relevant terminologies in Table.~\ref{tab:term} for convenient reference. 

\subsection{What is a system log?}
System logs are text files containing many single-line or multi-line log entries for modern computing and communication systems ranging from large-scale distributed clusters, supercomputers, end-user devices, and self-contained applications~\cite{he2020loghub}. Each entry records a specific runtime system event. 
There is no universal standard to indicate its constituting fields. A log entry can contain multiple fields such as timestamp, severity level, logger name, message-id, and the actual log message expressing a semantic meaning. These fields are separated with delimiters like spaces, colons, or equal signs. While most fields such as timestamps have relatively standard formats, the log message field is usually in a free-text format, defined by individual developers through the log printing statements as illustrated in Fig.~\ref{fig:example}.


\begin{table}[!tb]
\caption{Table of terminology and definitions}
\renewcommand{\arraystretch}{1}
\begin{center}
\resizebox{0.48\textwidth}{!}{
\begin{tabular}{|l||l|}
\hline
\bf Terminology & \bf Definition \\\hline\hline
\bf \multirow{2}{*}{Log file/System log} & A file of system execution records collected \\
& from any real-world systems. \\\hline
\bf \multirow{2}{*}{Log parsing} & The process of converting the unstructured \\
& entries in log files to structured event types. \\\hline
\bf \multirow{3}{*}{Log entry/record} & A single- or multi-line text record derived \\
& from a log printing statement. It is usually \\
& comprised of multiple fields. \\\hline
\bf \multirow{2}{*}{Log message} & The free-text field of a log entry. It describes \\
& a specific system event or status. \\\hline
\bf \multirow{2}{*}{Event type} & A notation marking a specific group of log \\
& entries from a log printing statement. \\\hline
\bf \multirow{2}{*}{Event template} & Denotes the format of an event type. It may \\
& consist of fixed tokens and variable tokens. \\\hline
\bf \multirow{2}{*}{Preprocessing} & The process of ruling out irrelevant raw log \\
& entries and tokenize the log messages. \\\hline
\bf \multirow{2}{*}{Data classification} & The process of organizing log messages \\
& based on specific metrics. \\\hline
\bf \multirow{2}{*}{Template extraction} & The process of identifying the correlative \\
& event template for each log message cluster. \\\hline
\end{tabular}}
\end{center}
\label{tab:term}
\end{table}%

\subsection{Log parsing process}\label{sec:lpp}
The basic idea of log parsing is classifying the input log entries based on specific procedures and extracting the correct event types.
We define a log file as a sequence of log entries: $L = (e_i: i=1,2,...)$, where each entry $l_i$ is generated by a log printing statement and can be represented as a sequence of tokens: $e_i = (t_j: j=1,2,...)$. A token can be any combination of alphanumeric and special characters. A set of predefined delimiters separates tokens in each log entry.
The length of each entry depends on the corresponding log printing statement and the parameters. Log entries generated by the same logging statement are of the same event type\footnote{Note that the event types here do not necessarily map to the same set of actual system events, which was solely defined by the system developers. Instead, they are just commonly used by researchers in the log parsing field to indicate the group of log messages generated by the same statement in the source code of system programs.}. 
The free-text message field of a log entry usually consists of constant tokens that stay fixed for all messages of the same event type and variable tokens corresponding to the parameters in the logging statement that may vary in each entry. A typical log template is extracted by keeping the constant parts and substituting the variable parts with predefined placeholders.

\begin{figure}[!tb]
\begin{center}
\includegraphics[width=0.45\textwidth]{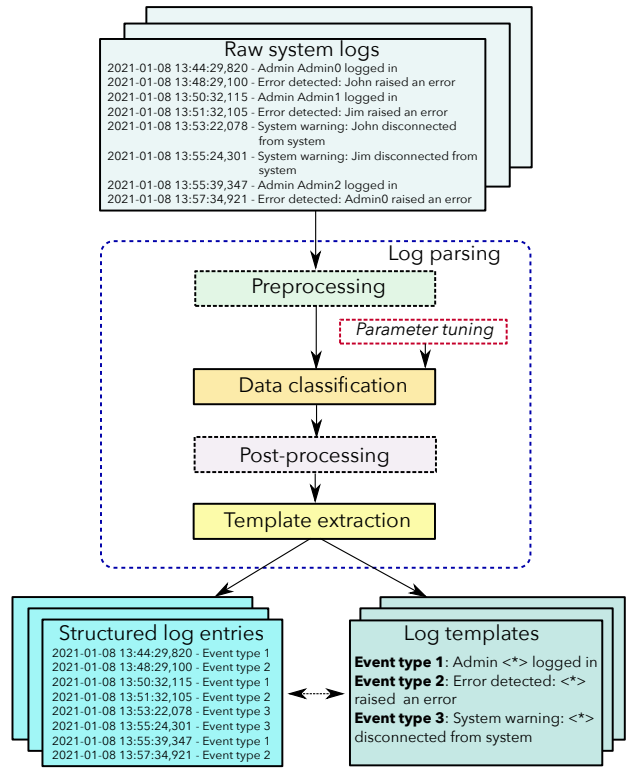}
\caption{A general overview of the log parsing process.}
\label{fig:log-parsing}
\end{center}
\end{figure}

As shown in Fig.~\ref{fig:log-parsing}, a log parser commonly comprises four steps: preprocessing, data classification, post-processing, and template extraction. In the preprocessing step, the input log entries can be filtered, deduplicated, converted, or tokenized based on bespoke rules and predefined delimiters. The rules can be composed based on domain-specific knowledge or regular expressions~\cite{astekin2019incremental}. We give a detailed discussion on preprocessing rules and delimiters in Sec.~\ref{sec:prep}. After the preprocessing step, the log message field for each entry is extracted for further parsing. 
Although some log parsers consider this step optional, most existing solutions still heavily rely on preprocessing to reduce the input size and processing noises. Some log parsers can significantly benefit from fine-grained preprocessing. 
In the data classification phase, log entries are encoded with tailor-made data structures (e.g., numerical vectors, trees, dictionaries) and matched or grouped based on predefined similarity metrics (e.g., cosine similarity~\cite{crnic2011introduction}, Jaccard similarity~\cite{tan2016introduction}, string edit distance~\cite{ristad1998learning}). Each resulting group constitutes a unique event type originating from a specific log printing statement. Many existing log parsers also optionally expose a set of tunable parameters to allow users to customize the subsequent log parsing process. For some solutions, parameter tuning can markedly improve their performance. We discuss the parameter tuning at length in Sec.~\ref{sec:tuning}.
As log messages usually have heterogeneous characteristics, the data classification phase may fail to adapt to all the possible formats and lead to over- or under-parsing. Some solutions thus also adopt a post-processing phase to adjust the existing data classification results and avoid bias. 
Finally, each cluster's log template (or signature) is extracted to represent the log message field for all the enclosed entries. This process entails identifying constant and variable tokens for log messages in the same group. 
After log parsing, the initially unstructured log entries are converted into structured events with associated types, which can be leveraged by the ensuing applications to generate vectors, matrices, or sequences for advanced data analysis, insight extraction, and decision making.


\section{Log parsing solutions: A taxonomy}\label{sec:taxonomy}
This section presents our taxonomy of existing log parsing solutions that employ different algorithms to interpret the free-text log messages and infer the corresponding event types.
Among the different log parsing phases, we opt to classify existing solutions based on the methods for classifying log messages and extracting templates since these are the primary building blocks for any log parser. 
Based on our literature review, we categorize existing log parsing solutions into four categories: clustering-based approaches, frequent pattern mining-based approaches, heuristic approaches, and program analysis.
Program analysis is a code-driven approach that primarily relies on the source code or the compiled executables to associate each log entry to the corresponding log printing statement, while solutions belonging to the other categories are data-driven. 
As most of the investigated log parsers mainly operate on the message field of each log entry, we thus refer to log message as a general representation of log entry in this section.
Due to the ample solution space, our classification method is not entirely non-overlapping, and a few log parsers might employ techniques from multiple categories. In this work, we classify them based on their {\em primary} log parsing method. 

\subsection*{Literature search}
We collected the related works from prevalent scientific publications databases, including Google Scholar, IEEE eXplore, ACM Digital Library, SpringerLink, ResearchGate, and arXiv. 
The literature search consisted of two steps. We first explored these databases using the following keywords: log parsing, log template extraction, and log analysis. For each paper matching the keyword search, we checked its coherence with the log parsing process defined in Sec.~\ref{sec:lpp}. In this way, we derived the initial set of related works. 
Then we checked the references for each selected paper to search for related works. We also checked the works citing this paper whenever applicable. For newly obtained papers, we repeated this process iteratively. After almost one year of literature search, our paper covers the most exhaustive set of the log parsing literature during the time of writing. 

\begin{table*}[!tb]
\caption{Existing log parsers by category.}
\begin{center}
\resizebox{\textwidth}{!}{
\begin{tabular}{|c|c|l|}
\hline
\bf Category & \bf Subcategory & \bf Existing research items \\\hline \hline
\multirow{5}{*}{\bf Clustering} & Hierarchical clustering &\it LogMine~\cite{hamooni2016logmine},  Lin et al.~\cite{lin2016log}, LPV~\cite{xiao2020lpv},  LogTree~\cite{tang2010logtree}, LogOHC~\cite{yang2019online}, METI$N$G~\cite{coustie2020meting}, HELO~\cite{gainaru2011event} \\ \cline{2-3}
& Density-based clustering & \it STE~\cite{kimura2014spatio}, LTE~\cite{ya2015automatic}, Pokharel et al.~\cite{pokharel2019hybrid},  Zou et al.~\cite{zou2016uilog}, HLAer~\cite{ning20141hlaer} \\\cline{2-3}
& \multirow{2}{*}{Online clustering} & \it LenMa~\cite{shima2016length}, Guo et al.~\cite{guo2018event}, LogSimilarity~\cite{kimura2018proactive}, Zhao et al.~\cite{zhao2018improvement}, StringMatch~\cite{aharon2009one}, FLP~\cite{zhong2018flp}, \\ 
&&  \it Joshi et al.~\cite{joshi2014intelligent}, One-to-one~\cite{chunyong2020log}  \\ \cline{2-3}
& Other clustering methods & \it LogSig~\cite{tang2011logsig}, LKE~\cite{fu2009execution}, Pylogabstract~\cite{studiawan2020automatic} \\ \hline
\multirow{3}{*}{\bf Frequent pattern mining} & Apriori-based approach &\it SLCT~\cite{vaarandi2003data}, LogHound~\cite{vaarandi2004breadth}, LogCluster~\cite{vaarandi2015logcluster}, LFA~\cite{nagappan2010abstracting}, ENG~\cite{tovarvnak2019normalization} \\ \cline{2-3}
& \multirow{2}{*}{Other approaches} & \it Signature Tree~\cite{qiu2010happened}, DLog~\cite{li2018dlog}, FT-tree~\cite{zhang2017syslog}, Craftsman~\cite{zhang2020efficient}, Prefix-Graph~\cite{chu2021prefix}, CAPRI~\cite{zulkernine2013capri} \\
&& \it Logram~\cite{dai2020logram}, Liu et al.~\cite{liu2020web}, Stearley~et~al.~\cite{stearley2004towards} \\\hline
\multirow{6}{*}{\bf Heuristic approaches} & LCS-based approach & \it Spell~\cite{du2016spell}, SwissLog~\cite{li2020swisslog}, Delog~\cite{agrawal2019delog}, Slop~\cite{zhao2018slop}, Logan~\cite{agrawal2019logan}, LTmatch~\cite{wang2021ltmatch} \\\cline{2-3}
& Parsing tree approach & \it Drain~\cite{he2018directed}, OLMPT~\cite{wen2020olmpt}, USTEP~\cite{vervaet2021ustep}, AECID-PG~\cite{wurzenberger2019aecid}, SHISO~\cite{mizutani2013incremental} \\ \cline{2-3}
& \multirow{2}{*}{ML-based approach} & \it NLP-LP~\cite{aussel2018improving}, Li et al.~\cite{li2005integrated}, Kobayashi et al.~\cite{kobayashi2014towards}, McLean et al.~\cite{mclean2020adaptive}, FastLogSim~\cite{liu2020fastlogsim}, NuLog~\cite{nedelkoski2020self} \\
&& \it LogParse~\cite{meng2020logparse}, Thaler et al.~\cite{thaler2017towards}, Rand et al.~\cite{rand2021automatic}, Ruecker et al.~\cite{ruecker2021flexparser}, LogDTL~\cite{nguyen2021logdtl}, LogStamp~\cite{tao2021logstamp} \\ \cline{2-3}
& \multirow{2}{*}{Other approaches} & \it Gao et al.~\cite{gao2018navigating}, Chuah et al.~\cite{chuah2010diagnosing}, LEARNPADS~\cite{zhu2010incremental}, Baler~\cite{taerat2011baler}, MoLFI~\cite{messaoudi2018search}, Lopper~\cite{liu2019lopper} \\
&& \it CLF~\cite{zhang2019efficient}, POP~\cite{he2017towards}, IPLoM~\cite{makanju2011lightweight}, Paddy~\cite{huang2020paddy}, AEL~\cite{jiang2008abstracting} \\\hline
\multirow{2}{*}{\bf Program analysis} & Code analysis & \it Xu et al.~\cite{xu2009detecting}, Yuan et al.~\cite{yuan2010sherlog}, Tak et al.~\cite{tak2016logan} \\\cline{2-3}
& Executable analysis &\it Genlog~\cite{zhang2017genlog}, Zhao et al.~\cite{zhao2014lprof} \\
\hline
\end{tabular}
}
\end{center}
\label{tab:all}
\end{table*}%

\subsection{Clustering-based log parsing}
Some log parsers are based on traditional clustering algorithms.
For a given sequence of log entries $L$, a clustering-based log parsing method clusters them into a set of $K$ clusters $\mathcal{C} = \{C_1, C_2, ..., C_K\}$, such that:
\begin{equation}
C_i \cap C_j = \emptyset, \forall i,j \in \{1,..., K\}, i \neq j
\end{equation}
\begin{equation}
C_i \neq \emptyset, \forall i \in \{1,..., K\}
\end{equation}
\begin{equation}
\bigcup\limits_{i=1}^{K} C_i = \mathcal{C}
\end{equation}

Log entries in the same cluster have high similarities, while entries across different clusters have low similarities. 
For each final cluster $C_i$, a log parsing method will extract a log template representing the event type of all the included entries.
Note that the $K$ can be specified beforehand or decided at runtime. According to our literature study, existing log parsers employ four clustering algorithms, namely hierarchical clustering, density-based clustering, online clustering, and other clustering methods.

\subsubsection{Hierarchical clustering}
A hierarchical clustering algorithm parses raw log messages into a hierarchy of clusters with different levels of similarities, i.e., dendrograms. Unlike other approaches that provide a definite set of clusters, it generates a collection of partitions with varying levels of details for users to select. Although the parsing results are more informative and flexible, it incurs a higher computation cost than the other clustering approaches~\cite{xu2015comprehensive}.
Hierarchical clustering approaches can be either {\it agglomerative} or {\it divisive}. 

Agglomerative clustering approaches begin with individual log messages and iteratively merge similar clusters until all the messages end up in the same cluster. 
{\bf LogMine~\cite{hamooni2016logmine}} is a typical solution embracing this approach. 
It firstly executes a one-pass clustering algorithm to scan all the log messages sequentially and generate a set of dense clusters based on a distance function. The first message in each cluster is selected as the pattern of that cluster. The resultant patterns form the bottom hierarchy. After this initial phase, LogMine repeats the clustering algorithm with a relaxed distance bound on the generated patterns. For each generated cluster, it employs a merging algorithm, which aligns and merges all the constituent patterns sequentially to generate a new pattern as a log template. These new patterns then form a new hierarchy level. The clustering and pattern recognition processes are iteratively invoked until the hierarchy is completed.
Similarly, {\bf Lin et al.~\cite{lin2016log}} convert raw log messages into sequences and iteratively merge them into new clusters using cosine similarity as the distance metric. The clustering process only terminates when all the newly merged clusters are far from each other. Afterward, the log message closest to the centroid of each cluster is selected as the template of that cluster. 
{\bf LPV~\cite{xiao2020lpv}}, {\bf LogTree~\cite{tang2010logtree}}, and {\bf LogOHC~\cite{yang2019online}} also belong to this category. LPV employs agglomerative hierarchical clustering (i.e., complete-linkage clustering) to incrementally group log messages based on Euclidean distance. LogTree employs single-linkage agglomerative clustering for multi-view event generation from system logs. LogOHC proposes a customized online hierarchical clustering algorithm to aggregate similar log messages.

On the contrary, a divisive clustering approach considers the entire input dataset as a cluster and iteratively partitions it until all the resulting clusters contain a single log message. 
{\bf METI$N$G~\cite{coustie2020meting}} and {\bf HELO~\cite{gainaru2011event}} follow divisive hierarchical clustering. 
METI$N$G constructs a dendrogram by recursively bisecting existing clusters. At each partitioning step, logs containing the most common n-grams of a cluster are separated into one sub-cluster. The bisection only stops if a cluster reaches adequate homogeneity, assessed using a customized criterion. 
Similarly, HELO recursively partitions existing clusters by columns until each cluster's log messages have $\geq 40\%$ common tokens. 

\subsubsection{Density-based clustering}
A density-based clustering algorithm explores the problem space and considers regions with {\em high point density} as clusters.
Density-based algorithms, especially the Density-Based Spatial Clustering (DBSCAN)~\cite{ester1996density}, have been widely used for log parsing.
For instance, based on the assumption that log messages of the same type tend to have identical static tokens appear in the same position, {\bf STE~\cite{kimura2014spatio}} defines a scoring function to evaluate the tendency of a token being static. Then it employs the DBSCAN algorithm to identify static tokens based on the function. The log template is extracted from the top clusters.
{\bf LTE~\cite{ya2015automatic}} has three modules: information filter, message clustering, and template extraction. The information filter rules out timestamps and IP fields of the raw log messages based on pre-composed regular expressions. Then the clustering module employs the DBSCAN algorithm to group together messages with similar formats. Finally, the template extraction module obtains event types for each group using the Latent Dirichlet Allocation (LDA)~\cite{blei2003latent} model integrated with the sampling algorithm.  
{\bf Pokharel et al.~\cite{pokharel2019hybrid}} convert each log message into bi-grams and cluster them directly using the DBSCAN algorithm.
{\bf Zou et al.~\cite{zou2016uilog}} also employ the DBSCAN algorithm along with a customized Levenshtein distance~\cite{levenshtein1966binary} to cluster log messages and extract the event templates. 
Besides the DBSCAN algorithm, {\bf HLAer~\cite{ning20141hlaer}} measures the similarities of the input logs and builds a clustering tree using the density-based OPTICS algorithm~\cite{ankerst1999optics}. In the final phase, it extracts log formats via a sequential alignment scheme for each node in the clustering tree.

\subsubsection{Online clustering}
Online incremental clustering is another commonly adopted approach. Solutions in this category use explicit similarity metrics to cluster the continuously arriving logs. 
The most representative example is {\bf LenMa~\cite{shima2016length}}, which incrementally clusters log messages based on positional token length. It converts each incoming log message into a vector of the constituent tokens' lengths. The vector is compared with the template of each existing cluster in terms of identical positional tokens and cosine similarity. The message is either appended to an existing cluster with the highest similarity or classified as a new cluster. 

Some solutions follow this procedure but use distinct similarity metrics. 
For instance, {\bf Guo et al.~\cite{guo2018event}} calculate message similarity based on the proportion of constant tokens.
{\bf LogSimilarity~\cite{kimura2018proactive}} does this according to the weighted ratio of shared tokens. 
{\bf Zhao et al.~\cite{zhao2016extracting,zhao2018improvement}} clusters log messages based on the ratio of position-wise identical tokens and common sequences. 
{\bf StringMatch~\cite{aharon2009one}} clusters log messages based on cosine similarity and employs token position entropy to adjust existing clusters incrementally.  
{\bf FLP~\cite{zhong2018flp}} incrementally clusters incoming logs based on message length, the first and last tokens, and a customized similarity metric.
{\bf Joshi et al.~\cite{joshi2014intelligent}} vectorize incoming logs with randomized hashing and employ a similarity search algorithm to cluster them through bitwise comparison incrementally. Messages with a common subsequence ratio beyond a threshold are clustered together.
{\bf One-to-one~\cite{chunyong2020log}} maintains a template list on-the-fly and follows three customized rules to incrementally cluster incoming logs. 

\subsubsection{Other clustering methods}
Besides the foregoing clustering approaches, there are three other clustering-based solutions. In particular, {\bf LogSig~\cite{tang2011logsig}} customizes K-Means algorithm~\cite{lloyd1982least}, a centroid-based clustering algorithm. LogSig converts input log messages into ordered word pairs and aggregates them into groups. Then it iterative moves log messages between groups to maximize the total number of common word pairs. Finally, it extracts the log template for each group based on common pairs.
{\bf LKE~\cite{fu2009execution}} clusters log messages based on string edit distance. It calculates weights using a Sigmoid function and assigns them to different token positions to prioritize the leading tokens. Two messages with a distance smaller than a threshold are grouped. A K-Means algorithm decides the threshold. Then it further splits each group by the least frequent token positions. The newly obtained groups constitute the final clustering results.
{\bf Pylogabstract~\cite{studiawan2020automatic}} employs a graph clustering approach. It firstly groups log messages by length and then employs Girvan-Newman community detection~\cite{girvan2002community} and modularity value for log clustering.  

\subsection{Frequent pattern mining}\label{sec:fpm}
Some log parsers employ Frequent Pattern Mining (FPM)~\cite{han2007frequent}, a traditional data mining approach to discover patterns that occur beyond a support value. 

Some solutions mimic the classical Apriori algorithm~\cite{agrawal1994fast} to extract frequent token sequences.  
{\bf SLCT~\cite{vaarandi2003data}} is the most representative solution adopting this approach. Its intuition originates from two fundamental properties of system logs: (i) most of the tokens occur only a few times, (ii) there are usually strong correlations between the frequent tokens. SLCT parses logs with three passes of the input dataset. The first pass extracts all the frequent words whose occurrences are larger than a predefined threshold. In the second pass, it builds cluster candidates by matching the frequent words on each line. Finally, candidates with enough frequent words are identified as clusters.

SLCT lays the foundation for several other solutions: 
{\bf LogHound~\cite{vaarandi2004breadth}} considers input logs as database transactions and employs a breadth-first algorithm to extract frequent messages using an in-memory tree, which is built by layers until it includes all the frequent itemsets.
{\bf LogCluster~\cite{vaarandi2015logcluster}} locates the frequent words using a hash table. Then it extracts all the frequent words from each log message to build or update a candidate group. Candidates with smaller support than the threshold are dropped as outliers, and the remaining ones are selected as final clusters. 
{\bf LFA~\cite{nagappan2010abstracting}} scans over the input logs to build a word frequency table recording the position-wise occurrence of each word. Then it parses the log file by line and retrieves the frequency for every word in their corresponding position. LFA identifies that log message's constant and variable parts based on the frequencies and builds an event type as a regular expression. {\bf ENG~\cite{tovarvnak2019normalization}} extends LFA to support multiple delimiters for tokenization and multi-word variables.

Aside from the Apriori-based approaches, some works rely on tailored data structures to explore the frequency properties. The most commonly used structure is the prefix tree (also known as a trie), which is sequentially constructed from the input tokens to reduce index search complexities and identify frequent log sequences. For instance, 
{\bf Signature Tree~\cite{qiu2010happened}} builds a prefix tree by recursively adding the most frequent combination of tokens as children until all the messages are associated with the tree. Then it prunes the tree by discarding all the nodes with more children than a threshold. After the pruning step, each remaining root-leaf path in the parsing tree constitutes a unique event template.
{\bf DLog~\cite{li2018dlog}} constructs a prefix tree by recursively picking the same beginning tokens and employing a hashmap to store token occurrences. Event templates are extracted by comparing the root node with subtree nodes.
{\bf FT-tree~\cite{zhang2017syslog}} scans the input logs to calculate token frequencies and uses a heuristic algorithm to construct a prefix tree. Newly learned log templates could also be incrementally added into the tree by only parsing the recently arrived logs.
{\bf Craftsman~\cite{zhang2020efficient}} employs a dynamic prefix tree to parse logs and extract templates. First, it scans the whole dataset to derive the token frequency list ranked in descending order. Then it parses each log message to incrementally add a new branch to the tree according to the frequency list and the common subsequence with existing nodes. The obtained tree is pruned following a node degree constraint.
{\bf Prefix-Graph~\cite{chu2021prefix}} extends a probabilistic graph structure from prefix tree. It begins with a directed acyclic graph and iteratively merges branches with similar frequency vectors. Finally, it uses a template extraction algorithm to retrieve the message signatures from the graph. 

Besides prefix trees, some solutions choose other ways to discover frequent sequences. 
{\bf CAPRI~\cite{zulkernine2013capri}} adopts type-casting technique and bitmap multiplication algorithm to extract log events with different frequency properties and support incremental log mining. It also generates rules to reflect the contextual relationship between sequential messages.
{\bf Logram~\cite{dai2020logram}} relies on $n$-gram dictionaries for log parsing. The n-grams with occurrence below a threshold are recursively transformed to $(n-1)$-gram until a list of infrequent 2-grams is obtained. Overlapping tokens in the list are identified as variables. 
{\bf Liu et al.~\cite{liu2020web}} propose an approach that scans the input logs to build a word-counting table. Subsequently, a log dictionary mapping each keyword to a set of clusters is constructed. Each log message is parsed by extracting the most frequent token from the word-counting table and retrieving the most related log templates from the dictionary. The message is added to the dictionary by measuring the edit distance with the templates. 
{\bf Stearley~et~al.~\cite{stearley2004towards}} maps each input token to an integer and employs a matching algorithm to locate all the patterns with a user-given specificity and support. Then it employs a set of conversions to classify log messages. 

\subsection{Heuristic approach}
Aside from the more conventional frequent pattern mining and clustering algorithms, many log parsers employ different heuristic algorithms and data structures for log encoding, data parsing, and template extraction. The most adopted approaches include {\it longest common subsequence (LCS)}, {\it Parsing tree} and {\it Machine Learning (ML)}, which are discussed in Sec.~\ref{sec:lcs}, Sec.~\ref{sec:tree}, and Sec.~\ref{sec:nlp}. Some other log parsing methods are based on more customized rules and data structures. We present them in Sec.~\ref{sec:rules}. 

\subsubsection{LCS-based approach}\label{sec:lcs} 
Longest Common Subsequence (LCS)~\cite{maier1978complexity} is a popular approach widely adopted in log processing. 
Given two log messages $l_1 = \{x_1, x_2, ..., x_n\}$ and $l_2 = \{y_1, y_2, ..., y_m\}$, with $x_i (1 \leq i \leq n)$ and $y_j (1 \leq j \leq m)$ being arbitrary tokens in each message. A token sequence $s = \{z_1, z_2, ..., z_k\}$ is considered a common sequence of $l_1$ and $l_2$ iff $s \subseteq l_1$ and $s \subseteq l_2$. Intuitively, any pair of log messages can have multiple common sequences. The LCS of $l_1$ and $l_2$ is defined as a sequence of tokens with the value of $k$ maximized.

{\bf Spell~\cite{du2016spell}} is a typical solution that embraces LCS-based approach. It maintains an LCS map for already parsed log entries, and each map consists of a group of log entry lineIDs, and the corresponding parsed LCS sequence (or message type).
Spell searches through the map for an incoming log entry $m$ to find the mapping $e$ whose sequence has the maximum LCS length with $m$. If the length is longer than $\frac{|m|}{2}$, the LCS sequence of $e$ is updated based on $m$ with the lineID of $m$ appended. Otherwise, a new mapping is created for $m$. 
In a later extension~{\bf \cite{du2018spell}}, the authors augmented Spell with more efficient search algorithms, parallel execution, and semantic recognition.

Similar to Spell, 
{\bf SwissLog~\cite{li2020swisslog}} relies on a dictionary to parse logs in four steps that involve multiple common heuristic approaches. It first tokenizes raw logs to build a valid {\em wordset}, which is then used to classify input logs. An LCS-based algorithm is then invoked to identify and mask the variable parts. Finally, it constructs a prefix tree to merge groups with common subsequences to avoid over-parsing.
{\bf Delog~\cite{agrawal2019delog}} adopts a hash-based searching and an LCS-matching algorithm to partition similar input logs into groups. Each group is adjusted using a sequence alignment algorithm to cope with the inaccuracy caused by messages of the same types with variable parameter lengths.
{\bf Slop~\cite{zhao2018slop}} partitions incoming logs by lengths. For each message, Slop matches it to existing logs in the same partition and employs an LCS-based algorithm to extract templates. As messages with the same type can have different lengths, Slop uses another algorithm to identify and merge existing templates from all the partitions.
{\bf Logan~\cite{agrawal2019logan}} defines a length-based method to rule out irrelevant templates and an LCS-based algorithm to match log messages with similar patterns. It also performs post-filtering constraints and periodical merges to improve parsing accuracy.
{\bf LTmatch~\cite{wang2021ltmatch}} also relies on LCS for its online processing pipeline. For each new log, LTmatch calculates its word matching rates with existing templates using an LCS-based algorithm. The log is added to the most matching group, and the corresponding template is updated using a proposed template extraction algorithm. 

\subsubsection{Parsing tree}\label{sec:tree}
Another commonly adopted heuristic method is the parsing tree. Unlike the prefix trees in Sec~\ref{sec:fpm}, a parsing tree approach employs bespoke encoding rules to match the incoming logs.
\textbf{Drain~\cite{he2017drain,he2018directed}} is an archetypal example in this vein. It relies on a fixed-depth parsing tree to cluster raw log messages into groups. Each leaf contains a set of log groups. Instead of comparing with all the groups, the tree structure effectively bounds the number of log groups each new message needs to traverse. A set of filtering rules are configured in the internal nodes to guide the search of the most suitable leaf node: the first-tier nodes match incoming messages by their lengths, and the following $n$ tier nodes match messages by their preceding $n$ tokens. When a message reaches a leaf node, it is assigned to the group with the highest per-token similarity. 

Similar to Drain, {\bf OLMPT~\cite{wen2020olmpt}} proposes a two-level parsing tree to match logs. Starting from the root, the first level children store the message length while the second level sequentially stores the initial token characters. OLMPT traverses the tree for each new log message and matches the most similar template by a character-wise score. 
Instead of fixed depth and encoding rules, {\bf USTEP~\cite{vervaet2021ustep}} relies on an evolving parsing tree for log parsing. It dynamically encodes rules on the intermediate nodes and incrementally matches each incoming log message to a leaf node.
{\bf AECID-PG~\cite{wurzenberger2018aecid,wurzenberger2019aecid}} constructs a parsing tree on-the-fly following four predefined rules. Each node is assigned a set of path frequencies to reflect different token sequences. It also allows defining a list of delimiters to flexibly tokenize the raw log messages.  
{\bf SHISO~\cite{mizutani2013incremental}} also adopts a parsing tree to classify incoming logs based on token-wise Euclidean distance. Based on the search result, the log message is either merged into an existing node or added as a new node. After that, a format search algorithm is instantiated to refine existing cluster templates. 

\subsubsection{Machine learning}\label{sec:nlp}
In recent years, Machine Learning (ML) has experienced unprecedented success in the field of Natural Language Processing (NLP). As logs have similar features to natural languages, some solutions explore the ML/NLP techniques.

{\bf NLP-LP~\cite{aussel2018improving}} applies tokenization, semantic processing, vectorization, model compression, and classification techniques to find the optimal combination. According to the experimental results, it combines Latent Dirichlet Allocation and bi-gram to achieve high-quality log parsing.
{\bf Li et al.~\cite{li2005integrated}} employs Hidden Markov Models (HMM) and a modified Naive Bayesian Model to classy logs and capture their temporal characteristics.
{\bf Kobayashi et al.~\cite{kobayashi2014towards}} employ the Conditional Random Fields~\cite{lafferty2001conditional} model to infer event templates by learning the structure of log message and exploring positional relations of words.
{\bf McLean et al.~\cite{mclean2020adaptive}} advocate adopting named entity recognition and NLP operations to train models on historical log data and incrementally learn the patterns and attributes therein. The trained models can then be utilized or customized to parse new logs.
{\bf FastLogSim~\cite{liu2020fastlogsim}} trains a TF-IDF model~\cite{salton1988term} to identify similar log patterns, which are then merged with their templates extracted. 
{\bf LogParse~\cite{meng2020logparse}} leverages the templates extracted by existing log parsers and uses a ML model (e.g., SVM) to train a classifier that can incrementally identify constant tokens and learn new log formats. 

Some solutions employ the more sophisticated Neural Network (NN) models to augment existing NLP models.
{\bf Thaler et al.~\cite{thaler2017towards}} employ a five-layer neural language model to rebuild the original characters and predict the constant/variable parts of log messages. 
Then in \cite{thaler2017unsupervised}, the same authors explore the recurrent neural networks (RNNs) with an LSTM encoder~\cite{hochreiter1997long}, which uses an RNN auto-encoder for log message embedding and an algorithm to classify messages in the embedding space and infer the event templates.
{\bf Rand et al.~\cite{rand2021automatic}} and {\bf Ruecker et al.~\cite{ruecker2021flexparser}} also employ LSTM to parse unseen log formats. In particular, FlexParser~\cite{ruecker2021flexparser} employs stateful LSTM to capture parsing patterns across the training epochs and extract templates from the evolving log messages.
{\bf LogDTL~\cite{nguyen2021logdtl}} constructs a deep transfer neural network model for log template generation. The model employs a transfer learning method to augment data training. 
{\bf LogStamp~\cite{tao2021logstamp}} treats log parsing as a sequential labeling problem. Given the historical logs, it employs a pre-trained bidirectional transformer to extract the relevant features. Then it employs a dual-path framework to extract the word embedding and labels, which are used to train a classifier that can perform online log parsing.  
{\bf NuLog~\cite{nedelkoski2020self}} embraces self-supervised learning. It employs masked-language modeling to randomly mask the input tokens, vectorized and positionally encoded to be fed to a two-layer NN encoder. A final linear layer takes the resultant matrix and maps the log messages to their vector representation. Then the model dynamically processes new logs by masking each token to identify variables and generate the corresponding event templates. 

\subsubsection{Other heuristic approaches}\label{sec:rules}
Besides the aforementioned LCS, parsing tree, and ML/NLP approaches, there is another group of log parsers that use other specialized heuristic approaches. As we observe, some of these solutions parse logs using a single heuristic algorithms, while others apply multiple heuristic algorithms to fully explore log features. We present all of them in this part. 

Several solutions parse logs using a single heuristic algorithm:  
{\bf Gao et al.~\cite{gao2018navigating}} propose a search-based algorithm to browse the raw messages and extract event templates from multi-line log messages;
{\bf Chuah et al.~\cite{chuah2010diagnosing}} propose to construct event templates based on the simple assumption that the variable parts of a log message are composed of alphanumerics; 
{\bf LEARNPADS~\cite{zhu2010incremental}} employs a learning algorithm to refine log formats iteratively until all the log messages are successfully parsed;
{\bf Baler~\cite{taerat2011baler}} employs a heuristic algorithm to extract log templates based on token-integer hash mapping and predefined attributes; 
{\bf MoLFI~\cite{messaoudi2018search}} models log parsing as a multi-objective optimization problem and solve it with an evolutionary algorithm. It employs a two-level encoding schema to represent the event templates, and applies uniform crossover and random mutations to obtain new templates. In the end, it returns a Pareto Optimal partition with different trade-offs for the users to choose.

Other solutions employ multi-step partitioning with different rules or heuristic algorithms to fully explore various characteristics of the input logs. For instance, 
{\bf Lopper~\cite{liu2019lopper}} first groups logs by length and then by the similarity ratio of identical positional tokens. Then the obtained groups are merged according to a similarity function. 
In the final step, the template for each group is extracted. Two templates are combined if they share the constant parts. 
{\bf CLF~\cite{zhang2019efficient}} follows a two-step approach to partition input logs by the initial token and length. Then it counts the occurrence of tokens on each position to identify the constant parts. Each partition splits the log messages based on the previously identified constant token positions and extracts the corresponding templates. 
{\bf POP~\cite{he2017towards}} first partitions input logs by length, and then it employs a heuristic method to partition by token position recursively. Templates for each group are generated by counting the positional distinct tokens.
{\bf IPLoM~\cite{makanju2009clustering,makanju2011lightweight}} employs a three-step approach for log parsing. 
First, it scans all the log messages and partitions them by message length. 
The resultant partitions are further divided by positional token frequency and a bijection search algorithm.
Finally, the template of each group is extracted by checking the tokens on each column. If a column has only a unique token, it is considered constant; otherwise, it is regarded as a variable piece.
{\bf Paddy~\cite{huang2020paddy}} parses logs using an inverted index dictionary that maps existing tokens to a list of log templates. Tokens of each new log message are used as reference keys to retrieve candidate templates from the inverted index. The candidates are then ranked and selected based on Jaccard similarity and length. 
{\bf AEL~\cite{jiang2008abstracting,jiang2008automated}} is based on the clone detection technique~\cite{kamiya2002ccfinder}. It employs several heuristics to identify variable tokens for each log message and group messages with a similar number of tokens and parameters into bins. Finally, it extracts the event template for each bin.

\subsection{Program analysis}
Besides the aforementioned data-driven solutions, program analysis is another broadly adopted approach by different log analysis applications~\cite{bushong2020matching,schipper2019tracing}. Although solutions based on program analysis are less practical than the data-driven ones, we include them in our survey for completeness. 
Some log parsers also extend {\it source code analysis} of system programs to pinpoint the related logging printing statements. For example,
{\bf Xu et al.~\cite{xu2009detecting}} convert raw logs into a schema with message type and message variables. Then they get all the logging statements from the source code to match each log message. 
{\bf Yuan et al.~\cite{yuan2010sherlog}} leverage source code with the related abstract syntax tree to generate regular expressions to parse log messages. Developers must annotate the logging statements and complex format strings to facilitate the process.
{\bf Tak et al.~\cite{tak2016logan}} employ code analysis to identify the logging statements and compose regular expressions for log parsing. They also devise a text clustering algorithm to expand the parsing coverage.
Although this approach renders optimistic results, the logging statements are not always easy to locate since developers may employ convoluted function calls.
Moreover, the source code of many systems is usually unavailable due to intellectual property restrictions.

To cope with these issues, some solutions further perform {\it executable analysis}. For instance,
{\bf Genlog~\cite{zhang2017genlog}} disassembles the target executable code and finds all the related log functions through a hybrid slicing approach. Then it reconstructs the log messages and employs data flow and taint propagation analysis to generate log templates. 
{\bf Zhao et al.~\cite{zhao2014lprof}} analyze the binary code and find all the logging statements through keyword search. The signature of each statement is represented as regular expressions that are used to identify the event template of each log message.


\section{Performance and operational features}\label{sec:perf}
Although the number of available log parsers is extensive, correctly choosing and configuring the most suitable solutions is non-trivial. As existing log parsers are implemented with different algorithms, they have divergent performance features and configuration complexities. Misconfigured parsers can lead to severe performance degradation for the ensuing log mining and analytic tasks. As a result, we devote this section to empirically analyzing the performance (Sec. \ref{sec:perf-char}) and operational features (Sec. \ref{sec:features}) of the existing solutions.

\subsection{Performance features}\label{sec:perf-char}
We review two key quantitative features in this work, namely the {\it parsing accuracy} and {\it execution time} of the existing solutions in Sec. \ref{sec:accuracy} and Sec.~\ref{sec:efficiency}.
Due to the large number of solutions and absence of source code for a subset of them, quantitatively comparing the performance of all the existing solutions is impractical. Therefore, we opt to conduct our empirical analysis in a hybrid fashion, combining the available literature results and the quantitative evaluation on the open-source log parsers.

\subsubsection{Parsing accuracy}\label{sec:accuracy}
Conventionally, there are three traditional metrics to measure the effectiveness of information retrieval, i.e., {\bf Precision, Recall}, and {\bf F-Measure (FM)}. The precision is the ratio of correctly parsed log pairs over the total pairs generated by a log parser. By definition, recall is the ratio of correctly parsed log pairs over the actual total log pairs. 
Precision and recall are calculated as follows:
\begin{equation}
Precision = \frac{TP}{TP+FP}
\end{equation}
\begin{equation}
Recall = \frac{TP}{TP+FN}
\end{equation}

In the context of log parsing, given $N$ raw input log entries (i.e., $\frac{N(N-1)}{2}$ message pairs), a true positive (TP) decision correctly identifies two log messages of the same type, a false positive (FP) decision arbitrates two log messages of different types to the same group. In contrast, a false negative (FN) decision parses two log messages of the same type to different groups. Precision and recall can reflect the effectiveness of existing solutions. In essence, under-parsing can result in more false positives and degrade precision, while over-parsing can lead to more false negatives and hurt recall. F-Measure (or F-score) is purposed to balance these two metrics, and it is calculated as follows: 
\begin{equation}
FM= \frac{(\beta^2+1)\cdot Precision \cdot Recall}{\beta^2 \cdot Precision + Recall}
\end{equation}
where $\beta$ is conventionally set to 1. 
Although FM can reflect log parsing accuracy, it only accounts for the correctly parsed log pairs, which is insufficient for log parsing. According to the definition of FM, two log messages of the same type are still considered a correct pair even if they are parsed into a group different from the rest of the same-typed log messages. Instead of FM, Du et al.~\cite{du2018spell} proposed the {\bf Parsing Accuracy (PA)} to account for the ratio of correctly parsed log messages over the total number of log messages. PA is formally calculated as: 
\begin{equation}\label{eq1}
PA = \frac{\# Correct{\_}Messages}{\# Total{\_}Messages}
\end{equation}
Note that, for the numerator in equation~\ref{eq1}, log messages belonging to the same event type are considered correctly parsed if and only if all of them are parsed to the same group; otherwise, none of these messages is regarded as correct. Therefore, PA is more strict than FM and is more suitable for evaluating log parsing accuracy. We thus choose PA as the accuracy metric for our experiments. 

\begin{figure}[!tb]
\begin{center}
\includegraphics[width=0.5\textwidth]{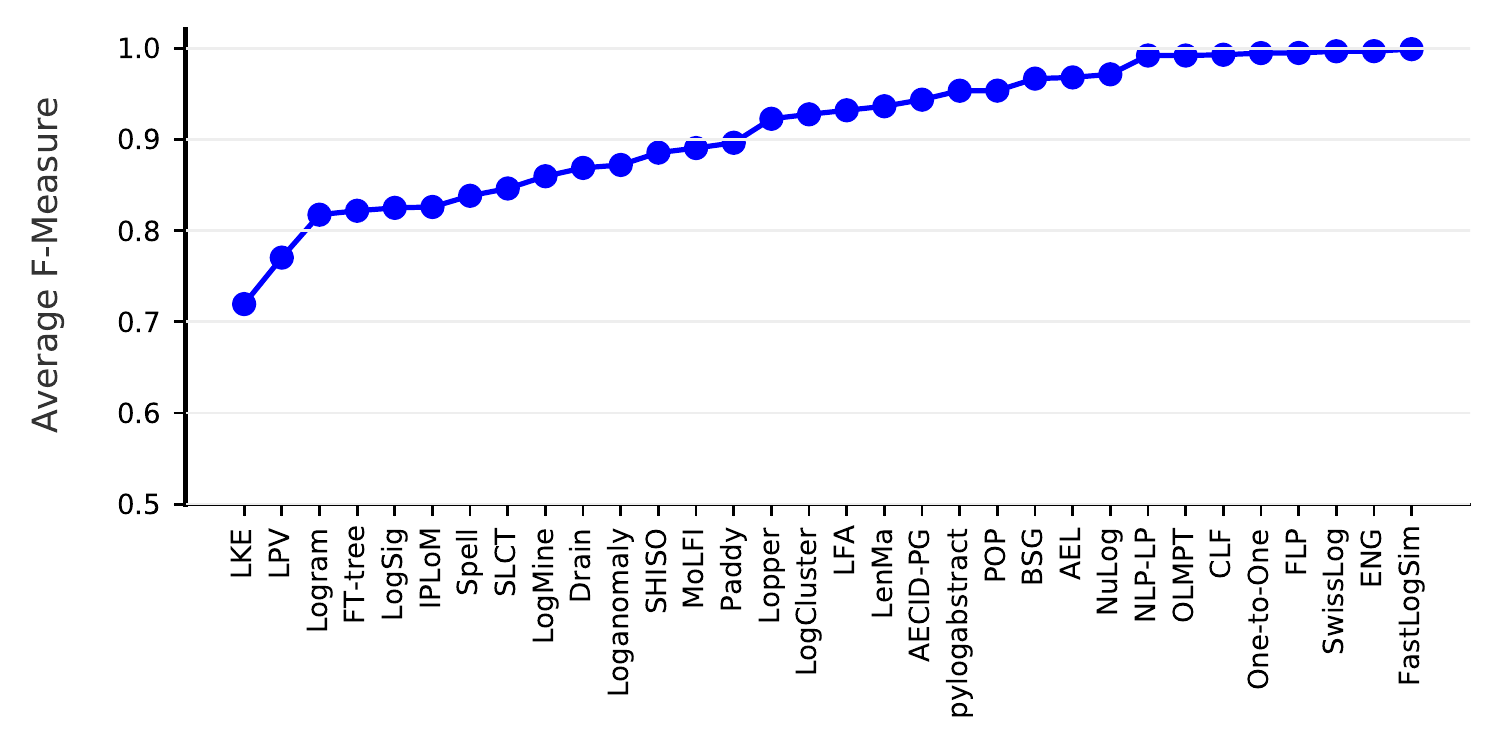}
\caption{The average F-Measure for some log parsers extracted from the literature}
\label{fig:f1}
\end{center}
\end{figure}

\begin{figure}[!tb]
\begin{center}
\includegraphics[width=0.5\textwidth]{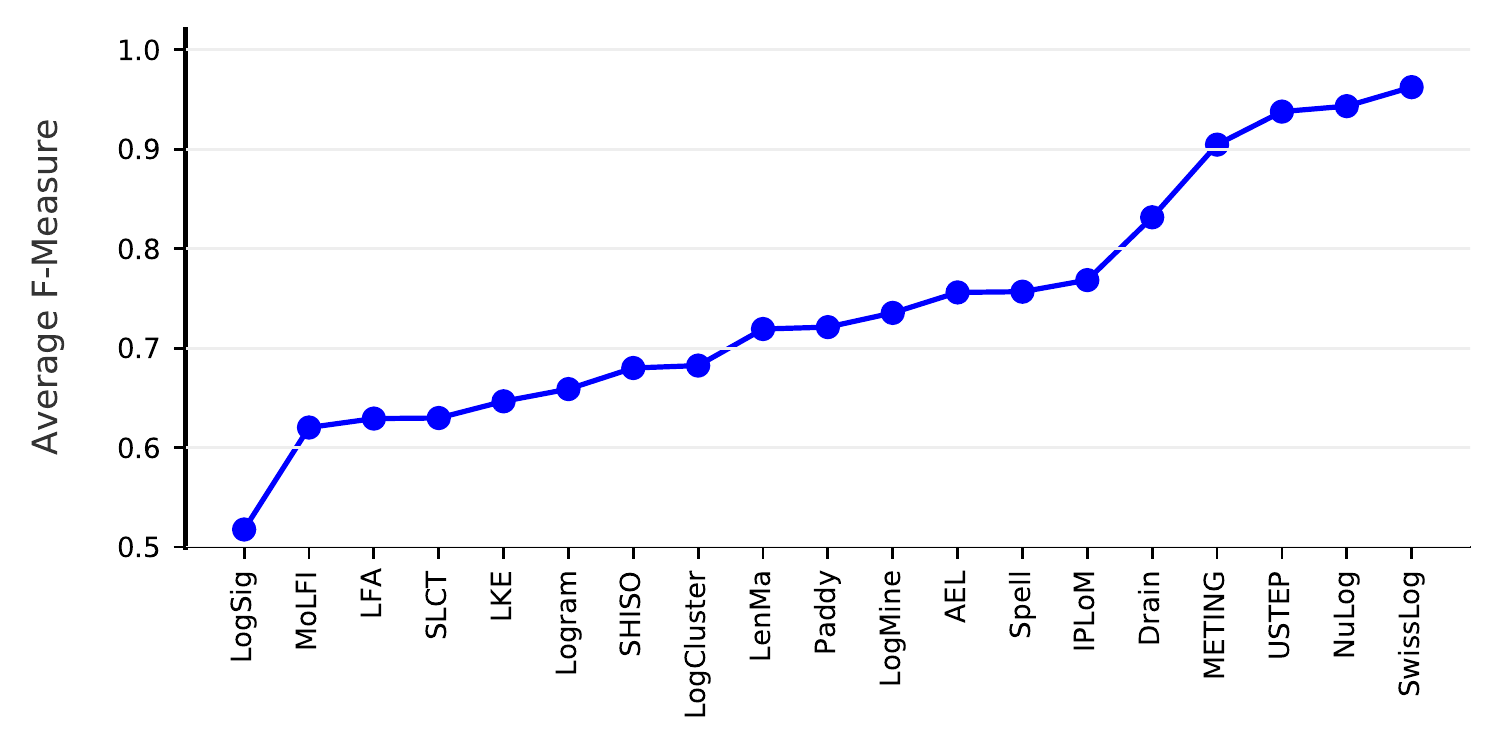}
\caption{The average PAs for some log parsers over heterogeneous logs evaluated in prior works}
\label{fig:pa}
\end{center}
\end{figure}

According to our literature review, approximately 50 papers evaluated the performance of different log parsers. These works used a variety of public/proprietary log datasets to compare the selected log parsers in terms of standard or customized performance metrics. Most of these related works used FM and PA to characterize the accuracy of existing log parsers\footnote{Note that some related works evaluated log parsing accuracy with {\em Rand Index (RI)}, is calculated as $\frac{TP + TN}{TP + FP + FN + TN}$. We omit RI in our study since it is not as commonly employed as the FM and PA. Please refer to \cite{zhang2019efficient,chunyong2020log,zhang2020efficient,tao2021logstamp,zhang2017syslog} for RI-related evaluation.}.
Therefore, we first capitalize on the numerical results in the literature to empirically understand the achievable accuracy for existing solutions. We thus collect all the numerical results and calculate the average FM and PA of existing log parsers in Fig.~\ref{fig:f1} and Fig.~\ref{fig:pa}. Although the comparisons are biased because the evaluation datasets and runtime configurations were not precisely the same for all the log parsers, they can still help us develop a general idea. 
As we can observe, most log parsers can achieve $\geq 80\%$ FM, and some of them can even reach $\geq 90\%$. 

Although PA is more strict than FM (as explained before), most of the parsers can still attain $\geq 60\%$ PA, which is reasonable given the heterogeneity of the validation logs. With proper parameter tuning and preprocessing (which will be discussed later), their accuracies are expected to improve. 
In general, state-of-the-art solutions such as SLCT~\cite{vaarandi2003data}, LogMine~\cite{hamooni2016logmine}, LogSig~\cite{tang2011logsig}, LKE~\cite{fu2009execution} generally perform worse than newly proposed ones, partially because they are more widely evaluated than the latter. Although some newly proposed solutions, such as FastLogSim~\cite{liu2020fastlogsim}, ENG~\cite{tovarvnak2019normalization}, FLP~\cite{zhong2018flp}, and One-to-One~\cite{chunyong2020log}, can achieve near $100\%$ FM, their high accuracies can be biased since they were only tested with a few datasets. 

\begin{table}[!tb]
\caption{Main characteristics of LogHub's 16 datasets}
\begin{center}
\resizebox{0.48\textwidth}{!}{
\begin{tabular}{|c|c|c|c|}
\hline
\multirow{2}{*}{\bf Dataset} & \multirow{2}{*}{\bf \#Log entries} & \bf Message length (2k) & \bf \#Templates  \\ 
&& (min/avg/max) & \bf (2k) \\ \hline
BGL & 4,747,963 & 14/47/409 & 120 \\\hline
HDFS & 11,175,629 & 8/57/425 & 14 \\\hline
HPC & 433,490 & 6/24/326 & 46 \\\hline
Proxifier & 10,108 & 32/52/125 & 8 \\\hline
Zookeeper & 74,380 & 7/42/295 & 50 \\\hline
Linux & 25,567 & 8/57/138 & 118 \\\hline
HealthApp & 253,395 & 7/48/141 & 75 \\\hline
Apache & 56,481& 22/47/62 & 6 \\\hline
Spark & 33,236,604 & 17/49/152 & 36 \\\hline
Hadoop & 394,308 & 12/82/437 & 114 \\\hline
OpenSSH & 655,146 & 23/64/106 & 27 \\\hline
OpenStack & 207,820 & 51/76/178 & 43 \\\hline
Windows& 114,608,388 & 16/88/297& 50 \\\hline
Android & 1,555,005 & 7/78/320 & 166 \\\hline
Thunderbird & 211,212,192 & 7/62/761& 149 \\\hline
Mac & 117,283 & 9/94/1138 & 341 \\\hline 
\end{tabular}}
\end{center}
\label{tab:dataset}
\vspace{-0.5cm}
\end{table}%

\begin{figure*}[!tb]
\begin{center}
\includegraphics[width=\textwidth]{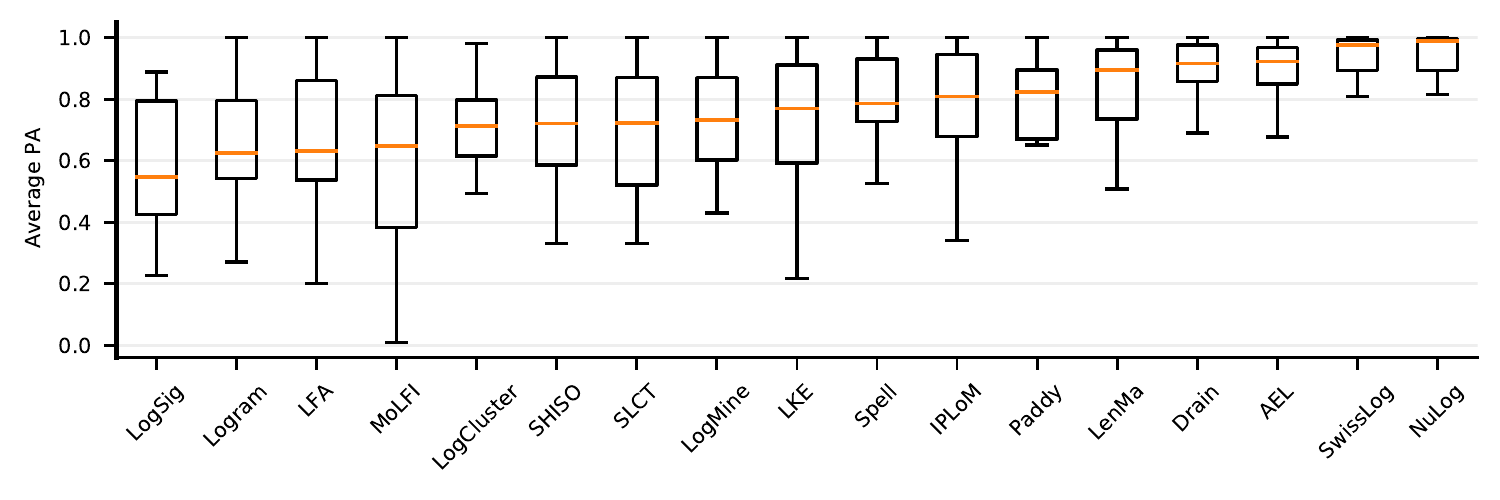}
\caption{The Parsing Accuracy distribution for some log parsers over heterogeneous logs.}
\label{fig:accuracy_dist}
\end{center}
\end{figure*}

Although the results are favorable, some solutions were only evaluated on a few datasets, and it is unclear if they can sustain similar effectiveness on other system logs. Therefore, we decide to complement existing works by thoroughly evaluating the open-source log parsers. We specifically measure the PA of 17 open-source log parsers, namely LogSig~\cite{tang2011logsig}, LKE~\cite{fu2009execution}, SLCT~\cite{vaarandi2003data}, LFA~\cite{nagappan2010abstracting}, MoLFI~\cite{messaoudi2018search}, SHISO~\cite{mizutani2013incremental}, LogClusterr~\cite{vaarandi2015logcluster}, LogMine~\cite{hamooni2016logmine}, AEL~\cite{jiang2008automated}, Spell~\cite{du2016spell}, LenMa~\cite{shima2016length}, IPLoM~\cite{makanju2011lightweight}, Drain~\cite{he2018directed}, Logram~\cite{dai2020logram}, Paddy~\cite{huang2020paddy}, NuLog~\cite{nedelkoski2020self}, and SwissLog~\cite{li2020swisslog}. The source code of the first 13 solutions are provided by the LogPAI team~\cite{zhu2019tools,he2016evaluation}. Note that other projects such as amuLog~\cite{kobayashi2020amulog} and LogParse~\cite{meng2020logparse} also provided implementations for some of these log parsers (e.g., Drain, LogSig). We decide to stick to LogPAI's solutions because they have already been widely employed by both industry and academia. The last four solutions were also implemented following the interface of LogPAI. 
We are aware of other open-source solutions (e.g., FLP~\cite{zhong2018flp}, Pylogabstract~\cite{studiawan2020automatic}, and LogDTL~\cite{nguyen2021logdtl}). We exclude them from our evaluation due to a lack of information or limited customizability, and integrating them is left for future work. 

We reuse the 16 datasets of LogHub~\cite{he2020loghub} to validate the PA of these solutions. The main characteristics of these datasets are shown in Table~\ref{tab:dataset}. Each dataset has 2k log entries randomly sampled from the original system logs, and the ground-truth event types and log templates have already been manually extracted for validation. These datasets cover a variety of ICT systems, including distributed computing, operating systems, supercomputers, and software apps, and performance evaluations across this dataset ensemble can reasonably validate the effectiveness of a log parser over heterogeneous logs~\cite{he2020loghub}. Following the approaches by Zhu et al.~\cite{zhu2019tools}, We measure the PAs of the 17 log parsers on all datasets. To ensure the optimality of the obtained results, we extensively tune the parameters of each log parser. 
For non-deterministic solutions like LKE, MoLFI, and NuLog, we repeat the experiments ten times to obtain the averages. NuLog was only evaluated on 10 of the 16 datasets because the authors did not provide the related regex rules and parameters. We will discuss preprocessing and parameter tuning at length in the next section.

We plot the distribution of PA all the log parsers under test in Fig.~\ref{fig:accuracy_dist}, ranked from left to right in ascending order of the median. In general, no solution can consistently prevail in all the test scenarios. Based on our observation, the accuracy of each log parser varies on different datasets, which is quite intuitive since each solution was designed to explore specific features of system logs. If the input logs deviate from the expected formats, the accuracy will decrease. 
For relatively simple datasets like Apache, almost all the solutions can achieve $100\%$ PA (except LogSig, which presents the lowest overall accuracy, as it is challenging to specify the cluster numbers beforehand). As the log complexities increase, the accuracies of these solutions begin to diverge. As we observe, the accuracy of IPLoM degrades on system logs with highly varied message lengths; SLCT and LFA cannot identify patterns below the threshold. The most extreme case is MoLFI, which achieves only $0.8\%$ PA on the Proxifier dataset since it has difficulty distinguishing event types with highly similar formats. MoLFI does achieve $74\%$ FM on the same dataset, which means it under-parses the identified messages, leading to extremely low PA. This case also validates the necessity of using PA as the accuracy metric over FM.
Even the overall performant solutions still possess deficiencies on specific logs, e.g., Drain performs poorly on datasets with many leading variable tokens (e.g., Mac). NuLog and SwissLog achieve the best overall accuracy based on the obtained result. 
Other heuristic solutions such as Drain and AEL also show promising results. In general, heuristic solutions outperform the clustering- and FPM-based approaches in log parsing accuracy. 

\begin{figure*}
    \centering
    \setkeys{Gin}{width=0.33\linewidth}
\subfloat[Apache]{\includegraphics{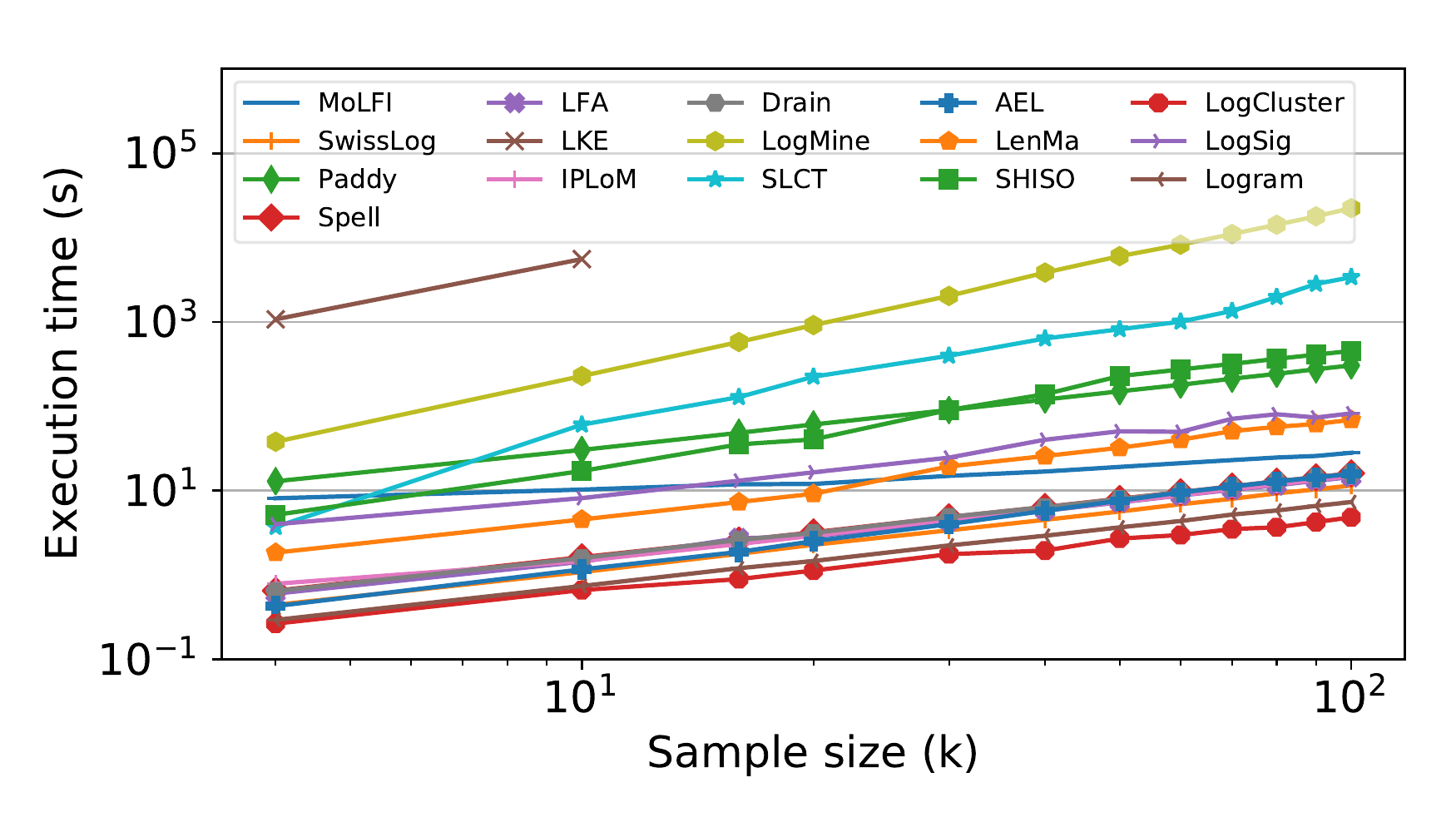}}
\hfill
\subfloat[OpenSSH]{\includegraphics{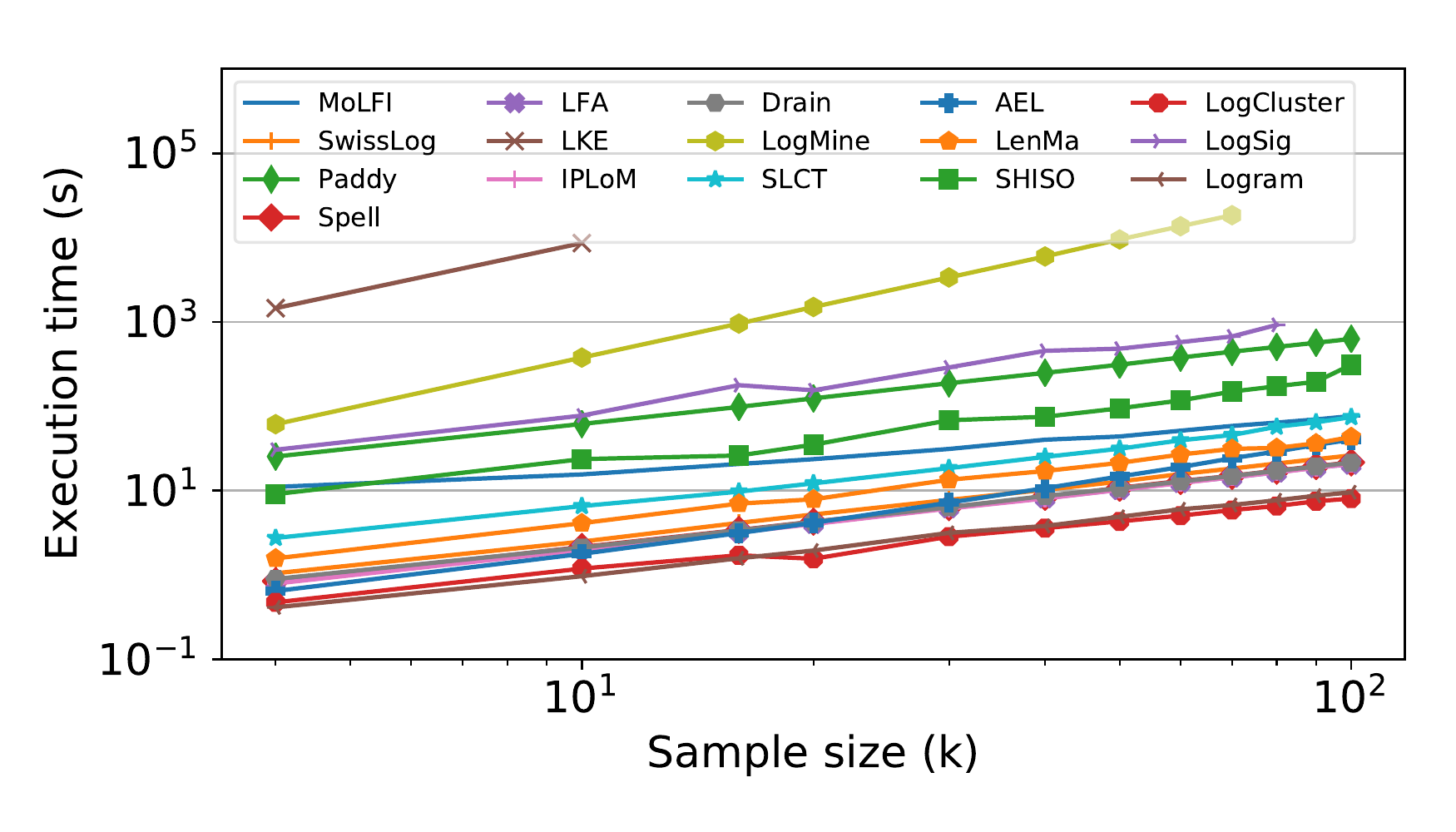}}
\hfill
\subfloat[OpenStack]{\includegraphics{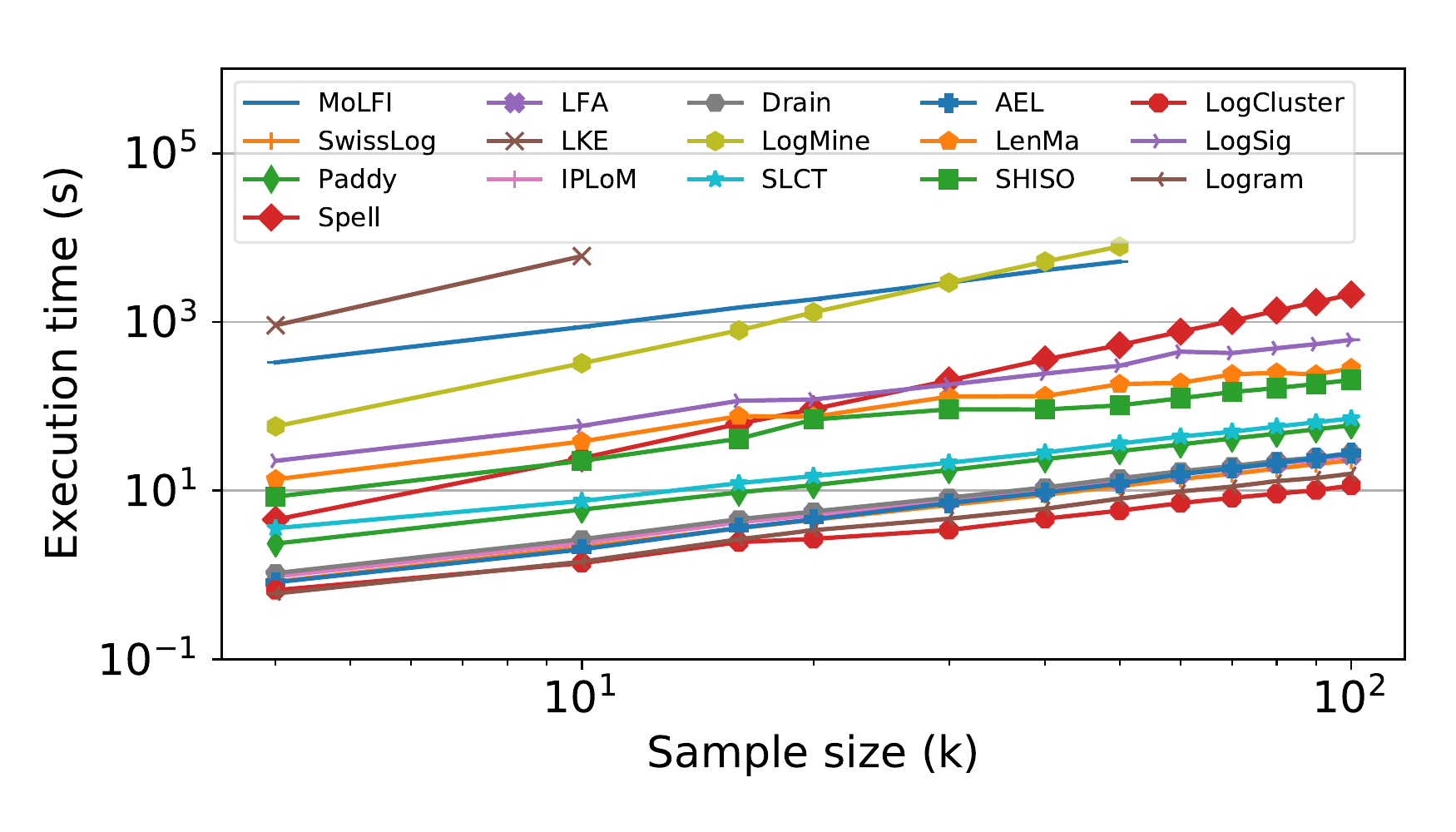}}
\hfill
\subfloat[BGL]{\includegraphics{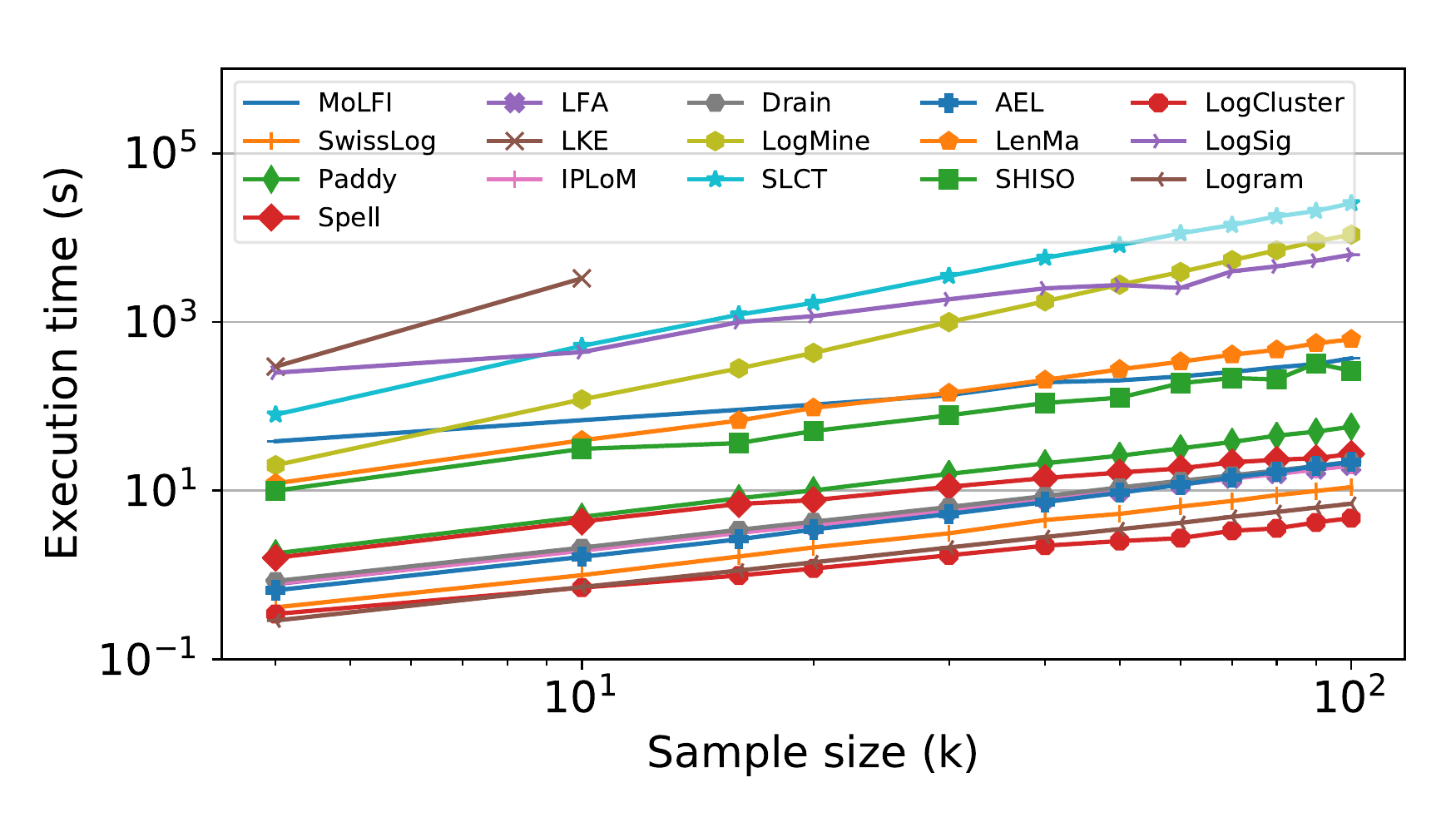}}
\hfill
\subfloat[Zookeeper]{\includegraphics{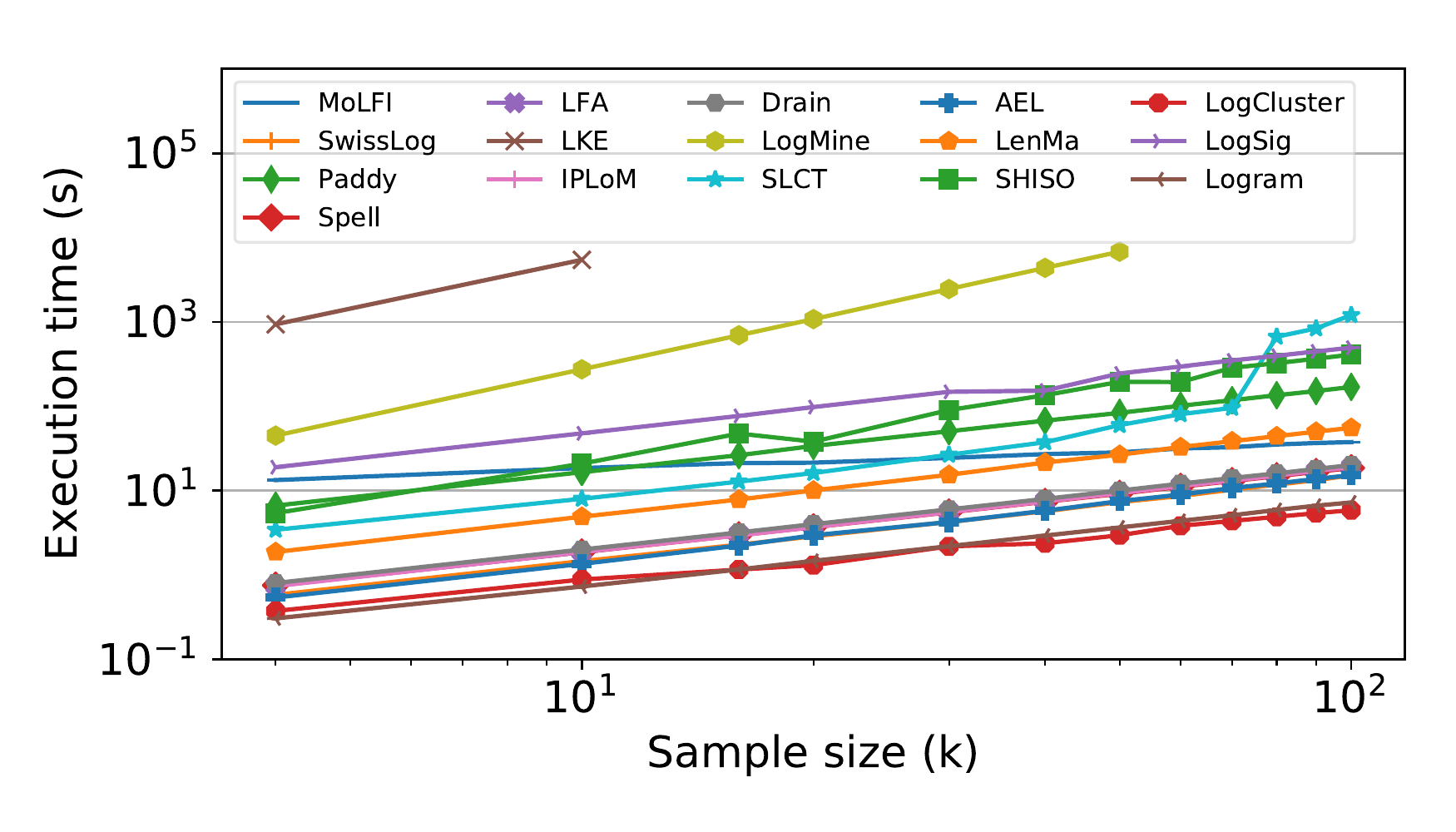}}
\hfill
\subfloat[Linux]{\includegraphics{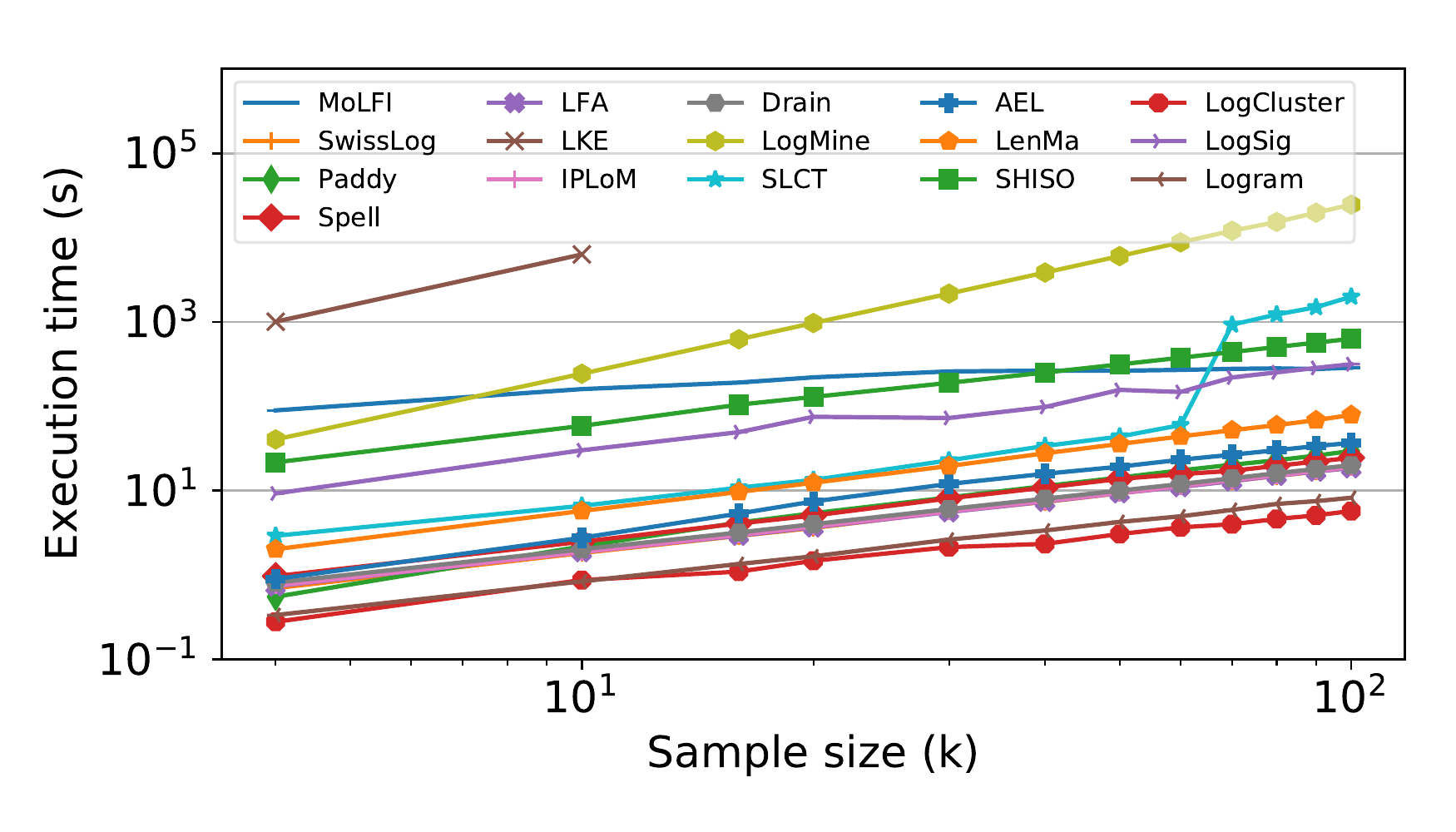}}
\hfill
\subfloat[HDFS]{\includegraphics{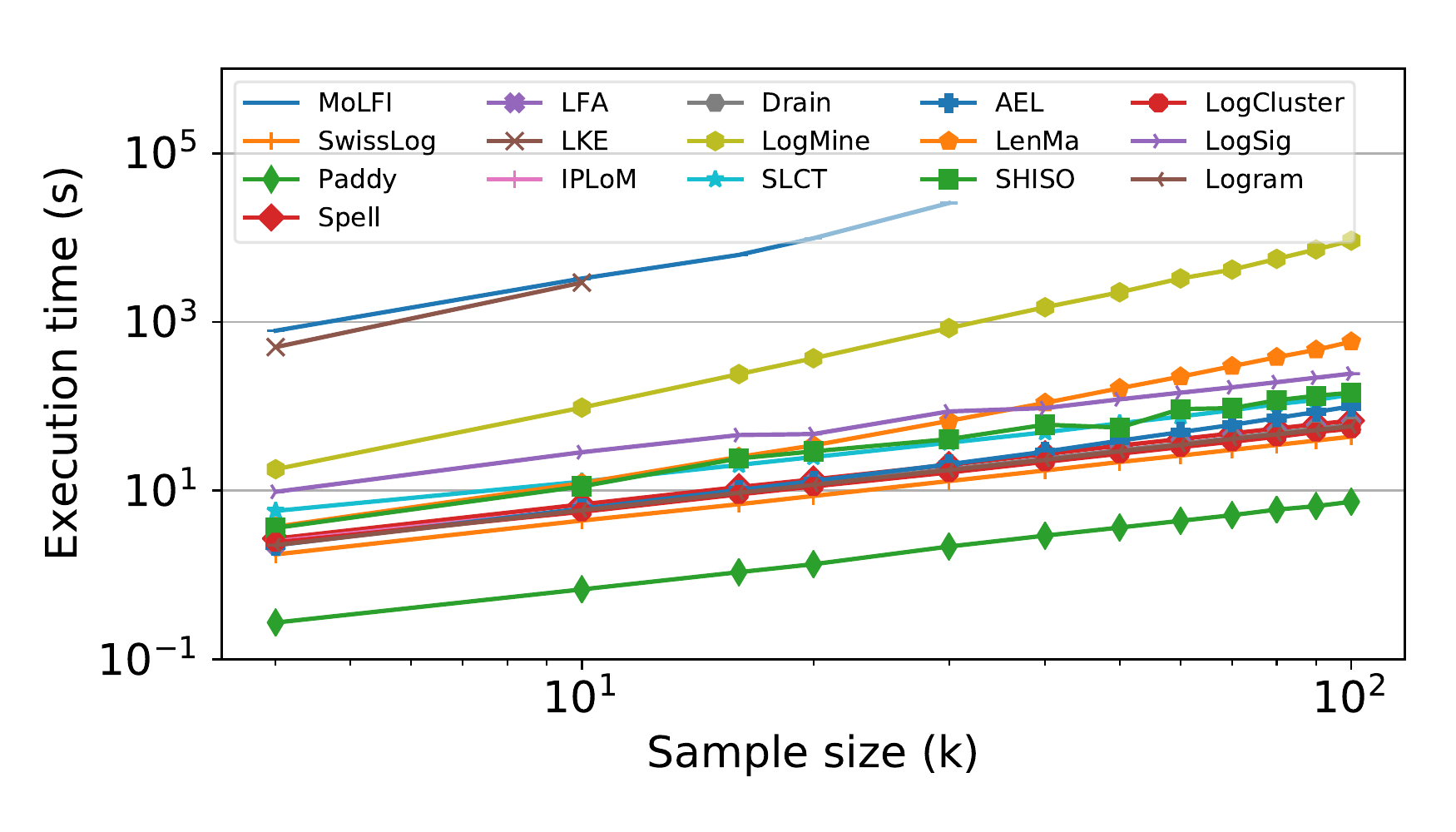}}
\hfill
\subfloat[Hadoop]{\includegraphics{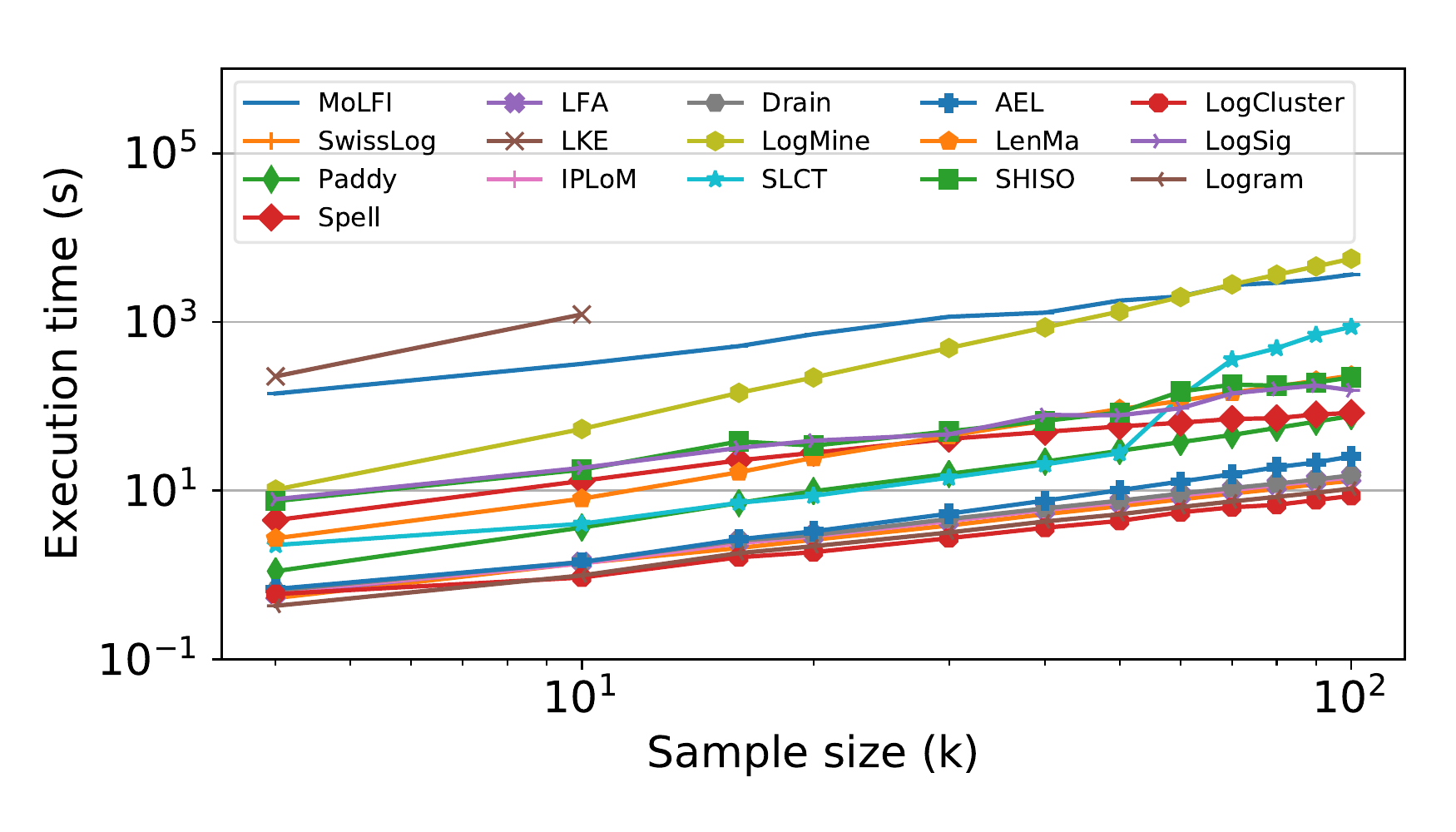}}
\hfill
\subfloat[Spark]{\includegraphics{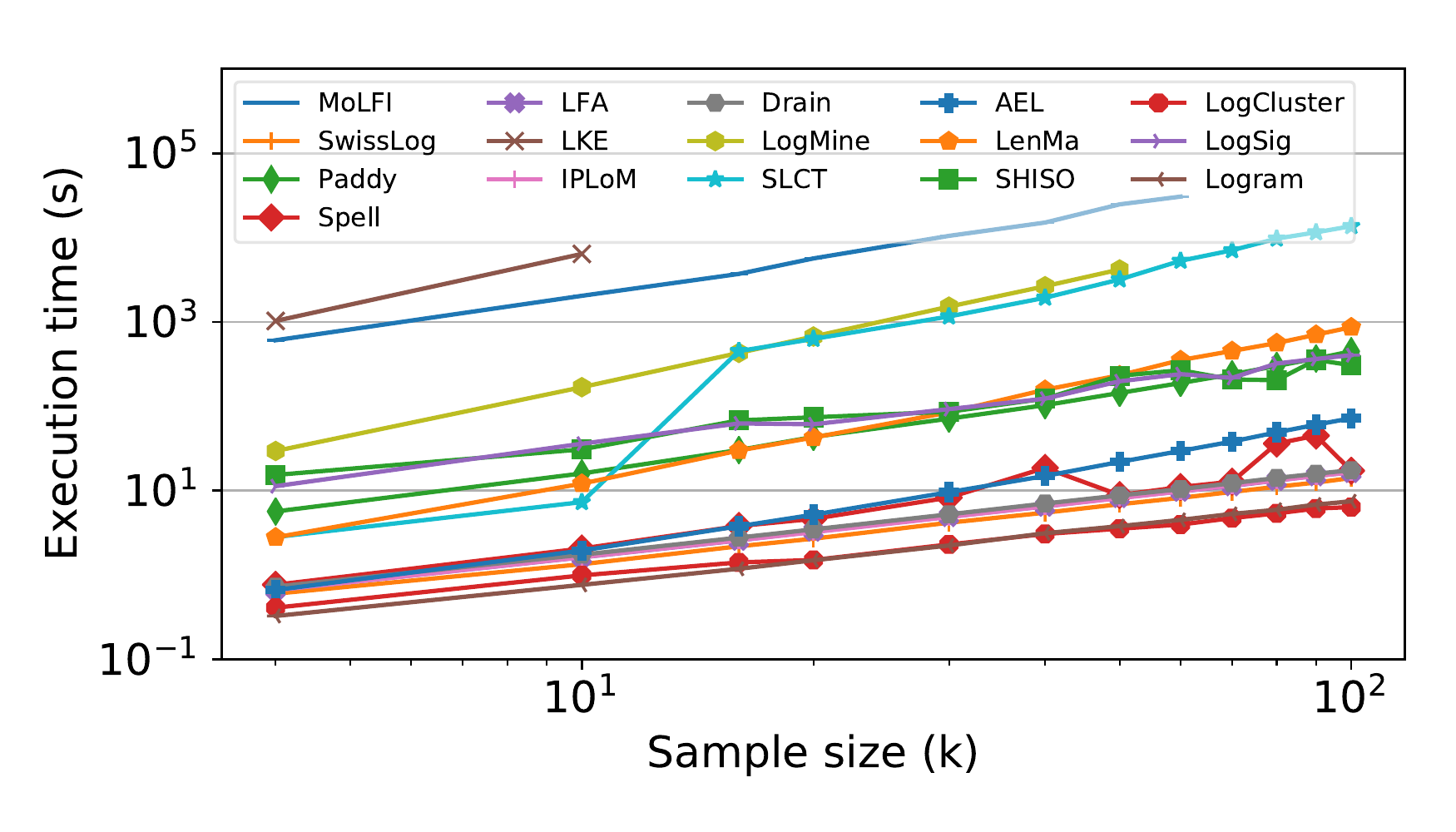}}
\hfill
\subfloat[HPC]{\includegraphics{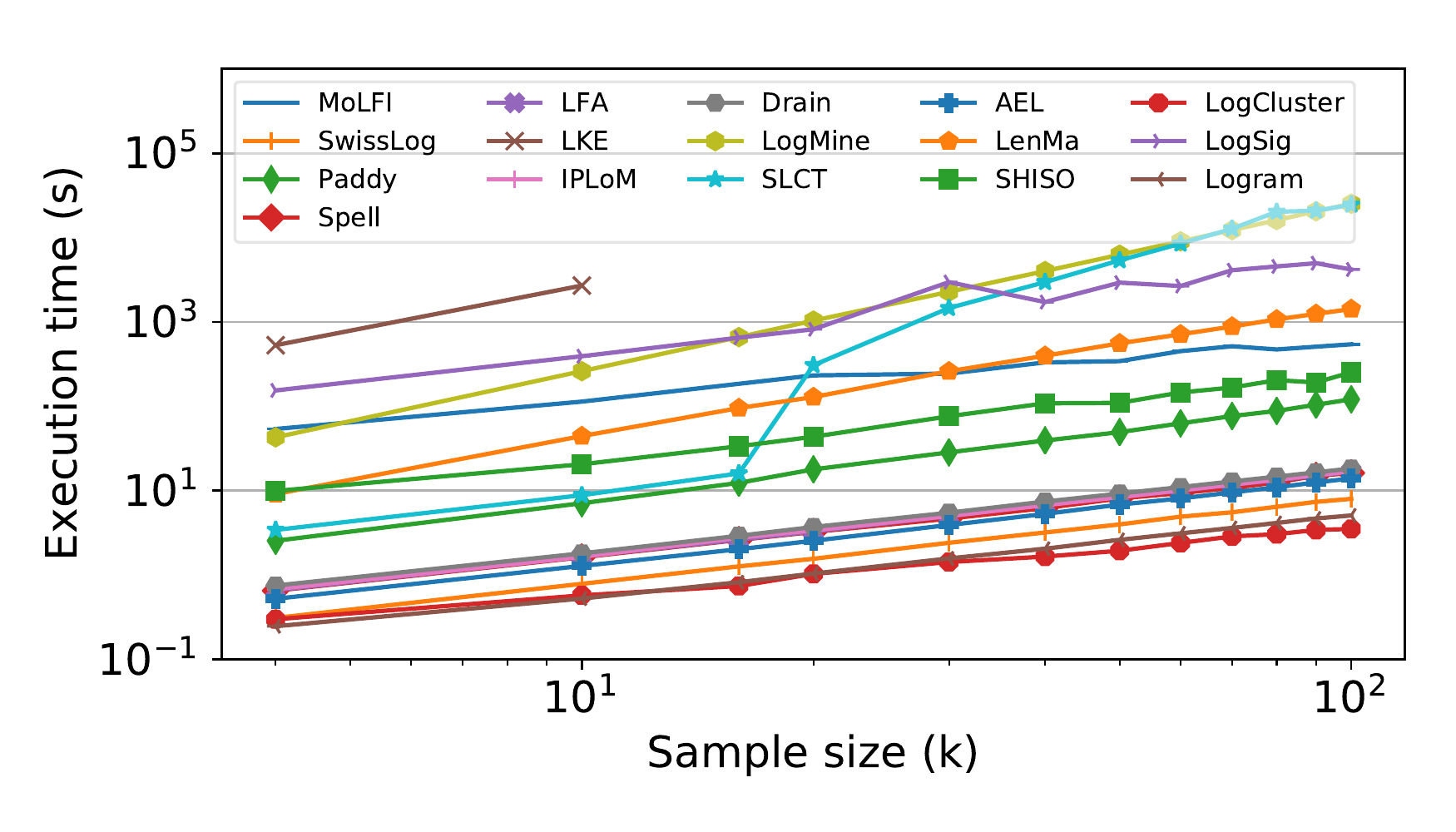}}
\hfill
\subfloat[Proxifier]{\includegraphics{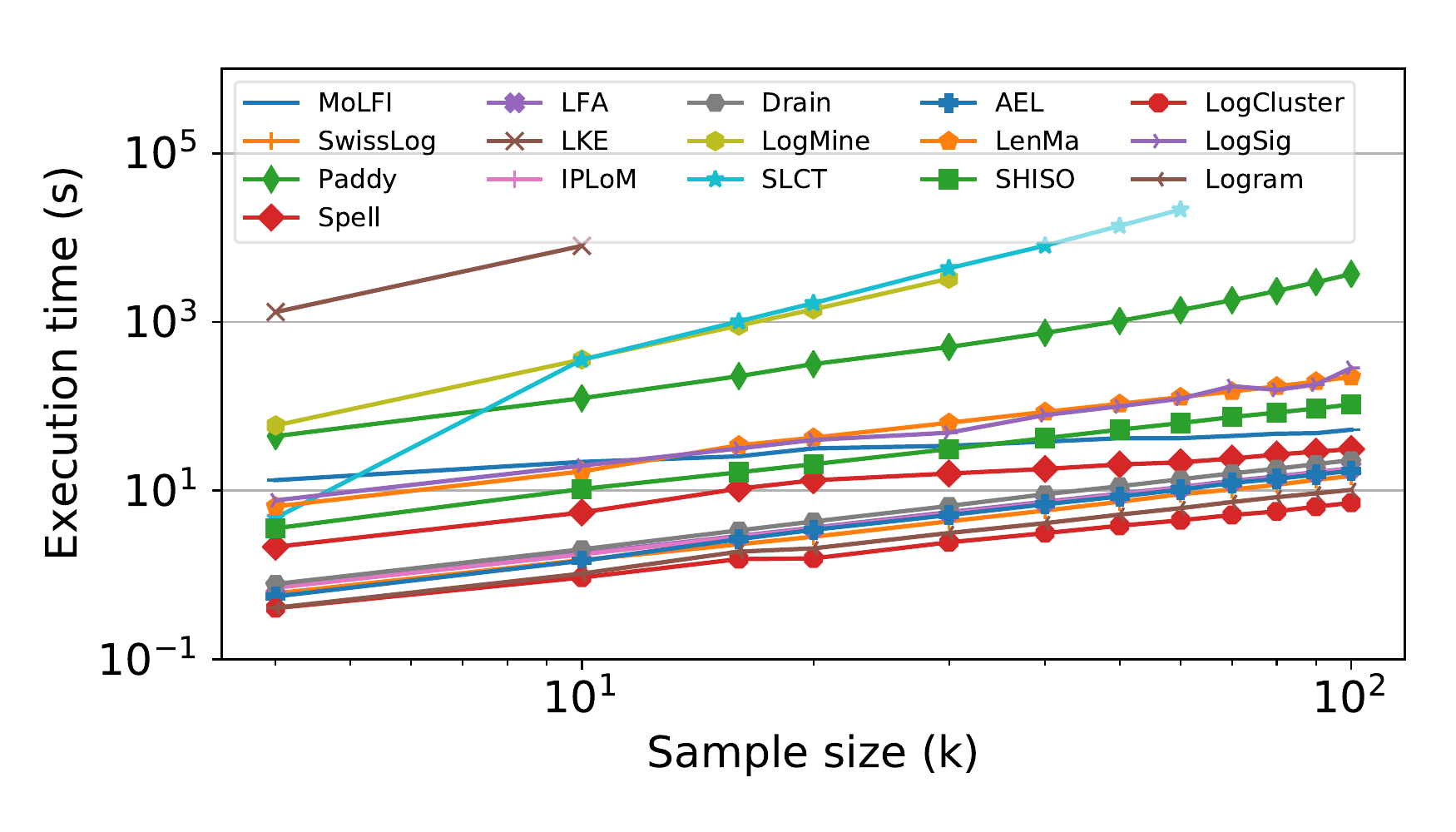}}
\hfill
\subfloat[HealthApp]{\includegraphics{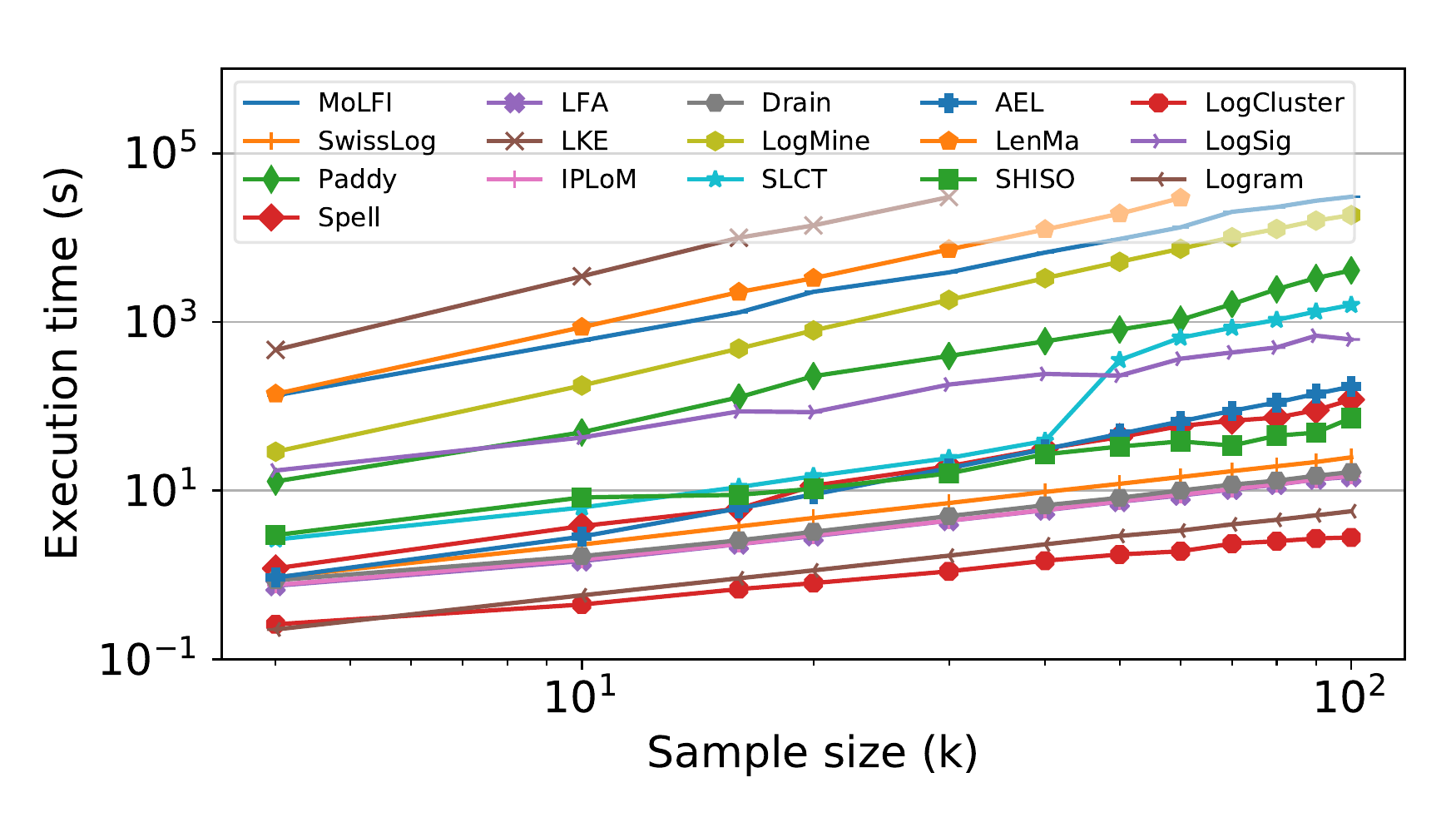}}
\caption{The log parsing time of 16 open-source log parsers on sample datasets with different number of log entries.}
\label{fig:efficiency-samples}
\end{figure*}

\subsubsection{Parsing time}\label{sec:efficiency}
Besides accuracy, parsing time is another critical performance criterion, especially for real-time log analysis tasks. Although users can estimate a log parser's run time by analyzing its time complexity with respect to the number of input log entries, this is inadequate in practice since there can be also other impactful factors for each log parser. 

We devote this section to characterizing the execution time of 16 open-source log parsers with real system logs. All the experiments are conducted on a commodity server equipped with Intel Xeon CPU E5-2670 @ 2.3GHz (48 cores across two NUMA nodes) and 64GB RAM. Each experiment is executed on an isolated CPU core to reduce system-level interference. Since the log parsers under test are all implemented as single-thread applications, we assign one CPU core to each solution for the latency test. Similar to prior works, we keep the default parameters and preprocessing rules. To ensure fairness, we use the online version of Logram (without the Spark acceleration). We also exclude NuLog from this evaluation as its neural network training is way slower than other solutions, even with GPU acceleration. 
We select 12 datasets from LogHub, i.e., the first 12 datasets in Table~\ref{tab:dataset}, and randomly sample log entries to compose the validation datasets. The sample sizes include 4k, 10k, 16k, 20k, 30k, 40k, 50k, 60k, 70k, 80k, 90k, and 100k. 
We measure the execution time for each solution on all the datasets and illustrate the results in Fig.~\ref{fig:efficiency-samples}. 
As some log parsers, especially LKE, require excessively long execution time to parse large log samples, we skip the experiments at some point to save time, yet the general trend still holds. 

As we observe from Fig.~\ref{fig:efficiency-samples}, the execution time of most solutions can approximately scale linearly with the sample size. 
The heuristic solutions achieve lower overall execution time than other solutions. In particular, IPLoM, SwissLog, AEL, and Spell consistently achieve low execution times, and they can finish parsing 100k log files in around 10 seconds. These heuristic solutions use simple data structures and control logic that significantly reduces the parsing time. FPM-based solutions, such as LFA, LogCluster, and Logram, manage to achieve comparable execution efficiency as the preceding heuristic solutions. 
The execution time of clustering-based solutions, i.e., LKE, LogMine, and LogSig, also scales linearly with sample size, but they take considerably longer to finish. LKE presents the highest parsing time due to its $O(n^2)$ time complexity, making it incapable of processing large datasets in a reasonable time. LogMine is also slow mainly because of its vast message merging overhead. LogSig performs slightly better, but its clustering method is still slow to converge, and its execution time also depends on the starting point.

The execution times of other solutions are divergent across different datasets even if the total number of log entries are the same, which means other factors can also impact their execution time. 
For instance, 
LenMa is slower than other heuristic solutions as it is less performant on datasets with highly varied message lengths.
MoLFI has comparable results to LenMa on most datasets, but its execution time increases significantly on the distributed systems datasets (e.g., Hadoop, HDFS, Spark, and OpenStack) with more diversified log formats.
SLCT is the only solution that does not scale linearly on most datasets because it relies on the frequent tokens to match the log messages for pattern mining. As the sample size increases, so is its frequent vocabulary set, which consequently increases the computation overhead of SLCT. This phenomenon is particularly obvious on complex datasets that contain large vocabularies, which explains the SLCT's sudden jump on complex datasets such as Spark, Hadoop, and HPC. 

\subsection{Operational features}\label{sec:features}
Besides the performance indicators, there are also several noteworthy operational features, i.e., the parsing mode, preprocessing, and parameter tuning. The parsing mode alludes a log parser's compatibility with the ensuing log analytics applications, while preprocessing and parameter tuning reveal a log parser's accessibility to average users. In this section, we summarize and discuss these features combining the benchmark results for the open-source solutions. When applicable, we also discuss some alternative methods we derived from the related works. 

\subsubsection{Parsing mode}
Existing log parsers can operate in three different modes, namely {\it online}, {\it offline}, and {\it hybrid}.
Offline log parsers need to process all the log messages in batches. 
Most relatively early log parsers such as SLCT, LKE, LogMine, and IPLoM operate in offline, batch processing mode.
Intuitively, offline parsing mode should lead to satisfactory performance as it allows log parsers to scan all the messages and parse the logs with a global view. 
However, offline log parsers cannot allow for real-time log analytics, making them ill-suited for hyper-scale distributed systems. 

Many log parsers embrace the online streaming mode to cope with this challenge.
In fact, with the emergence of more specialized heuristics, online log parsers have managed to attain comparable (if not better) performance than their offline counterparts~\cite{he2017drain,du2018spell,dai2020logram}.
Online parsers such as Spell, Drain, and SHISO operate on streams and can be readily adapted for data-driven analysis in real-time. They have two significant advantages over offline solutions. First, online parsers can interpret newly collected logs on the fly and incrementally refine their internal parsing results without going through the offline training phase, making them ideal for real-time tasks such as system monitoring and fault diagnosis.
Offline log parsers generally fall short in this regard. Some solutions, such as LKE and LogSig, can take days to parse large datasets. Second, unlike offline solutions online parsers do not need to load the entire input data (which can be prohibitively huge) into the memory space, making them more accessible to users that do not possess enough resources. 

Besides these two standard modes, some log parsers operate in a hybrid mode, which entails an offline training phase and an online parsing phase, just like a typical machine learning pipeline. For instance, NuLog requires offline training to populate the model parameters via backward propagation. After that, it serves the model and parses the input logs in online mode. This self-supervised approach can better learn the characteristics of logs from different sources and overcome the limitations of heuristic approaches that fail to generalize for unobserved log formats.
According to our benchmarking results, NuLog presents the highest overall parsing accuracy on ten datasets. 
Such advantages have also been observed by other hybrid-mode log parsers~\cite{coustie2020meting,xiao2020lpv,nguyen2021logdtl}. Although more extensive evaluations are still needed to prove the validity of these solutions, hybrid mode presents an intriguing future direction.

\subsubsection{Parameter tuning}\label{sec:tuning}
Many existing log parsers expose some parameters to allow users to fine-tune the performance. Nonetheless, for three reasons, correctly tuning these parameters is a non-trivial task. First off, the exposed parameters usually have distinct sensitivities on the input logs, which can only be understood through extensive benchmarking. For example, SLCT exposes a support threshold to locate frequent words from the input logs. An overly strict threshold cannot identify all the relevant tokens and their associated patterns, while an overly loose threshold furthers the computation overhead and may cause overfitting~\cite{vaarandi2015logcluster}. Second, the tunable parameters may have an implicit impact on each other, which requires a joint evaluation campaign to find the most suitable combination of parameters. Third, the tuning process must be repeated for new logs to guarantee the best log parsing results~\cite{zhu2019tools}. 
\begin{figure}[!tb]
\begin{center}
\includegraphics[width=0.5\textwidth]{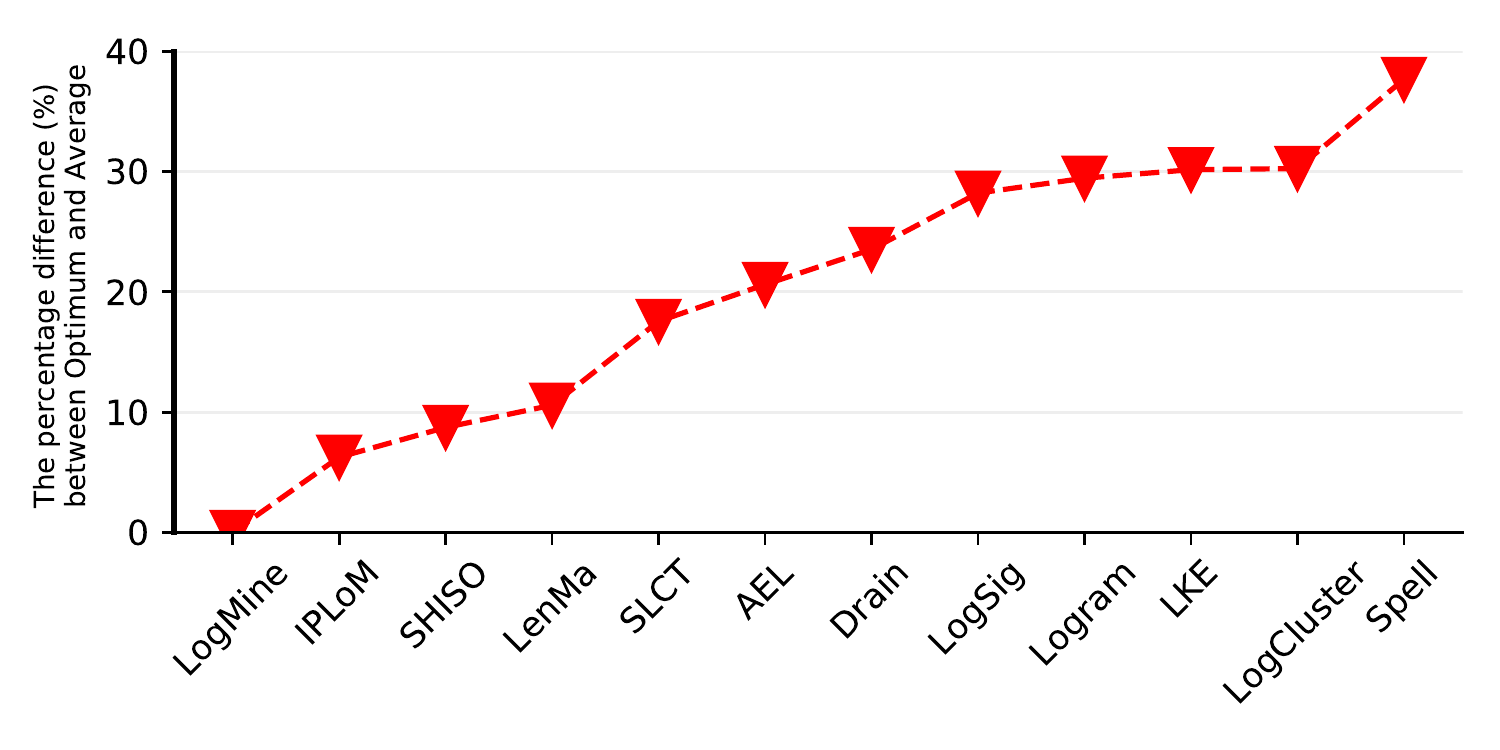}
\caption{The general impact of parameter tuning.}
\label{fig:tuning}
\end{center}
\end{figure}

We devote this section to investigating the impact of parameter tuning. Staring from the default values, we sensibly perturb the parameters for each solution and collect the obtained PA. The same procedure is repeated on all the labeled 2k datasets in Sec.~\ref{sec:accuracy}. To reflect the impact in a general sense, we calculate the ratio of optimal PA over the average PA, as illustrated in Fig.~\ref{fig:tuning}. We notice that some solutions from the LogPAI team have a slight difference from the original papers. In this case, we opt to use LogPAI's implementations. 
According to our study, more tunable parameters do not necessarily mean higher tuning overhead. Some log parsers are relatively easy to tune because their parameters have negligible sensitivity. 
For instance, SHISO exposes four tunable parameters, i.e., the maximal child nodes and three format thresholds, that have little impact on the accuracy.
IPLoM has five tunable parameters, but these parameters also have a minimal impact. 

Conversely, some log parsers only expose a few parameters that require considerate tuning.
For instance, Spell's message type threshold can strikingly impact accuracy ($\approx 40\%$). According to our experience, NuLog has the highest parameter tuning overhead. It exposes three parameters, i.e., \#epochs, $k$, \#samples, that can cover a vast range of values. The \#epoch value has to be carefully tuned across different datasets, and $k$ can only be fixed via cross-validation. We failed to attain even $20\%$ PA on the Linux dataset. 
These solutions require a deep understanding of their design internals and data characteristics, which can be highly challenging for average users. 
One way to reduce this overhead is to tune parameters on small samples and directly apply them to larger datasets. Although this transferred parameter tuning performs decently in some cases, many solutions still fail to sustain satisfactory performance on large system logs with more disparate data attributes~\cite{he2016evaluation}. 

Some solutions are designed with this difficulty in mind and thus do not require manual parameter tuning. For example, LFA, SwissLog, and Paddy automatically shield users from this potentially overwhelming undertaking. Other solutions have also embraced such an approach: 
Slop~\cite{zhao2018slop} defines a non-linear threshold criterion that can adapt to several system logs and thus does not require any manual parameter tuning; LogSimilarity~\cite{kimura2018proactive} employed an online classification algorithm to adjust the involved parameters incrementally.
Although present research on automatic parameter tuning is still limited, it is a promising direction to implement new solutions that can be adaptive to different settings without human intervention.

\begin{figure}[!tb]
\begin{center}
\includegraphics[width=0.5\textwidth]{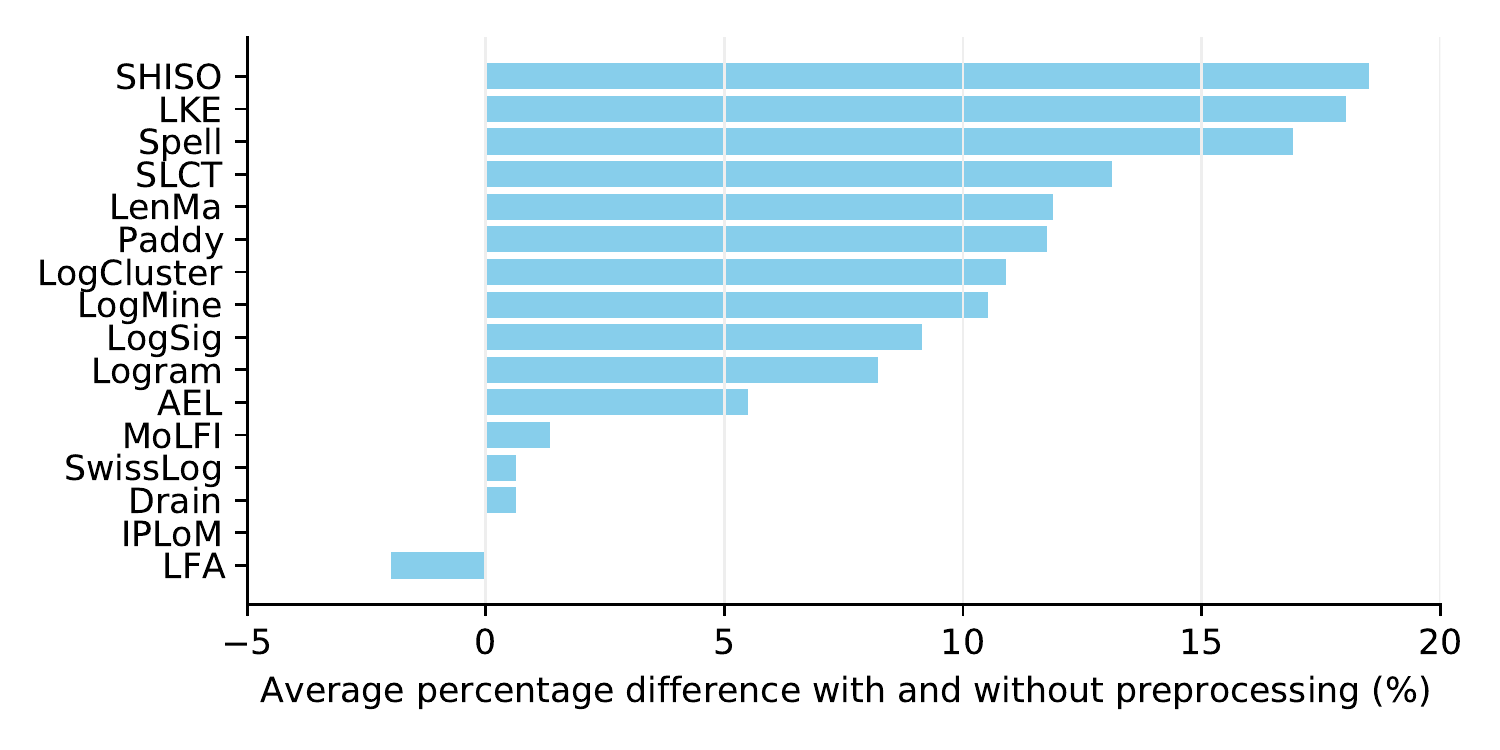}
\caption{The general impact of preprocessing}
\label{fig:preprocessing}
\end{center}
\end{figure}

\begin{table*}[!tb]
\renewcommand{\arraystretch}{1}
\centering
\footnotesize
\caption{A summary of the 17 open-source log parsers under evaluation}
\resizebox{0.99\textwidth}{!}{
\begin{tabular}{|c|c||cc|cc||ccc|cc|c|}
\multicolumn{2}{c}{} & \multicolumn{4}{c}{\bf Performance features} & \multicolumn{6}{c}{\bf Operational features}  \\
\hline 
\multirow{1}{*}{} & \multirow{2}{*}{\bf Log parsers}  & \multicolumn{2}{c|}{\bf Average accuracy} & \multicolumn{2}{c||}{\bf Parsing time} & \multicolumn{3}{c|}{\bf Parsing mode} & \multicolumn{2}{c|}{\bf Parameter Tuning} & \bf Preprocess. \\  \cline{3-11}  
\multirow{1}{*}{}  & & FM (\%) & PA (\%) & Complex. & Latency & Offline & Online & Hybrid & Parameters & Impact & Impact \\\hline \hline
\multirow{4}{*}{\rotatebox[origin=c]{90}{FPM}}& \bf SLCT~\cite{vaarandi2003data} & 88.98 & 62.94 &  $O(n)$ & \em medium & \checkmark & & &  \em support & \em medium & \em high \\ \cline{2-12}
& \bf LFA~\cite{nagappan2010abstracting} & 93.20 & 64.33 & $O(n)$  & \em low & \checkmark &  &  & -- & \em zero & \em high  \\ \cline{2-12}
& \bf LogCluster \cite{vaarandi2015logcluster} & 92.77 & 65.25 &$O(n)$ & \em low & \checkmark &  &  & \em rsupport & \em high & \em high \\ \cline{2-12}
& \bf Logram~\cite{dai2020logram}  & 81.73 & 53.57 & $O(n)$ & \em low &  &  & \checkmark & \em double-/tri-Threshold & \em high & \em medium \\ 
 \hline \hline
\multirow{4}{*}{\rotatebox[origin=c]{90}{Clustering}} & \bf LogSig~\cite{tang2011logsig}  & 90.03 & 52.93 & $O(n)$ & \em medium &  \checkmark &  &  & \em groupNum & \em high & \em medium \\ \cline{2-12}
& \bf LKE~\cite{fu2009execution} & 78.80 & 61.33 & $O(n^2)$ & \em high & \checkmark &  &  & \em split\_threshold & \em high & \em high \\ \cline{2-12}
& \bf LenMa~\cite{shima2016length} & 93.63 &  76.64 & $O(n)$ & \em medium &  & \checkmark &  & \em threshold & \em medium & \em high \\ \cline{2-12}
& \bf LogMine \cite{hamooni2016logmine}  & 93.62 & 73.52 & $O(n)$ & \em high & \checkmark  &  &  & \em max\_dist,k,level & \em low & \em medium \\ \hline \hline
 \multirow{14}{*}{\rotatebox[origin=c]{90}{Heuristic approaches}} & \bf \multirow{2}{*}{AEL~\cite{jiang2008abstracting,jiang2008automated}}  & \multirow{2}{*}{96.81} & \multirow{2}{*}{79.12} & \multirow{2}{*}{$O(n)$} & \em \multirow{2}{*}{low} & \multirow{2}{*}{\checkmark} &  &  & \em minEventCount, & \em \multirow{2}{*}{low} & \em \multirow{2}{*}{medium} \\ 
& & & & & &  &  &  & \em merge\_percent &&  \\ \cline{2-12}
& \bf MoLFI~\cite{messaoudi2018search}  & 88.94 & 60.67 &  $O(n^2)$ & \em high & \checkmark &  &  & -- & \em zero & \em low \\ \cline{2-12}
& \multirow{2}{*}{\bf IPLoM~\cite{makanju2011lightweight}}  & \multirow{2}{*}{96.76} & \multirow{2}{*}{75.66} & \multirow{2}{*}{$O(n)$} & \em \multirow{2}{*}{low} & \multirow{2}{*}{\checkmark} &  &  &  \em CT,FT,PT &\em \multirow{2}{*}{low} & \em \multirow{2}{*}{low} \\ 
& & & & & &  &  &  & \em lowerbound/upperbound & &  \\ \cline{2-12}
& \bf \multirow{4}{*}{SHISO~\cite{mizutani2013incremental}}  & \multirow{4}{*}{92.09}& \multirow{4}{*}{67.80} & \multirow{4}{*}{$O(n)$} & \em \multirow{4}{*}{medium} &  & \multirow{4}{*}{\checkmark} &  & \em maxChildNum & \em \multirow{4}{*}{low} & \em \multirow{4}{*}{high} \\ 
& & & & & &  &  &  & \em mergeThreshold & &  \\
& & & & & &  &  &  & \em formatLookupThreshold & &  \\
&& & & & &  &  &  & \em superFormatThreshold & &  \\\cline{2-12}
&\bf Drain~\cite{he2017drain}  & 97.74 & 86.54 & $O(n)$ & \em low &  & \checkmark  &  & \em st, depth & \em medium & \em low \\ \cline{2-12}
& \bf Spell~\cite{du2016spell} & 96.07 & 79.26 & $O(n)$ & \em medium &  & \checkmark &  & \em tau & \em high & \em high \\ \cline{2-12}
&\bf NuLog~\cite{nedelkoski2020self}  & 97.13 & 94.31 & -- & \em high &  &  & \checkmark & \em \#epochs, \#samples, k & \em high & \em high \\ \cline{2-12}
&\bf Paddy~\cite{huang2020paddy} & 89.63 & 71.43 & $O(n)$ &  \em medium &  & \checkmark  & & -- & \em zero& \em high \\ \cline{2-12}
& \bf SwissLog~\cite{li2020swisslog} & 99.56 & 93.29 & $O(n)$ & \em low &  &  & \checkmark & -- & \em zero & \em low \\ \hline
\end{tabular}}
\label{tab:feature}
\end{table*}

\subsubsection{Preprocessing}\label{sec:prep}
Besides parameter tuning, existing solutions also heavily rely on human involvement for preprocessing, which may include interpreting raw log messages, removing irrelevant segments, and substituting redundant tokens (e.g., timestamps, IP addresses, unique identifiers) to reduce the parsing noise and enhance the performance. 
All the open-source solutions we have evaluated rely on empirically composing regular expressions for preprocessing. 
According to prior research, log parsers can generally benefit from fine-grained preprocessing based on the characteristic of the input logs. However, this has to be conducted with caution, as incorrect preprocessing may inversely lead to performance degradation~\cite{liu2020fastlogsim,he2016evaluation,he2017towards}. It is thus necessary to understand the impact of preprocessing on existing log parsers. 

In this part, we rerun the accuracy tests without the preprocessing process on all the 2k datasets and compare the difference with the corresponding original accuracy. The regular expressions were empirically composed by the LogPAI team for each dataset. These rules only perform basic textual processing and can exhibit the bottom-line impact on each solution.
The average PA difference with and without preprocessing for each solution is depicted in Fig.~\ref{fig:preprocessing}. According to the results, preprocessing has little impact on Drain, SwissLog, and IPLoM, making them more accessible to average users without prior knowledge. On the contrary, SHISO, LKE, Spell show the highest overall difference ($\geq 15\%$). SLCT, Paddy, Lenma, and LogCluster are also sensitive to preprocessing ($\geq 10\%$). In particular, Lenma and LogCluster can only achieve $0.1\%$ PA without preprocessing on the Proxifier dataset. NuLog's preprocessing requires specialized filters for the tokenization and masking, which calls for prior knowledge of the log formats. We do not present NuLog's result in Fig.~\ref{fig:tuning}, as we are uncertain of the filters for all test datasets.
These solutions require more specialized preprocessing rules to unleash their full potential. We have also found a negative result for LFA's accuracy, reflecting its particular preprocessing rules requirement. 

Although existing solutions widely use regular expressions, they are too general and fail to provide customized preprocessing for individual log parsers. We have even observed sightly decreased accuracy in some rare cases due to improper preprocessing (e.g., LFA). 
To this end, some other log parsers propose more advanced preprocessing procedures.
For example, POP~\cite{he2017towards} provides two types of preprocessing functions for users to trim redundant fields and prioritize specific log events.
Aside from the empirical understanding of input data, some solutions can benefit from detailed knowledge of the target systems.
For instance, LogTree~\cite{tang2010logtree} relies on grammar parsers to preprocess log data and build the hierarchical message segments. These grammar parsers are highly dependent on specific system programs and usually require prior knowledge of the relevant domains. Users can also specify different clustering algorithms based on their needs, e.g., LEARNPADS~\cite{zhu2010incremental} requires a user-specified log format description to preprocess log entries. 
Although these solutions enable more sophisticated parsing by consolidating user experiences and domain-specific knowledge, they are also more difficult for average users to master.

Another critical aspect of preprocessing, as pointed out by existing works~\cite{el2020systematic}, is how to handle the delimiters for heterogeneous logs. Most existing solutions, such as Spell, consider the commonly used signs such as space and equal as delimiters. Nonetheless, such an assumption does not always hold due to the free-text nature of system log messages, which can result in unexpected parsing errors~\cite{makanju2011lightweight}. Some solutions try to tackle this issue by simultaneously considering multiple signs as delimiters~\cite{li2020swisslog,wurzenberger2019aecid,tovarvnak2019normalization}. For instance, SwissLog employs a set of 5 delimiters (\{, . ; : "\}) to tokenize the raw log messages more precisely. 
One-to-One allows users to specify the delimiters based on knowledge of the input log formats. 
Although these methods can alleviate the tokenization mistake, they cannot altogether avoid it since log entries can always employ more distinctive characters as delimiters.
To overcome this limitation, {Wurzenberger et al.~\cite{wurzenberger2020creating}} proposed a character-based log parser that performs a character-wise comparison to evaluate the similarity of two log messages, avoiding the entire tokenization step. Although their evaluation results were promising, this novel approach is still in its infancy, and more extensive benchmarking is needed to validate its applicability and performance thoroughly. 

\subsubsection*{Summary} 
To sum up, we epitomize all the relevant performance and operational features of the 17 open-source solutions in Table~\ref{tab:feature}. In particular, for the parsing accuracy, we list the average FM and PA for each solution. For the parsing time, we list both the time complexity and the general latency (which is derived from the evaluation of Sec.~\ref{sec:efficiency}). For parameter tuning, we enumerate all the parameters and their overall impact on the final results. We have also shown the impact of preprocessing on the final outcome.
Although we cannot comprehensively evaluate all the existing solutions due to lack of information, we believe our benchmarking campaign can provide general guidance to choose the most suitable open-source solutions. Whenever applicable, we have also discussed the operational features of some closed-source solutions, which can be reused to implement new solutions.


\section{Challenges and future directions}\label{sec:future}
Despite the manifold solutions and overall robust outcomes, log parsers still face some limitations that prevent them from pervasive deployment in a production environment.
This section contemplates future challenges for log parsing technology and discusses the potentially relevant directions from the systems perspective. 

\subsection{The scarcity of public datasets}
Like any data-driven approach, log parsers require abundant data to thoroughly validate and optimize the performance, especially labeled datasets. 
Unfortunately, 
due to the privacy concerns with personal and commercial data, there are still only a few public log datasets~\cite{xu2009detecting,oliner2007supercomputers,he2020loghub,nguyen2021logdtl,studiawan2020automatic}. 
Although these datasets have tremendously boosted the advancement of log parsing, real-world system logs (especially those from production environments) are still sought to increase the diversity of available data~\cite{he2016evaluation}. 
Since the performance of log parsers fluctuates substantially across datasets, the lack of data diversity hinders the continuous improvement of log parsing solutions. 
He et al.~\cite{he2017towards} tried to mitigate this problem by procedurally generating derivative samples with diversified properties from original public datasets, and Nguyen et al.~\cite{nguyen2021logdtl} explored transfer learning techniques to alleviate data scarcity. However, these are not sustainable solutions in the long run. 
It is critical to continuously collect, label, and disclose more logs from real systems, services, and applications to facilitate the design and validation of novel log parsers, which calls for a concerted effort from both industry and academia. 

\subsection{Limited generalizability}
Albeit existing log parsers can generally achieve sound outcomes, they still inevitably suffer from inaccuracies on most log datasets. As discussed in Sec.~\ref{sec:accuracy}, these artifacts are caused by the intrinsic limitations of each existing solution with specific log properties such as the log size, message length, event distribution, vocabulary size, and so on. 
Although the accuracy of a log parser can always be improved with more specialized heuristic algorithms and optimization techniques, it is challenging to keep pace with the rapid emergence of new log types induced by system requirement evolution.  

As each log parser has specific performance characteristics on different datasets, one possible direction is to combine multiple parsers to compensate for the drawbacks of a single solution and thus enhance the overall performance. 
For instance, Xie et al.~\cite{xie2019pvalue} proposed a p-value guided approach that aggregates all the templates extracted by four state-of-the-art log parsers, including IPLoM~\cite{makanju2011lightweight}, LogCluster~\cite{vaarandi2015logcluster}, AEL~\cite{jiang2008automated}, and Spell~\cite{du2016spell} to improve the effectiveness of anomaly detection for the industrial IoT systems. According to their evaluation results, the proposed method achieved higher accuracy than any single log parser. 

Another orientation is to exploit advanced Machine Learning (especially Deep Learning (DL)) techniques for sustainable performance. Our taxonomy involves some ML-based log parsers, as discussed in Sec.~\ref{sec:nlp}. These solutions have already shown encouraging outcomes. In particular, as one of the 17 evaluated open-source solutions, NuLog~\cite{nedelkoski2020self} achieves the highest accuracy on ten public datasets. Although the effectiveness of these solutions still requires further validation, we believe they can deliver more promising results with the rapid advancement of DL techniques. 
Based on our experience with NuLog, two potential drawbacks for the DL-based solutions are the high configuration difficulty and slow training process. They can be alleviated via system-level automation and acceleration, which will be thoroughly discussed in the following two subsections.

\subsection{Lack of automation}
Although existing log parsing solutions aim at enabling automatic log analysis, most of them still heavily rely on human intervention to achieve satisfactory performance. 
As discussed in the previous section, empirical and domain-specific knowledge can play a decisive role for a log parser, especially during the parameter tuning and preprocessing phases.
Also, as modern ICT systems are highly dynamic and constantly evolving, log parsers are expected to deal with a perpetual shift in concepts and data contents. Existing solutions usually neglect this level of automation, which hinders their adoption in real systems. 

In AI/ML domain, there are several popular frameworks such as Airflow~\cite{airflow}, Kubeflow~\cite{kubeflow}, and MLflow~\cite{mlflow}, that automatically manage the end-to-end provisioning of machine learning pipelines. As log parsers have a similar workflow, they can significantly benefit from a functionally equivalent framework that automates their delivery process, including log collection, data preprocessing, parameter tuning, algorithm training, continuous validation, and incremental deployment. Such a framework can significantly expedite the integration of log parsers in the production environment to facilitate advanced analysis and management. 

\subsection{Insufficient system-level acceleration}
Existing log parsers have incorporated many algorithmic techniques to accelerate processing. However, they do not give sufficient consideration to system-level acceleration to scale up the operation in real-world scenarios. 

As the volume of system logs increases, so does the resource footprint for log parsers. We believe log parsing can further benefit from system-level support for handling big data~\cite{blanas2010comparison}. 
For example, LogMine augmented with MapReduce~\cite{dean2008mapreduce} can achieve up to 5$\times$ speedup with multiple parallel workers~\cite{hamooni2016logmine}.
Similarly, POP~\cite{he2017towards}, Logram~\cite{dai2020logram}, Delog~\cite{agrawal2019delog}, and Logan~\cite{tak2016logan} were built on Spark clusters~\cite{zaharia2010spark} to benefit from the large-scale data processing capabilities.
Ren et al.~\cite{ren2019parallel} computed the weighted edit distance for LKE using GPUs and reduced the processing time by roughly $90\%$.
With the explosion of log data and the urgent need for real-time analytics, log parsers must extend the necessary support for heterogeneous accelerators  (e.g., APIs, SDKs, and GUIs). Such an extension can also smooth their integration with the ensuing log analytics applications already popularly deployed on GPU and Big Data clusters.

\section{Conclusion}\label{sec:conclusion}

This paper aims at providing a comprehensive survey of log parsers. According to their log data classification and template extraction methods, we exhaustively investigate existing solutions and organize them in an easily-accessible taxonomy. 
Then we systematically analyze their performance metrics (accuracy and parsing time) and operational features (parsing mode, parameter tuning, and preprocessing), extracting a consistent set of benchmark results on the most prevalent open-source solutions. We believe this survey provides a reasoned first-hand account on the whole solution space of log parsing. 
This work can also help practitioners select the most appropriate open-source solutions for their needs. We also envision the future directions to promote the continued development of novel accurate and scalable parsing techniques that may befit the evolution of log analysis requirements in modern ICT systems. 

\ifCLASSOPTIONcaptionsoff
 \newpage
\fi



\bibliographystyle{IEEEtran}
\bibliography{IEEEabrv,biblio}

\begin{thebibliography}{100}
\providecommand{\url}[1]{#1}
\csname url@samestyle\endcsname
\providecommand{\newblock}{\relax}
\providecommand{\bibinfo}[2]{#2}
\providecommand{\BIBentrySTDinterwordspacing}{\spaceskip=0pt\relax}
\providecommand{\BIBentryALTinterwordstretchfactor}{4}
\providecommand{\BIBentryALTinterwordspacing}{\spaceskip=\fontdimen2\font plus
\BIBentryALTinterwordstretchfactor\fontdimen3\font minus
  \fontdimen4\font\relax}
\providecommand{\BIBforeignlanguage}[2]{{%
\expandafter\ifx\csname l@#1\endcsname\relax
\typeout{** WARNING: IEEEtran.bst: No hyphenation pattern has been}%
\typeout{** loaded for the language `#1'. Using the pattern for}%
\typeout{** the default language instead.}%
\else
\language=\csname l@#1\endcsname
\fi
#2}}
\providecommand{\BIBdecl}{\relax}
\BIBdecl

\bibitem{hwang2013distributed}
K.~Hwang, J.~Dongarra, and G.~C. Fox, \emph{Distributed and cloud computing:
  from parallel processing to the internet of things}.\hskip 1em plus 0.5em
  minus 0.4em\relax Morgan Kaufmann, 2013.

\bibitem{jiang2008abstracting}
Z.~M. Jiang, A.~E. Hassan, P.~Flora, and G.~Hamann, ``Abstracting execution
  logs to execution events for enterprise applications (short paper),'' in
  \emph{2008 The Eighth International Conference on Quality Software}.\hskip
  1em plus 0.5em minus 0.4em\relax IEEE, 2008, pp. 181--186.

\bibitem{lonvick2001bsd}
C.~Lonvick, ``{The BSD syslog protocol},'' 2001.

\bibitem{fu2014developers}
Q.~Fu, J.~Zhu, W.~Hu, J.-G. Lou, R.~Ding, Q.~Lin, D.~Zhang, and T.~Xie, ``Where
  do developers log? an empirical study on logging practices in industry,'' in
  \emph{Companion Proceedings of the 36th International Conference on Software
  Engineering}, 2014, pp. 24--33.

\bibitem{mi2013toward}
H.~Mi, H.~Wang, Y.~Zhou, M.~R.-T. Lyu, and H.~Cai, ``Toward fine-grained,
  unsupervised, scalable performance diagnosis for production cloud computing
  systems,'' \emph{IEEE Transactions on Parallel and Distributed Systems},
  vol.~24, no.~6, pp. 1245--1255, 2013.

\bibitem{vaarandi2014using}
R.~Vaarandi and M.~Pihelgas, ``Using security logs for collecting and reporting
  technical security metrics,'' in \emph{2014 IEEE Military Communications
  Conference}.\hskip 1em plus 0.5em minus 0.4em\relax IEEE, 2014, pp. 294--299.

\bibitem{hamooni2016logmine}
H.~Hamooni, B.~Debnath, J.~Xu, H.~Zhang, G.~Jiang, and A.~Mueen, ``Logmine:
  Fast pattern recognition for log analytics,'' in \emph{Proceedings of the
  25th ACM International on Conference on Information and Knowledge
  Management}, 2016, pp. 1573--1582.

\bibitem{splunk}
\url{https://www.splunk.com/}.

\bibitem{prelert}
\url{http://www.prelert.com/docs/products/0.3/accessing-logfiles.html}.

\bibitem{logstash}
\url{https://www.elastic.co/logstash/}.

\bibitem{sumo}
\url{https://www.sumologic.com/}.

\bibitem{elasticsearch}
\url{https://www.elastic.co/elasticsearch/}.

\bibitem{logrhythm}
\url{https://logrhythm.com/}.

\bibitem{logsurfer}
\url{http://logsurfer.sourceforge.net/}.

\bibitem{loggly}
\url{https://www.loggly.com/}.

\bibitem{logentries}
\url{https://logentries.com/}.

\bibitem{graylog}
\url{https://www.graylog.org/}.

\bibitem{ossim}
\url{https://cybersecurity.att.com/products/ossim}.

\bibitem{braun2016syslog}
U.~Braun, Y.~Zaslavsky, and Y.~Teitz, ``Syslog parser,'' Nov.~1 2016, uS Patent
  9,483,583.

\bibitem{jackson1986introduction}
P.~Jackson, ``Introduction to expert systems,'' 1986.

\bibitem{xu2010system}
W.~Xu, ``System problem detection by mining console logs,'' Ph.D. dissertation,
  UC Berkeley, 2010.

\bibitem{prewett2003analyzing}
J.~E. Prewett, ``Analyzing cluster log files using logsurfer,'' in
  \emph{Proceedings of the 4th Annual Conference on Linux Clusters}.\hskip 1em
  plus 0.5em minus 0.4em\relax Citeseer, 2003.

\bibitem{mclean2020adaptive}
L.~McLean, ``Adaptive parsing and normalizing of logs at mssp,'' Mar.~24 2020,
  uS patent 10,599,668.

\bibitem{tang2010logtree}
L.~Tang and T.~Li, ``Logtree: A framework for generating system events from raw
  textual logs,'' in \emph{2010 IEEE International Conference on Data
  Mining}.\hskip 1em plus 0.5em minus 0.4em\relax IEEE, 2010, pp. 491--500.

\bibitem{van2008process}
W.~M. van~der Aalst and H.~E. Verbeek, ``Process mining in web services: The
  websphere case.'' \emph{IEEE Data Eng. Bull.}, vol.~31, no.~3, pp. 45--48,
  2008.

\bibitem{de2005web}
W.~De~Pauw, M.~Lei, E.~Pring, L.~Villard, M.~Arnold, and J.~F. Morar, ``Web
  services navigator: Visualizing the execution of web services,'' \emph{IBM
  Systems Journal}, vol.~44, no.~4, pp. 821--845, 2005.

\bibitem{dai2020logram}
H.~Dai, H.~Li, W.~Shang, T.-H. Chen, and C.-S. Chen, ``Logram: Efficient log
  parsing using n-gram dictionaries,'' \emph{arXiv preprint arXiv:2001.03038},
  2020.

\bibitem{chu2021prefix}
G.~Chu, J.~Wang, Q.~Qi, H.~Sun, S.~Tao, and J.~Liao, ``Prefix-graph: A
  versatile log parsing approach merging prefix tree with probabilistic
  graph,'' in \emph{2021 IEEE 37th International Conference on Data Engineering
  (ICDE)}.\hskip 1em plus 0.5em minus 0.4em\relax IEEE, 2021, pp. 2411--2422.

\bibitem{tao2021logstamp}
S.~Tao, W.~Meng, Y.~Chen, Y.~Zhu, Y.~Liu, C.~Du, T.~Han, Y.~Zhao, X.~Wang, and
  H.~Yang, ``Logstamp: Automatic online log parsing based on sequence
  labelling,'' \emph{Interface}, vol.~19, no.~03, p.~03, 2021.

\bibitem{tan2008salsa}
J.~Tan, X.~Pan, S.~Kavulya, R.~Gandhi, and P.~Narasimhan, ``Salsa: Analyzing
  logs as state machines.'' \emph{WASL}, vol.~8, pp. 6--6, 2008.

\bibitem{hellerstein2002discovering}
J.~L. Hellerstein, S.~Ma, and C.-S. Perng, ``Discovering actionable patterns in
  event data,'' \emph{IBM Systems Journal}, vol.~41, no.~3, pp. 475--493, 2002.

\bibitem{peng2007event}
W.~Peng, C.~Perng, T.~Li, and H.~Wang, ``Event summarization for system
  management,'' in \emph{Proceedings of the 13th ACM SIGKDD international
  conference on Knowledge discovery and data mining}, 2007, pp. 1028--1032.

\bibitem{beschastnikh2011leveraging}
I.~Beschastnikh, Y.~Brun, S.~Schneider, M.~Sloan, and M.~D. Ernst, ``Leveraging
  existing instrumentation to automatically infer invariant-constrained
  models,'' in \emph{Proceedings of the 19th ACM SIGSOFT symposium and the 13th
  European conference on Foundations of software engineering}, 2011, pp.
  267--277.

\bibitem{li2005integrated}
T.~Li, F.~Liang, S.~Ma, and W.~Peng, ``An integrated framework on mining logs
  files for computing system management,'' in \emph{Proceedings of the eleventh
  ACM SIGKDD international conference on Knowledge discovery in data mining},
  2005, pp. 776--781.

\bibitem{luotonen1995common}
A.~Luotonen, ``The common log file format,'' 1995.

\bibitem{hallam1996extended}
P.~M. Hallam-Baker and B.~Behlendorf, ``{Extended Log File Format: W3C Working
  Draft WD-logfile-960323},'' \emph{http://www.w3.org/TR/WD-logfile.html},
  1996.

\bibitem{team2016apache}
C.~Team, ``Apache commons logging-overview,'' 2016.

\bibitem{bae2020improving}
C.-s. Bae and S.-c. Goh, ``For improving security log big data analysis
  efficiency, a firewall log data standard format proposed,'' \emph{Journal of
  the Korea Institute of Information Security \& Cryptology}, vol.~30, no.~1,
  pp. 157--167, 2020.

\bibitem{ogle2004canonical}
D.~Ogle, H.~Kreger, A.~Salahshour, J.~Cornpropst, E.~Labadie, M.~Chessell,
  B.~Horn, J.~Gerken, J.~Schoech, and M.~Wamboldt, ``Canonical situation data
  format: The common base event v1. 0.1,'' \emph{IBM Corporation}, 2004.

\bibitem{topol2003automating}
B.~Topol, D.~Ogle, D.~Pierson, J.~Thoensen, J.~Sweitzer, M.~Chow, M.~A.
  Hoffmann, P.~Durham, R.~Telford, S.~Sheth \emph{et~al.}, ``Automating problem
  determination: A first step toward self-healing computing systems,''
  \emph{IBM white paper}, 2003.

\bibitem{fu2009execution}
Q.~Fu, J.-G. Lou, Y.~Wang, and J.~Li, ``Execution anomaly detection in
  distributed systems through unstructured log analysis,'' in \emph{2009 ninth
  IEEE international conference on data mining}.\hskip 1em plus 0.5em minus
  0.4em\relax IEEE, 2009, pp. 149--158.

\bibitem{oliner2008alert}
A.~J. Oliner, A.~Aiken, and J.~Stearley, ``Alert detection in system logs,'' in
  \emph{2008 Eighth IEEE International Conference on Data Mining}.\hskip 1em
  plus 0.5em minus 0.4em\relax IEEE, 2008, pp. 959--964.

\bibitem{li2020swisslog}
X.~Li, P.~Chen, L.~Jing, Z.~He, and G.~Yu, ``Swisslog: Robust and unified deep
  learning based log anomaly detection for diverse faults,'' in \emph{2020 IEEE
  31st International Symposium on Software Reliability Engineering
  (ISSRE)}.\hskip 1em plus 0.5em minus 0.4em\relax IEEE, 2020, pp. 92--103.

\bibitem{meng2019loganomaly}
W.~Meng, Y.~Liu, Y.~Zhu, S.~Zhang, D.~Pei, Y.~Liu, Y.~Chen, R.~Zhang, S.~Tao,
  P.~Sun \emph{et~al.}, ``Loganomaly: Unsupervised detection of sequential and
  quantitative anomalies in unstructured logs.'' in \emph{IJCAI}, 2019, pp.
  4739--4745.

\bibitem{chen2020logtransfer}
R.~Chen, S.~Zhang, D.~Li, Y.~Zhang, F.~Guo, W.~Meng, D.~Pei, Y.~Zhang, X.~Chen,
  and Y.~Liu, ``Logtransfer: Cross-system log anomaly detection for software
  systems with transfer learning,'' in \emph{2020 IEEE 31st International
  Symposium on Software Reliability Engineering (ISSRE)}.\hskip 1em plus 0.5em
  minus 0.4em\relax IEEE, 2020, pp. 37--47.

\bibitem{du2017deeplog}
M.~Du, F.~Li, G.~Zheng, and V.~Srikumar, ``Deeplog: Anomaly detection and
  diagnosis from system logs through deep learning,'' in \emph{Proceedings of
  the 2017 ACM SIGSAC Conference on Computer and Communications Security},
  2017, pp. 1285--1298.

\bibitem{lin2016log}
Q.~Lin, H.~Zhang, J.-G. Lou, Y.~Zhang, and X.~Chen, ``Log clustering based
  problem identification for online service systems,'' in \emph{2016 IEEE/ACM
  38th International Conference on Software Engineering Companion
  (ICSE-C)}.\hskip 1em plus 0.5em minus 0.4em\relax IEEE, 2016, pp. 102--111.

\bibitem{yuan2010sherlog}
D.~Yuan, H.~Mai, W.~Xiong, L.~Tan, Y.~Zhou, and S.~Pasupathy, ``Sherlog: error
  diagnosis by connecting clues from run-time logs,'' in \emph{Proceedings of
  the fifteenth International Conference on Architectural support for
  programming languages and operating systems}, 2010, pp. 143--154.

\bibitem{kobayashi2017mining}
S.~Kobayashi, K.~Otomo, K.~Fukuda, and H.~Esaki, ``Mining causality of network
  events in log data,'' \emph{IEEE Transactions on Network and Service
  Management}, vol.~15, no.~1, pp. 53--67, 2017.

\bibitem{zhang2017syslog}
S.~Zhang, W.~Meng, J.~Bu, S.~Yang, Y.~Liu, D.~Pei, J.~Xu, Y.~Chen, H.~Dong,
  X.~Qu \emph{et~al.}, ``Syslog processing for switch failure diagnosis and
  prediction in datacenter networks,'' in \emph{2017 IEEE/ACM 25th
  International Symposium on Quality of Service (IWQoS)}.\hskip 1em plus 0.5em
  minus 0.4em\relax IEEE, 2017, pp. 1--10.

\bibitem{salfner2007using}
F.~Salfner and M.~Malek, ``Using hidden semi-markov models for effective online
  failure prediction,'' in \emph{2007 26th IEEE International Symposium on
  Reliable Distributed Systems (SRDS 2007)}.\hskip 1em plus 0.5em minus
  0.4em\relax IEEE, 2007, pp. 161--174.

\bibitem{kimura2018proactive}
T.~Kimura, A.~Watanabe, T.~Toyono, and K.~Ishibashi, ``Proactive failure
  detection learning generation patterns of large-scale network logs,''
  \emph{IEICE Transactions on Communications}, 2018.

\bibitem{li2017flap}
T.~Li, Y.~Jiang, C.~Zeng, B.~Xia, Z.~Liu, W.~Zhou, X.~Zhu, W.~Wang, L.~Zhang,
  J.~Wu \emph{et~al.}, ``Flap: An end-to-end event log analysis platform for
  system management,'' in \emph{Proceedings of the 23rd ACM SIGKDD
  International Conference on Knowledge Discovery and Data Mining}, 2017, pp.
  1547--1556.

\bibitem{debnath2018loglens}
B.~Debnath, M.~Solaimani, M.~A.~G. Gulzar, N.~Arora, C.~Lumezanu, J.~Xu,
  B.~Zong, H.~Zhang, G.~Jiang, and L.~Khan, ``Loglens: A real-time log analysis
  system,'' in \emph{2018 IEEE 38th International Conference on Distributed
  Computing Systems (ICDCS)}.\hskip 1em plus 0.5em minus 0.4em\relax IEEE,
  2018, pp. 1052--1062.

\bibitem{shang2013assisting}
W.~Shang, Z.~M. Jiang, H.~Hemmati, B.~Adams, A.~E. Hassan, and P.~Martin,
  ``Assisting developers of big data analytics applications when deploying on
  hadoop clouds,'' in \emph{2013 35th International Conference on Software
  Engineering (ICSE)}.\hskip 1em plus 0.5em minus 0.4em\relax IEEE, 2013, pp.
  402--411.

\bibitem{makanju2011lightweight}
A.~Makanju, A.~N. Zincir-Heywood, and E.~E. Milios, ``A lightweight algorithm
  for message type extraction in system application logs,'' \emph{IEEE
  Transactions on Knowledge and Data Engineering}, vol.~24, no.~11, pp.
  1921--1936, 2011.

\bibitem{meng2020summarizing}
W.~Meng, F.~Zaiter, Y.~Huang, Y.~Liu, S.~Zhang, Y.~Zhang, Y.~Zhu, T.~Zhang,
  E.~Wang, Z.~Ren \emph{et~al.}, ``Summarizing unstructured logs in online
  services,'' \emph{arXiv preprint arXiv:2012.08938}, 2020.

\bibitem{kobayashi2020amulog}
S.~Kobayashi, Y.~Yamashiro, K.~Otomo, and K.~Fukuda, ``amulog: A general log
  analysis framework for diverse template generation methods,'' in \emph{2020
  16th International Conference on Network and Service Management
  (CNSM)}.\hskip 1em plus 0.5em minus 0.4em\relax IEEE, 2020, pp. 1--5.

\bibitem{liu2019logzip}
J.~Liu, J.~Zhu, S.~He, P.~He, Z.~Zheng, and M.~R. Lyu, ``Logzip: Extracting
  hidden structures via iterative clustering for execution log compression,''
  in \emph{The 34th IEEE/ACM International Conference on Automated Software
  Engineering}, 2019.

\bibitem{he2016evaluation}
P.~He, J.~Zhu, S.~He, J.~Li, and M.~R. Lyu, ``An evaluation study on log
  parsing and its use in log mining,'' in \emph{2016 46th annual IEEE/IFIP
  international conference on dependable systems and networks (DSN)}.\hskip 1em
  plus 0.5em minus 0.4em\relax IEEE, 2016, pp. 654--661.

\bibitem{landauer2020system}
M.~Landauer, F.~Skopik, M.~Wurzenberger, and A.~Rauber, ``System log clustering
  approaches for cyber security applications: A survey,'' \emph{Computers \&
  Security}, vol.~92, p. 101739, 2020.

\bibitem{svacina2020vulnerability}
J.~Svacina, J.~Raffety, C.~Woodahl, B.~Stone, T.~Cerny, M.~Bures, D.~Shin,
  K.~Frajtak, and P.~Tisnovsky, ``On vulnerability and security log analysis: A
  systematic literature review on recent trends,'' in \emph{Proceedings of the
  International Conference on Research in Adaptive and Convergent Systems},
  2020, pp. 175--180.

\bibitem{he2020survey}
S.~He, P.~He, Z.~Chen, T.~Yang, Y.~Su, and M.~R. Lyu, ``A survey on automated
  log analysis for reliability engineering,'' \emph{arXiv preprint
  arXiv:2009.07237}, 2020.

\bibitem{bhanage2021infrastructure}
D.~A. Bhanage, A.~V. Pawar, and K.~Kotecha, ``It infrastructure anomaly
  detection and failure handling: A systematic literature review focusing on
  datasets, log preprocessing, machine \& deep learning approaches and
  automated tool,'' \emph{IEEE Access}, 2021.

\bibitem{skopik2021online}
F.~Skopik, M.~Landauer, and M.~Wurzenberger, ``Online log data analysis with
  efficient machine learning: A review,'' \emph{IEEE Security \& Privacy},
  no.~01, pp. 2--12, 2021.

\bibitem{zheng2019survey}
L.~Zheng, L.~Tao, and W.~Junchang, ``A survey on event mining for ict network
  infrastructure management,'' \emph{ZTE Communications}, vol.~14, no.~2, pp.
  47--55, 2019.

\bibitem{el2020systematic}
D.~El-Masri, F.~Petrillo, Y.-G. Gu{\'e}h{\'e}neuc, A.~Hamou-Lhadj, and
  A.~Bouziane, ``A systematic literature review on automated log abstraction
  techniques,'' \emph{Information and Software Technology}, vol. 122, p.
  106276, 2020.

\bibitem{zhu2019tools}
J.~Zhu, S.~He, J.~Liu, P.~He, Q.~Xie, Z.~Zheng, and M.~R. Lyu, ``Tools and
  benchmarks for automated log parsing,'' in \emph{2019 IEEE/ACM 41st
  International Conference on Software Engineering: Software Engineering in
  Practice (ICSE-SEIP)}.\hskip 1em plus 0.5em minus 0.4em\relax IEEE, 2019, pp.
  121--130.

\bibitem{copstein2021log}
R.~Copstein, J.~Schwartzentruber, N.~Zincir-Heywood, and M.~Heywood, ``Log
  abstraction for information security: Heuristics and reproducibility,'' in
  \emph{The 16th International Conference on Availability, Reliability and
  Security}, 2021, pp. 1--10.

\bibitem{he2020loghub}
S.~He, J.~Zhu, P.~He, and M.~R. Lyu, ``Loghub: a large collection of system log
  datasets towards automated log analytics,'' \emph{arXiv preprint
  arXiv:2008.06448}, 2020.

\bibitem{astekin2019incremental}
M.~Astekin, S.~{\"O}zcan, and H.~S{\"o}zer, ``Incremental analysis of
  large-scale system logs for anomaly detection,'' in \emph{2019 IEEE
  International Conference on Big Data (Big Data)}.\hskip 1em plus 0.5em minus
  0.4em\relax IEEE, 2019, pp. 2119--2127.

\bibitem{crnic2011introduction}
J.~Crnic, ``Introduction to modern information retrieval,'' \emph{Library
  Management}, 2011.

\bibitem{tan2016introduction}
P.-N. Tan, M.~Steinbach, and V.~Kumar, \emph{Introduction to data
  mining}.\hskip 1em plus 0.5em minus 0.4em\relax Pearson Education India,
  2016.

\bibitem{ristad1998learning}
E.~S. Ristad and P.~N. Yianilos, ``Learning string-edit distance,'' \emph{IEEE
  Transactions on Pattern Analysis and Machine Intelligence}, vol.~20, no.~5,
  pp. 522--532, 1998.

\bibitem{xiao2020lpv}
T.~Xiao, Z.~Quan, Z.-J. Wang, K.~Zhao, and X.~Liao, ``Lpv: A log parser based
  on vectorization for offline and online log parsing,'' in \emph{2020 IEEE
  International Conference on Data Mining (ICDM)}.\hskip 1em plus 0.5em minus
  0.4em\relax IEEE, 2020, pp. 1346--1351.

\bibitem{yang2019online}
R.~Yang, D.~Qu, Y.~Qian, Y.~Dai, and S.~Zhu, ``An online log template
  extraction method based on hierarchical clustering,'' \emph{EURASIP Journal
  on Wireless Communications and Networking}, vol. 2019, no.~1, pp. 1--12,
  2019.

\bibitem{coustie2020meting}
O.~Cousti{\'e}, J.~Mothe, O.~Teste, and X.~Baril, ``Meting: A robust log parser
  based on frequent n-gram mining,'' in \emph{2020 IEEE International
  Conference on Web Services (ICWS)}.\hskip 1em plus 0.5em minus 0.4em\relax
  IEEE, 2020, pp. 84--88.

\bibitem{gainaru2011event}
A.~Gainaru, F.~Cappello, S.~Trausan-Matu, and B.~Kramer, ``Event log mining
  tool for large scale hpc systems,'' in \emph{European Conference on Parallel
  Processing}.\hskip 1em plus 0.5em minus 0.4em\relax Springer, 2011, pp.
  52--64.

\bibitem{kimura2014spatio}
T.~Kimura, K.~Ishibashi, T.~Mori, H.~Sawada, T.~Toyono, K.~Nishimatsu,
  A.~Watanabe, A.~Shimoda, and K.~Shiomoto, ``Spatio-temporal factorization of
  log data for understanding network events,'' in \emph{IEEE INFOCOM 2014-IEEE
  Conference on Computer Communications}.\hskip 1em plus 0.5em minus
  0.4em\relax IEEE, 2014, pp. 610--618.

\bibitem{ya2015automatic}
J.~Ya, T.~Liu, H.~Zhang, J.~Shi, and L.~Guo, ``An automatic approach to extract
  the formats of network and security log messages,'' in \emph{MILCOM 2015-2015
  IEEE Military Communications Conference}.\hskip 1em plus 0.5em minus
  0.4em\relax IEEE, 2015, pp. 1542--1547.

\bibitem{pokharel2019hybrid}
P.~Pokharel, R.~Pokhrel, and B.~Joshi, ``A hybrid approach for log signature
  generation,'' \emph{Applied Computing and Informatics}, 2019.

\bibitem{zou2016uilog}
D.-Q. Zou, H.~Qin, and H.~Jin, ``Uilog: Improving log-based fault diagnosis by
  log analysis,'' \emph{Journal of computer science and technology}, vol.~31,
  no.~5, pp. 1038--1052, 2016.

\bibitem{ning20141hlaer}
X.~Ning, G.~Jiang, H.~Chen, and K.~Yoshihira, ``{HLAer: A} system for
  heterogeneous log analysis,'' 2014.

\bibitem{shima2016length}
K.~Shima, ``Length matters: Clustering system log messages using length of
  words,'' \emph{arXiv preprint arXiv:1611.03213}, 2016.

\bibitem{guo2018event}
S.~Guo, Z.~Liu, W.~Chen, and T.~Li, ``Event extraction from streaming system
  logs,'' in \emph{International Conference on Information Science and
  Applications}.\hskip 1em plus 0.5em minus 0.4em\relax Springer, 2018, pp.
  465--474.

\bibitem{zhao2018improvement}
Y.~Zhao, X.~Wang, H.~Xiao, and X.~Chi, ``Improvement of the log pattern
  extracting algorithm using text similarity,'' in \emph{2018 IEEE
  International Parallel and Distributed Processing Symposium Workshops
  (IPDPSW)}.\hskip 1em plus 0.5em minus 0.4em\relax IEEE, 2018, pp. 507--514.

\bibitem{aharon2009one}
M.~Aharon, G.~Barash, I.~Cohen, and E.~Mordechai, ``One graph is worth a
  thousand logs: Uncovering hidden structures in massive system event logs,''
  in \emph{Joint European Conference on Machine Learning and Knowledge
  Discovery in Databases}.\hskip 1em plus 0.5em minus 0.4em\relax Springer,
  2009, pp. 227--243.

\bibitem{zhong2018flp}
Y.~Zhong, Y.~Guo, and C.~Liu, ``{FLP: A} feature-based method for log
  parsing,'' \emph{Electronics Letters}, vol.~54, no.~23, pp. 1334--1336, 2018.

\bibitem{joshi2014intelligent}
B.~Joshi, U.~Bista, and M.~Ghimire, ``Intelligent clustering scheme for log
  data streams,'' in \emph{International Conference on Intelligent Text
  Processing and Computational Linguistics}.\hskip 1em plus 0.5em minus
  0.4em\relax Springer, 2014, pp. 454--465.

\bibitem{chunyong2020log}
Z.~Chunyong and X.~Meng, ``Log parser with one-to-one markup,'' in \emph{2020
  3rd International Conference on Information and Computer Technologies
  (ICICT)}.\hskip 1em plus 0.5em minus 0.4em\relax IEEE, 2020, pp. 251--257.

\bibitem{tang2011logsig}
L.~Tang, T.~Li, and C.-S. Perng, ``Logsig: Generating system events from raw
  textual logs,'' in \emph{Proceedings of the 20th ACM international conference
  on Information and knowledge management}, 2011, pp. 785--794.

\bibitem{studiawan2020automatic}
H.~Studiawan, F.~Sohel, and C.~Payne, ``Automatic event log abstraction to
  support forensic investigation,'' in \emph{Proceedings of the Australasian
  Computer Science Week Multiconference}, 2020, pp. 1--9.

\bibitem{vaarandi2003data}
R.~Vaarandi, ``A data clustering algorithm for mining patterns from event
  logs,'' in \emph{Proceedings of the 3rd IEEE Workshop on IP Operations \&
  Management (IPOM 2003)(IEEE Cat. No. 03EX764)}.\hskip 1em plus 0.5em minus
  0.4em\relax IEEE, 2003, pp. 119--126.

\bibitem{vaarandi2004breadth}
------, ``A breadth-first algorithm for mining frequent patterns from event
  logs,'' in \emph{International Conference on Intelligence in Communication
  Systems}.\hskip 1em plus 0.5em minus 0.4em\relax Springer, 2004, pp.
  293--308.

\bibitem{vaarandi2015logcluster}
R.~Vaarandi and M.~Pihelgas, ``Logcluster-a data clustering and pattern mining
  algorithm for event logs,'' in \emph{2015 11th International conference on
  network and service management (CNSM)}.\hskip 1em plus 0.5em minus
  0.4em\relax IEEE, 2015, pp. 1--7.

\bibitem{nagappan2010abstracting}
M.~Nagappan and M.~A. Vouk, ``Abstracting log lines to log event types for
  mining software system logs,'' in \emph{2010 7th IEEE Working Conference on
  Mining Software Repositories (MSR 2010)}.\hskip 1em plus 0.5em minus
  0.4em\relax IEEE, 2010, pp. 114--117.

\bibitem{tovarvnak2019normalization}
D.~Tovar{\v{n}}{\'a}k and T.~Pitner, ``Normalization of unstructured log data
  into streams of structured event objects,'' in \emph{2019 IFIP/IEEE Symposium
  on Integrated Network and Service Management (IM)}.\hskip 1em plus 0.5em
  minus 0.4em\relax IEEE, 2019, pp. 671--676.

\bibitem{qiu2010happened}
T.~Qiu, Z.~Ge, D.~Pei, J.~Wang, and J.~Xu, ``What happened in my network:
  mining network events from router syslogs,'' in \emph{Proceedings of the 10th
  ACM SIGCOMM conference on Internet measurement}, 2010, pp. 472--484.

\bibitem{li2018dlog}
T.~Li, J.~Ma, and C.~Sun, ``Dlog: diagnosing router events with syslogs for
  anomaly detection,'' \emph{The Journal of Supercomputing}, vol.~74, no.~2,
  pp. 845--867, 2018.

\bibitem{zhang2020efficient}
S.~Zhang, Y.~Liu, W.~Meng, J.~Bu, S.~Yang, Y.~Sun, D.~Pei, J.~Xu, Y.~Zhang,
  L.~Song \emph{et~al.}, ``Efficient and robust syslog parsing for network
  devices in datacenter networks,'' \emph{IEEE Access}, vol.~8, pp.
  30\,245--30\,261, 2020.

\bibitem{zulkernine2013capri}
F.~Zulkernine, P.~Martin, W.~Powley, S.~Soltani, S.~Mankovskii, and
  M.~Addleman, ``Capri: a tool for mining complex line patterns in large log
  data,'' in \emph{Proceedings of the 2nd International Workshop on Big Data,
  Streams and Heterogeneous Source Mining: Algorithms, Systems, Programming
  Models and Applications}, 2013, pp. 47--54.

\bibitem{liu2020web}
X.~Liu, Y.~Zhu, and S.~Ji, ``Web log analysis in genealogy system,'' in
  \emph{2020 IEEE International Conference on Knowledge Graph (ICKG)}.\hskip
  1em plus 0.5em minus 0.4em\relax IEEE, 2020, pp. 536--543.

\bibitem{stearley2004towards}
J.~Stearley, ``Towards informatic analysis of syslogs,'' in \emph{2004 IEEE
  International Conference on Cluster Computing (IEEE Cat. No. 04EX935)}.\hskip
  1em plus 0.5em minus 0.4em\relax IEEE, 2004, pp. 309--318.

\bibitem{du2016spell}
M.~Du and F.~Li, ``Spell: Streaming parsing of system event logs,'' in
  \emph{2016 IEEE 16th International Conference on Data Mining (ICDM)}.\hskip
  1em plus 0.5em minus 0.4em\relax IEEE, 2016, pp. 859--864.

\bibitem{agrawal2019delog}
A.~Agrawal, A.~Dixit, N.~A. Shettar, D.~Kapadia, V.~Agrawal, R.~Gupta, and
  R.~Karlupia, ``Delog: A high-performance privacy preserving log filtering
  framework,'' in \emph{2019 IEEE International Conference on Big Data (Big
  Data)}.\hskip 1em plus 0.5em minus 0.4em\relax IEEE, 2019, pp. 1739--1748.

\bibitem{zhao2018slop}
Z.~Zhao, C.~Wang, and W.~Rao, ``Slop: Towards an efficient and universal
  streaming log parser,'' in \emph{International Conference on Information and
  Communications Security}.\hskip 1em plus 0.5em minus 0.4em\relax Springer,
  2018, pp. 325--341.

\bibitem{agrawal2019logan}
A.~Agrawal, R.~Karlupia, and R.~Gupta, ``Logan: A distributed online log
  parser,'' in \emph{2019 IEEE 35th International Conference on Data
  Engineering (ICDE)}.\hskip 1em plus 0.5em minus 0.4em\relax IEEE, 2019, pp.
  1946--1951.

\bibitem{wang2021ltmatch}
X.~Wang, Y.~Zhao, H.~Xiao, X.~Wang, and X.~Chi, ``Ltmatch: A method to abstract
  pattern from unstructured log,'' \emph{Applied Sciences}, vol.~11, no.~11, p.
  5302, 2021.

\bibitem{he2018directed}
P.~He, J.~Zhu, P.~Xu, Z.~Zheng, and M.~R. Lyu, ``A directed acyclic graph
  approach to online log parsing,'' \emph{arXiv preprint arXiv:1806.04356},
  2018.

\bibitem{wen2020olmpt}
P.~Wen, Z.~Zhang, and B.~Deng, ``Olmpt: Research on online log parsing method
  based on prefix tree,'' in \emph{Proceedings of the 3rd International
  Conference on Information Technologies and Electrical Engineering}, 2020, pp.
  55--59.

\bibitem{vervaet2021ustep}
A.~Vervaet, R.~Chiky, and M.~Callau-Zori, ``Ustep: Unfixed search tree for
  efficient log parsing,'' in \emph{2021 IEEE International Conference on Data
  Mining (ICDM)}.\hskip 1em plus 0.5em minus 0.4em\relax IEEE, 2021, pp.
  659--668.

\bibitem{wurzenberger2019aecid}
M.~Wurzenberger, M.~Landauer, F.~Skopik, and W.~Kastner, ``Aecid-pg: A
  tree-based log parser generator to enable log analysis,'' in \emph{2019
  IFIP/IEEE Symposium on Integrated Network and Service Management (IM)}.\hskip
  1em plus 0.5em minus 0.4em\relax IEEE, 2019, pp. 7--12.

\bibitem{mizutani2013incremental}
M.~Mizutani, ``Incremental mining of system log format,'' in \emph{2013 IEEE
  International Conference on Services Computing}.\hskip 1em plus 0.5em minus
  0.4em\relax IEEE, 2013, pp. 595--602.

\bibitem{aussel2018improving}
N.~Aussel, Y.~Petetin, and S.~Chabridon, ``Improving performances of log mining
  for anomaly prediction through nlp-based log parsing,'' in \emph{2018 IEEE
  26th International Symposium on Modeling, Analysis, and Simulation of
  Computer and Telecommunication Systems (MASCOTS)}.\hskip 1em plus 0.5em minus
  0.4em\relax IEEE, 2018, pp. 237--243.

\bibitem{kobayashi2014towards}
S.~Kobayashi, K.~Fukuda, and H.~Esaki, ``Towards an nlp-based log template
  generation algorithm for system log analysis,'' in \emph{Proceedings of The
  Ninth International Conference on Future Internet Technologies}, 2014, pp.
  1--4.

\bibitem{liu2020fastlogsim}
W.~Liu, X.~Liu, X.~Di, and B.~Cai, ``Fastlogsim: A quick log pattern parser
  scheme based on text similarity,'' in \emph{International Conference on
  Knowledge Science, Engineering and Management}.\hskip 1em plus 0.5em minus
  0.4em\relax Springer, 2020, pp. 211--219.

\bibitem{nedelkoski2020self}
S.~Nedelkoski, J.~Bogatinovski, A.~Acker, J.~Cardoso, and O.~Kao,
  ``Self-supervised log parsing,'' \emph{arXiv preprint arXiv:2003.07905},
  2020.

\bibitem{meng2020logparse}
W.~Meng, Y.~Liu, F.~Zaiter, S.~Zhang, Y.~Chen, Y.~Zhang, Y.~Zhu, E.~Wang,
  R.~Zhang, S.~Tao \emph{et~al.}, ``Logparse: Making log parsing adaptive
  through word classification,'' in \emph{2020 29th International Conference on
  Computer Communications and Networks (ICCCN)}.\hskip 1em plus 0.5em minus
  0.4em\relax IEEE, 2020, pp. 1--9.

\bibitem{thaler2017towards}
S.~Thaler, V.~Menkonvski, and M.~Petkovic, ``Towards a neural language model
  for signature extraction from forensic logs,'' in \emph{2017 5th
  International Symposium on Digital Forensic and Security (ISDFS)}.\hskip 1em
  plus 0.5em minus 0.4em\relax IEEE, 2017, pp. 1--6.

\bibitem{rand2021automatic}
J.~Rand and A.~Miranskyy, ``On automatic parsing of log records,'' in
  \emph{2021 IEEE/ACM 43rd International Conference on Software Engineering:
  New Ideas and Emerging Results (ICSE-NIER)}.\hskip 1em plus 0.5em minus
  0.4em\relax IEEE, 2021, pp. 41--45.

\bibitem{ruecker2021flexparser}
N.~Ruecker and A.~Maier, ``Flexparser--the adaptive log file parser for
  continuous results in a changing world,'' \emph{arXiv preprint
  arXiv:2106.03170}, 2021.

\bibitem{nguyen2021logdtl}
T.~Nguyen, S.~Kobayashi, and K.~Fukuda, ``Logdtl: Network log template
  generation with deep transfer learning,'' in \emph{2021 IFIP/IEEE
  International Symposium on Integrated Network Management (IM)}.\hskip 1em
  plus 0.5em minus 0.4em\relax IEEE, 2021, pp. 848--853.

\bibitem{gao2018navigating}
Y.~Gao, S.~Huang, and A.~Parameswaran, ``Navigating the data lake with
  datamaran: Automatically extracting structure from log datasets,'' in
  \emph{Proceedings of the 2018 International Conference on Management of
  Data}, 2018, pp. 943--958.

\bibitem{chuah2010diagnosing}
E.~Chuah, S.-h. Kuo, P.~Hiew, W.-C. Tjhi, G.~Lee, J.~Hammond, M.~T.
  Michalewicz, T.~Hung, and J.~C. Browne, ``Diagnosing the root-causes of
  failures from cluster log files,'' in \emph{2010 International Conference on
  High Performance Computing}.\hskip 1em plus 0.5em minus 0.4em\relax IEEE,
  2010, pp. 1--10.

\bibitem{zhu2010incremental}
K.~Q. Zhu, K.~Fisher, and D.~Walker, ``Incremental learning of system log
  formats,'' \emph{ACM SIGOPS Operating Systems Review}, vol.~44, no.~1, pp.
  85--90, 2010.

\bibitem{taerat2011baler}
N.~Taerat, J.~Brandt, A.~Gentile, M.~Wong, and C.~Leangsuksun, ``Baler:
  deterministic, lossless log message clustering tool,'' \emph{Computer
  Science-Research and Development}, vol.~26, no. 3-4, p. 285, 2011.

\bibitem{messaoudi2018search}
S.~Messaoudi, A.~Panichella, D.~Bianculli, L.~Briand, and R.~Sasnauskas, ``A
  search-based approach for accurate identification of log message formats,''
  in \emph{Proceedings of the 26th Conference on Program Comprehension}, 2018,
  pp. 167--177.

\bibitem{liu2019lopper}
J.~Liu, Z.~Hou, and Y.~Li, ``Lopper: An efficient method for online log pattern
  mining based on hybrid clustering tree,'' in \emph{International Conference
  on Database and Expert Systems Applications}.\hskip 1em plus 0.5em minus
  0.4em\relax Springer, 2019, pp. 63--78.

\bibitem{zhang2019efficient}
L.~Zhang, X.~Xie, K.~Xie, Z.~Wang, Y.~Lu, and Y.~Zhang, ``An efficient log
  parsing algorithm based on heuristic rules,'' in \emph{International
  Symposium on Advanced Parallel Processing Technologies}.\hskip 1em plus 0.5em
  minus 0.4em\relax Springer, 2019, pp. 123--134.

\bibitem{he2017towards}
P.~He, J.~Zhu, S.~He, J.~Li, and M.~R. Lyu, ``Towards automated log parsing for
  large-scale log data analysis,'' \emph{IEEE Transactions on Dependable and
  Secure Computing}, vol.~15, no.~6, pp. 931--944, 2017.

\bibitem{huang2020paddy}
S.~Huang, Y.~Liu, C.~Fung, R.~He, Y.~Zhao, H.~Yang, and Z.~Luan, ``Paddy: An
  event log parsing approach using dynamic dictionary,'' in \emph{NOMS
  2020-2020 IEEE/IFIP Network Operations and Management Symposium}.\hskip 1em
  plus 0.5em minus 0.4em\relax IEEE, 2020, pp. 1--8.

\bibitem{xu2009detecting}
W.~Xu, L.~Huang, A.~Fox, D.~Patterson, and M.~I. Jordan, ``Detecting
  large-scale system problems by mining console logs,'' in \emph{Proceedings of
  the ACM SIGOPS 22nd symposium on Operating systems principles}, 2009, pp.
  117--132.

\bibitem{tak2016logan}
B.~C. Tak, S.~Tao, L.~Yang, C.~Zhu, and Y.~Ruan, ``Logan: Problem diagnosis in
  the cloud using log-based reference models,'' in \emph{2016 IEEE
  International Conference on Cloud Engineering (IC2E)}.\hskip 1em plus 0.5em
  minus 0.4em\relax IEEE, 2016, pp. 62--67.

\bibitem{zhang2017genlog}
M.~Zhang, Y.~Zhao, and Z.~He, ``Genlog: Accurate log template discovery for
  stripped x86 binaries,'' in \emph{2017 IEEE 41st Annual Computer Software and
  Applications Conference (COMPSAC)}, vol.~1.\hskip 1em plus 0.5em minus
  0.4em\relax IEEE, 2017, pp. 337--346.

\bibitem{zhao2014lprof}
X.~Zhao, Y.~Zhang, D.~Lion, M.~F. Ullah, Y.~Luo, D.~Yuan, and M.~Stumm,
  ``lprof: A non-intrusive request flow profiler for distributed systems,'' in
  \emph{11th $\{$USENIX$\}$ Symposium on Operating Systems Design and
  Implementation ($\{$OSDI$\}$ 14)}, 2014, pp. 629--644.

\bibitem{xu2015comprehensive}
D.~Xu and Y.~Tian, ``A comprehensive survey of clustering algorithms,''
  \emph{Annals of Data Science}, vol.~2, no.~2, pp. 165--193, 2015.

\bibitem{ester1996density}
M.~Ester, H.-P. Kriegel, J.~Sander, X.~Xu \emph{et~al.}, ``A density-based
  algorithm for discovering clusters in large spatial databases with noise.''
  in \emph{Kdd}, vol.~96, no.~34, 1996, pp. 226--231.

\bibitem{blei2003latent}
D.~M. Blei, A.~Y. Ng, and M.~I. Jordan, ``Latent dirichlet allocation,''
  \emph{Journal of machine Learning research}, vol.~3, no. Jan, pp. 993--1022,
  2003.

\bibitem{levenshtein1966binary}
V.~I. Levenshtein, ``Binary codes capable of correcting deletions, insertions,
  and reversals,'' in \emph{Soviet physics doklady}, vol.~10, no.~8, 1966, pp.
  707--710.

\bibitem{ankerst1999optics}
M.~Ankerst, M.~M. Breunig, H.-P. Kriegel, and J.~Sander, ``{OPTICS}: ordering
  points to identify the clustering structure,'' \emph{ACM Sigmod record},
  vol.~28, no.~2, pp. 49--60, 1999.

\bibitem{zhao2016extracting}
Y.~Zhao and H.~Xiao, ``Extracting log patterns from system logs in large,'' in
  \emph{2016 IEEE International Parallel and Distributed Processing Symposium
  Workshops (IPDPSW)}.\hskip 1em plus 0.5em minus 0.4em\relax IEEE, 2016, pp.
  1645--1652.

\bibitem{lloyd1982least}
S.~Lloyd, ``Least squares quantization in pcm,'' \emph{IEEE transactions on
  information theory}, vol.~28, no.~2, pp. 129--137, 1982.

\bibitem{girvan2002community}
M.~Girvan and M.~E. Newman, ``Community structure in social and biological
  networks,'' \emph{Proceedings of the national academy of sciences}, vol.~99,
  no.~12, pp. 7821--7826, 2002.

\bibitem{han2007frequent}
J.~Han, H.~Cheng, D.~Xin, and X.~Yan, ``Frequent pattern mining: current status
  and future directions,'' \emph{Data mining and knowledge discovery}, vol.~15,
  no.~1, pp. 55--86, 2007.

\bibitem{agrawal1994fast}
R.~Agrawal, R.~Srikant \emph{et~al.}, ``Fast algorithms for mining association
  rules,'' in \emph{Proc. 20th int. conf. very large data bases, VLDB}, vol.
  1215.\hskip 1em plus 0.5em minus 0.4em\relax Citeseer, 1994, pp. 487--499.

\bibitem{maier1978complexity}
D.~Maier, ``The complexity of some problems on subsequences and
  supersequences,'' \emph{Journal of the ACM (JACM)}, vol.~25, no.~2, pp.
  322--336, 1978.

\bibitem{du2018spell}
M.~Du and F.~Li, ``Spell: Online streaming parsing of large unstructured system
  logs,'' \emph{IEEE Transactions on Knowledge and Data Engineering}, vol.~31,
  no.~11, pp. 2213--2227, 2018.

\bibitem{he2017drain}
P.~He, J.~Zhu, Z.~Zheng, and M.~R. Lyu, ``Drain: An online log parsing approach
  with fixed depth tree,'' in \emph{2017 IEEE International Conference on Web
  Services (ICWS)}.\hskip 1em plus 0.5em minus 0.4em\relax IEEE, 2017, pp.
  33--40.

\bibitem{wurzenberger2018aecid}
M.~Wurzenberger, F.~Skopik, G.~Settanni, and R.~Fiedler, ``Aecid: A
  self-learning anomaly detection approach based on light-weight log parser
  models.'' 2018.

\bibitem{lafferty2001conditional}
J.~Lafferty, A.~McCallum, and F.~C. Pereira, ``Conditional random fields:
  Probabilistic models for segmenting and labeling sequence data,'' 2001.

\bibitem{salton1988term}
G.~Salton and C.~Buckley, ``Term-weighting approaches in automatic text
  retrieval,'' \emph{Information processing \& management}, vol.~24, no.~5, pp.
  513--523, 1988.

\bibitem{thaler2017unsupervised}
S.~Thaler, V.~Menkovski, and M.~Petkovic, ``Unsupervised signature extraction
  from forensic logs,'' in \emph{Joint European Conference on Machine Learning
  and Knowledge Discovery in Databases}.\hskip 1em plus 0.5em minus 0.4em\relax
  Springer, 2017, pp. 305--316.

\bibitem{hochreiter1997long}
S.~Hochreiter and J.~Schmidhuber, ``Long short-term memory,'' \emph{Neural
  computation}, vol.~9, no.~8, pp. 1735--1780, 1997.

\bibitem{makanju2009clustering}
A.~Makanju, A.~N. Zincir-Heywood, and E.~E. Milios, ``Clustering event logs
  using iterative partitioning,'' in \emph{Proceedings of the 15th ACM SIGKDD
  international conference on Knowledge discovery and data mining}, 2009, pp.
  1255--1264.

\bibitem{jiang2008automated}
Z.~M. Jiang, A.~E. Hassan, G.~Hamann, and P.~Flora, ``An automated approach for
  abstracting execution logs to execution events,'' \emph{Journal of Software
  Maintenance and Evolution: Research and Practice}, vol.~20, no.~4, pp.
  249--267, 2008.

\bibitem{kamiya2002ccfinder}
T.~Kamiya, S.~Kusumoto, and K.~Inoue, ``Ccfinder: a multilinguistic token-based
  code clone detection system for large scale source code,'' \emph{IEEE
  Transactions on Software Engineering}, vol.~28, no.~7, pp. 654--670, 2002.

\bibitem{bushong2020matching}
V.~Bushong, R.~Sanders, J.~Curtis, M.~Du, T.~Cerny, K.~Frajtak, M.~Bures,
  P.~Tisnovsky, and D.~Shin, ``On matching log analysis to source code: A
  systematic mapping study,'' in \emph{Proceedings of the International
  Conference on Research in Adaptive and Convergent Systems}, 2020, pp.
  181--187.

\bibitem{schipper2019tracing}
D.~Schipper, M.~Aniche, and A.~van Deursen, ``Tracing back log data to its log
  statement: from research to practice,'' in \emph{2019 IEEE/ACM 16th
  International Conference on Mining Software Repositories (MSR)}.\hskip 1em
  plus 0.5em minus 0.4em\relax IEEE, 2019, pp. 545--549.

\bibitem{wurzenberger2020creating}
M.~Wurzenberger, G.~Hold, M.~Landauer, F.~Skopik, and W.~Kastner, ``Creating
  character-based templates for log data to enable security event
  classification,'' in \emph{Proceedings of the 15th ACM Asia Conference on
  Computer and Communications Security}, 2020, pp. 141--152.

\bibitem{oliner2007supercomputers}
A.~Oliner and J.~Stearley, ``What supercomputers say: A study of five system
  logs,'' in \emph{37th Annual IEEE/IFIP International Conference on Dependable
  Systems and Networks (DSN'07)}.\hskip 1em plus 0.5em minus 0.4em\relax IEEE,
  2007, pp. 575--584.

\bibitem{xie2019pvalue}
X.~Xie, Z.~Wang, X.~Xiao, L.~Yang, S.~Huang, and T.~Li, ``A pvalue-guided
  anomaly detection approach combining multiple heterogeneous log parser
  algorithms on iiot systems,'' \emph{arXiv preprint arXiv:1907.02765}, 2019.

\bibitem{airflow}
``{Apache Airflow},'' \url{https://airflow.apache.org/}, accessed on March 16,
  2022.

\bibitem{kubeflow}
``Kubeflow,'' \url{https://www.kubeflow.org/}, accessed on March 16, 2022.

\bibitem{mlflow}
``An open source platform for the machine learning lifecycle,''
  \url{https://mlflow.org/}, accessed on March 16, 2022.

\bibitem{blanas2010comparison}
S.~Blanas, J.~M. Patel, V.~Ercegovac, J.~Rao, E.~J. Shekita, and Y.~Tian, ``A
  comparison of join algorithms for log processing in mapreduce,'' in
  \emph{Proceedings of the 2010 ACM SIGMOD International Conference on
  Management of data}, 2010, pp. 975--986.

\bibitem{dean2008mapreduce}
J.~Dean and S.~Ghemawat, ``Mapreduce: simplified data processing on large
  clusters,'' \emph{Communications of the ACM}, vol.~51, no.~1, pp. 107--113,
  2008.

\bibitem{zaharia2010spark}
M.~Zaharia, M.~Chowdhury, M.~J. Franklin, S.~Shenker, I.~Stoica \emph{et~al.},
  ``Spark: Cluster computing with working sets.'' \emph{HotCloud}, vol.~10, no.
  10-10, p.~95, 2010.

\bibitem{ren2019parallel}
X.~Ren, L.~Zhang, K.~Xie, and Q.~Dong, ``A parallel approach of weighted edit
  distance calculation for log parsing,'' in \emph{2019 IEEE 2nd International
  Conference on Computer and Communication Engineering Technology
  (CCET)}.\hskip 1em plus 0.5em minus 0.4em\relax IEEE, 2019, pp. 101--104.

\end{thebibliography}
%
%
%

%

\begin{IEEEbiography}[{\includegraphics[width=1in,height=1.25in,clip,keepaspectratio]{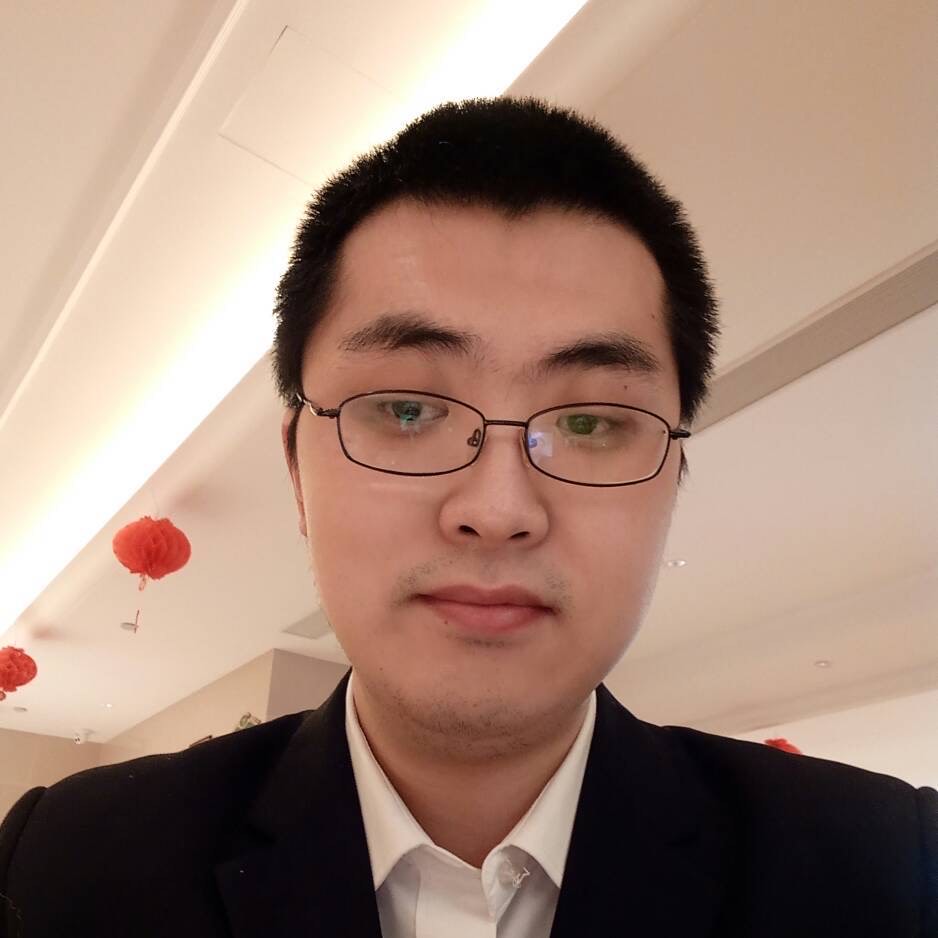}}]{Tianzhu Zhang} (Member, IEEE)
is a Research Engineer at Nokia Bell Labs and an Associate Member of the Laboratory of Information, Networking and Communication Sciences (LINCS). He received his B.S. degree from Huazhong University of Science and Technology, Wuhan, China, in 2012. 
Afterward, he received the M.S. degree in 2014, and the Ph.D. degree in 2017, both from Politecnico di Torino, Turin, Italy. From 2017 to 2019, he was a postdoc researcher at Telecom Paris and LINCS, under a grant from Cisco Systems. He joined Nokia Bell Labs in August 2020. His research interests include Computer Networks and Artificial Intelligence.
\end{IEEEbiography}

\begin{IEEEbiography}[{\includegraphics[width=1in,height=1.25in,clip,keepaspectratio]{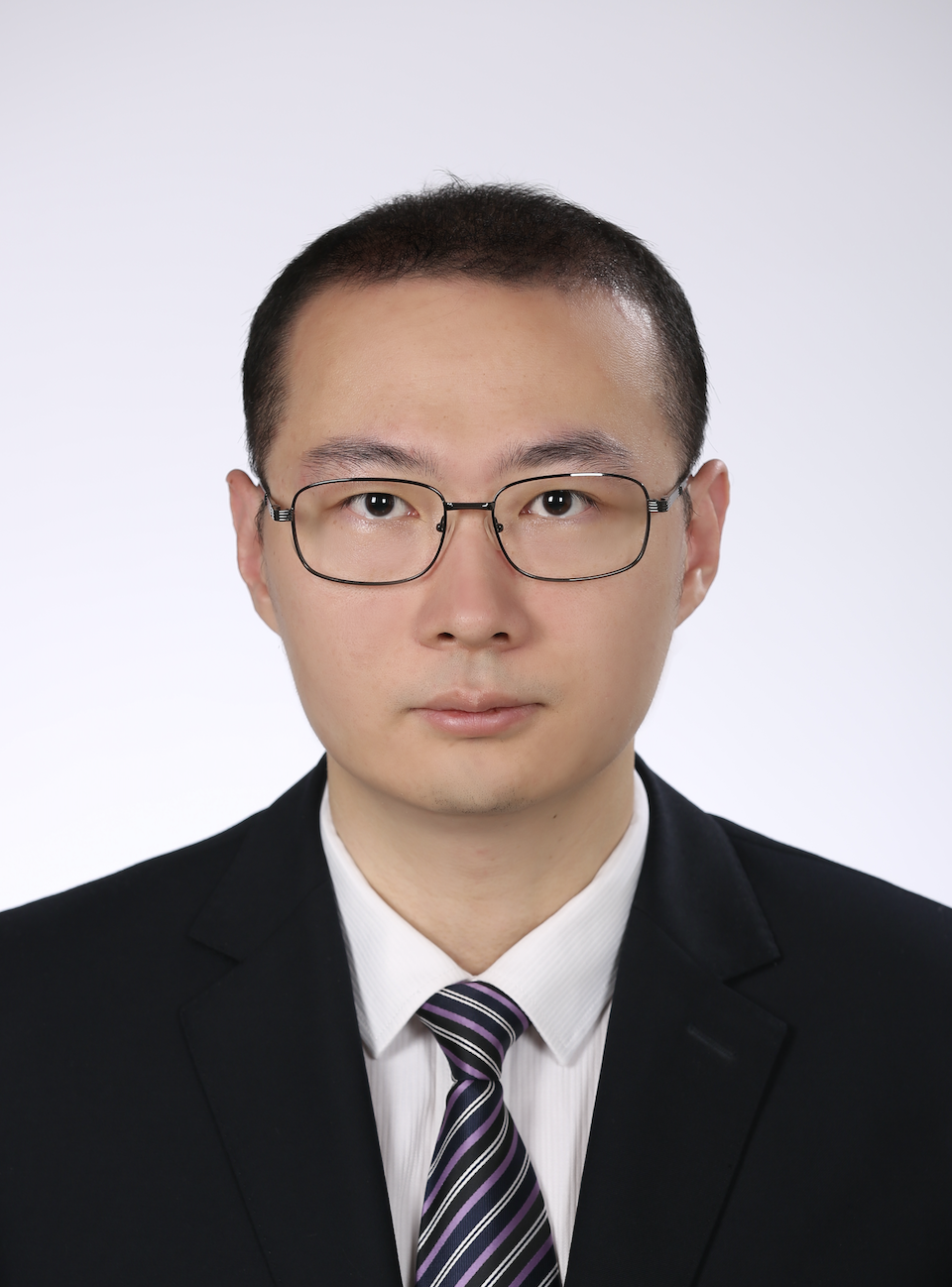}}]{Han Qiu}
received the B.E. degree from the Beijing University of Posts and Telecommunications, Beijing, China, in 2011, the M.S. degree from Telecom-ParisTech (Institute Eurecom), Biot, France, in 2013, and the Ph.D. degree in computer science from the Department of Networks and Computer Science, Telecom-ParisTech, Paris, France, in 2017. He worked as a postdoc and a research engineer with Telecom Paris and LINCS Lab from 2017 to 2020. Currently, he is an assistant professor at Institute for Network Sciences and Cyberspace, Tsinghua University, Beijing, China. His research interests include AI security, big data security, and the security of intelligent transportation systems.
\end{IEEEbiography}

\begin{IEEEbiography}[{\includegraphics[width=1in,height=1.25in,clip,keepaspectratio]{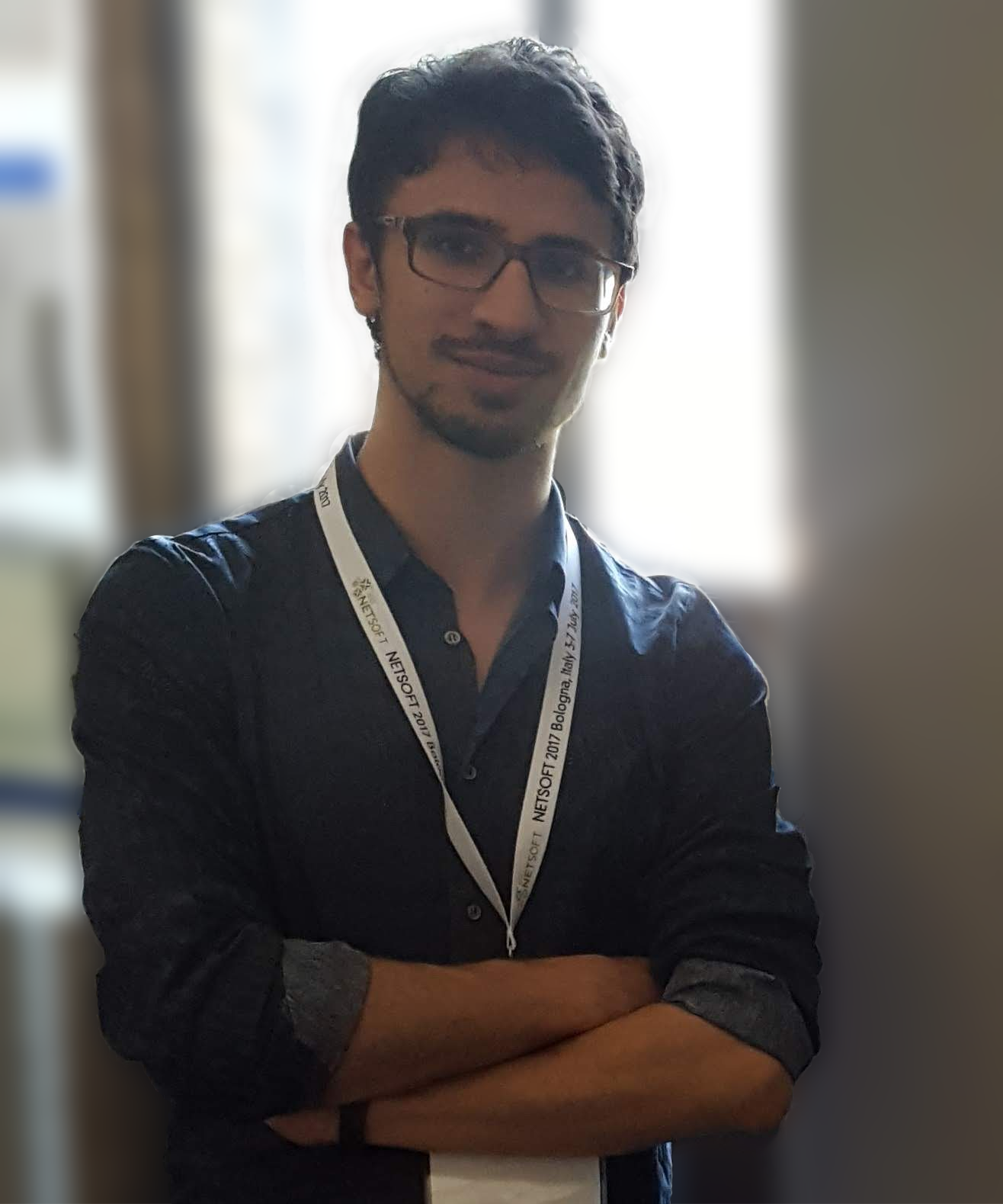}}]{Gabriele Castellano}
is a Postdoctoral researcher at Nokia Bell Labs and Inria. He obtained  his Ph.D. degree at Politecnico  di  Torino, Italy, where he received his Master’s degree in Computer Engineering in 2016. During his Ph.D. career, he  spent  five  months as visiting  student  at  Saint  Louis  University (Saint Louis) and three months at Telefonica Research (Barcelona).  His  research  interests  include  service virtualization, resource orchestration, distributed algorithms, and artificial intelligence.
\end{IEEEbiography}

\begin{IEEEbiography}[{\includegraphics[width=1in,height=1.25in,clip,keepaspectratio]{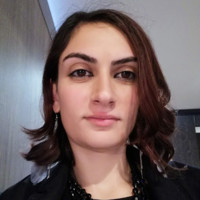}}]{Myriana Rifai} received the B.S. in Information Technology from Global University, Lebanon in 2012, M.S in Ubiquitous Computing from Polytech Nice,France in 2014 and PhD. degree in Next Generation SDN Networks in 2017 from UCA, France. She then worked for a year as a consultant for Orange. She joined Nokia Bell labs in 2018 as a research engineer. Her main research interest include networking architecture, protocols and software. She is a Member of the Laboratory of Information, Networking and Communication Sciences (LINCS), France. 
\end{IEEEbiography}

\begin{IEEEbiography}[{\includegraphics[width=1in,height=1.25in,clip,keepaspectratio]{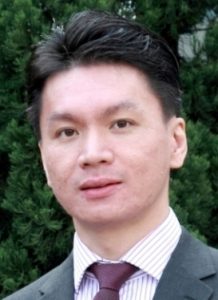}}]{Chung Shue Chen}(Senior Member, IEEE)
received the B.Eng., M.Phil., and Ph.D. degrees in information engineering from the Chinese University of Hong Kong (CUHK), Hong Kong, in 1999, 2001, and 2005, respectively. He is a DMTS at Nokia Bell Labs. Prior to joining Bell Labs, he worked at INRIA, in the research group on Network Theory and Communications (TREC, INRIA-ENS). He was an Assistant Professor with CUHK. He was an ERCIM Alain Bensoussan Fellow with the Norwegian University of Science and Technology (NTNU), Norway, and the National Centre for Mathematics and Computer Science (CWI), The Netherlands. He worked at CNRS in Lorraine on Real-Time and Embedded Systems. His research interests include wireless networks, communications, optimization, machine learning, 5G/6G, IoT, and intelligent systems. Dr. Chen was a recipient of the Sir Edward Youde Memorial Fellowship and the ERCIM Fellowship. He was a TPC in international conferences, including IEEE ICC, Globecom, WCNC, PIMRC, VTC, CCNC, and WiOpt (a TPC Vice Chair). He is an Editor of the Transactions on Emerging Telecommunications Technologies (ETT). He is a Permanent Member of the Laboratory of Information, Networking and Communication Sciences (LINCS), France. 
\end{IEEEbiography}

\begin{IEEEbiography}[{\includegraphics[width=1in,height=1.25in,clip,keepaspectratio]{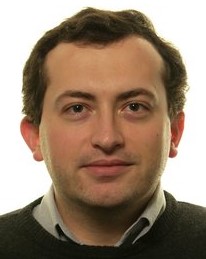}}]{Fabio Pianese}(Member, IEEE)
leads the {\it Augmented Machine Interaction} team at Nokia Bell Labs in Paris-Saclay, France. He holds a Ph.D. degree in Computer Science and received his B.S. in EE from Politecnico di Torino and M.S. in 2004 (CS) \& 2008 (EE) from University of Nice - Sophia Antipolis and Politecnico di Torino, respectively. Prior to joining Bell Labs in 2009, he was with France Telecom R\&D (presently Orange Labs). Dr. Pianese is the author of more than 25 papers published in peer-reviewed international journals and conferences (one best paper award) and the co-inventor of 8 granted patents. His main research interests are in networking, distributed systems, and applied machine learning. He is Associate Member of the Laboratory of Information, Networking and Communication Sciences (LINCS), France.
\end{IEEEbiography}

%
%



\enlargethispage{-2in}

\end{document}